%% file: PIP_100_Rocha_v2.tex
\documentclass[longauth,twocolumn,traditabstract]{aa}
\usepackage{graphicx,amsmath,amsfonts,amssymb}
\usepackage{txfonts}
\usepackage{url}
\usepackage{color}
\usepackage{fancyvrb} 
\usepackage{morefloats}
\usepackage{natbib}
\bibpunct{(}{)}{;}{a}{}{,} 
\usepackage{pdfcolmk}
\usepackage[breaklinks,colorlinks,citecolor=blue]{hyperref}

\providecommand{\Planck}{\textit{Planck}}
\providecommand{\planck}{\Planck}
\def\WMAP{\textit{WMAP}}
\def\RECFAST{{\tt RECFAST}}

\newcommand{\WP}{WP}
\newcommand{\highL}{highL}

\newcommand{\HighL}{\highL}
\newcommand{\HST}{{\it HST\/}}
\newcommand{\hatn}{\vec{\hat{n\,}}}

\input Planck.tex

\def\Mpc{{\rm Mpc}}

\newcommand{\dd}{{\rm d}}
\newcommand{\elec}{{\rm e}}
\newcommand{\proton}{{\rm p}}
\newcommand{\helium}{{\rm He}}
\newcommand{\heliumUn}{{\rm HeI}}
\newcommand{\heliumDeux}{{\rm HeII}}
\newcommand{\hydro}{{\rm H}}
\newcommand{\aem}{\alpha}

\newcommand{\troisj}[6]{\left(\begin{array}{ccc}
      #1 & #2 & #3\\
      #4 & #5 & #6\end{array}\right)}
\newcommand{\prodtroisj}[6]{\troisj{#1}{#2}{#3}{#4}{#5}{#6}\troisj{#1}{#2}{#3}{0}{0}{0}}
\newcommand{\As}{A_{\rm s}}

\newcommand{\ns}{n_{\rm s}}

\newcommand{\yhe}{Y_{\text{p}}}

\newcommand{\mnu}{\sum m_\nu}

\newcommand{\me}{m_{\rm e}}
\newcommand{\mezero}{m_{\rm e0}}

\newcommand{\TT}{$TT$}
\newcommand{\TE}{$TE$}
\newcommand{\EE}{$EE$}

\graphicspath{{Figures/alpha/}}

\begin{document}
\title{\textit{Planck\/} intermediate results.  XXIV.\\
Constraints on variation of fundamental constants}
\titlerunning{Fundamental constants}
\input{PIP_100_Rocha_authors_and_institutes.tex}
\authorrunning{\Planck\ Collaboration}  


\abstract{Any variation of the fundamental physical constants, and more
particularly of the fine structure constant, $\alpha$, or of the mass of the
electron, $\me$, would affect the recombination history of the Universe and
cause an imprint on the cosmic microwave background angular power
spectra.  We show that the \Planck\ data allow one to improve the constraint
on the time variation of the fine structure constant at redshift $z\sim 10^3$
by about a factor of 5 compared to \WMAP\ data, as well as to break
the degeneracy with the Hubble constant, $H_0$.  In addition to $\alpha$, we
can set a constraint on the variation of the mass of the electron, $m_{\rm e}$,
and on the simultaneous variation of the two constants.  We examine in detail
the degeneracies between fundamental constants and
the cosmological parameters, in order to compare the
limits obtained from \Planck\ and \WMAP\ and to determine the constraining
power gained by including other cosmological probes.
We conclude that independent time variations of the fine structure constant
and of the mass of the electron are constrained by \Planck\ to
${\Delta\alpha}/{\alpha}= (3.6\pm 3.7)\times10^{-3}$ and
${\Delta m_{\rm e}}/{m_{\rm e}}= (4 \pm 11)\times10^{-3}$ at the 68\,\%
confidence level.  We also investigate the possibility of a spatial variation
of the fine structure constant.  The relative amplitude of a dipolar spatial
variation of $\alpha$ (corresponding to a gradient across our Hubble volume)
is constrained to be $\delta\alpha/\alpha=(-2.4\pm 3.7)\times 10^{-2}$.}

\keywords{Cosmology: observations -- cosmic background radiation --
cosmological parameters -- Atomic data}

\maketitle

\section{Introduction}\label{sec1}

The construction of the standard cosmological model, known as $\Lambda$CDM, relies on the assumption that General Relativity offers a good description of gravity on astrophysical scales. The existence of a dark sector, including both dark matter and dark energy, has motivated an important activity in testing General Relativity on astrophysical scales~\citep[see,][for reviews]{ugrg,jain13} in order to better quantify and extend its domain of validity. As part of this programme, testing for the constancy of fundamental constants offers a unique window on the Einstein equivalence principle~(see \citealt{dicke64} and~\citealt{jpu-revue,jpu-llr} for reviews) and thus on General Relativity, as well as on other metric theories of gravity. 

Various systems, spanning different time scales and physical environments, are now used to set constraints on a possible variation of the fundamental constants. This includes the comparison of atomic clocks in the laboratory at $z=0$~\citep{rosenband,cingoz,peik,bize}, the Oklo phenomenon at a redshift $z\simeq0.14$~\citep{kuroda,shly,dady,fujii,gould}, meteorite dating~\citep{wilk,dyson,fujii2,kids}, quasar absorption spectra observations~\citep{savedoff,webb01,vlt1,vlt2}, molecular absorption lines~\citep{carilli00,kanekar05}, clusters of galaxies \citep{galli13}, population~III stars~\citep{livio,sylvia,sylvia2}, cosmic microwave background (CMB) anisotropies (see below for details), and big bang nucleosynthesis at $z\sim10^8$~\citep{berg,muller,cnouv,cduv}.  Several reviews
\citep[e.g.,][]{jpu-revue,jpu-llr,Martins2003,Flambaum2007} provide detailed
discussions of the
various methods and constraints. The claim that the fine structure constant may have been smaller in the past~\citep{webb01}, drawn from observations of specific quasar absorption spectra by the Keck telescope, has not been confirmed by independent studies. All VLT observations of quasar absorption spectra~\citep{vlt1,vlt2} and observations of molecular absorption lines~\citep{kanekar05}, are compatible with no variation \citep[see also][and references therein]{bonifacio13}.
Despite the lack of definitive empirical evidence of any
constants having a different value in the past, there are many theoretical
ideas to motivate the continued search for variations in cosmological data.

The observation of CMB temperature anisotropies has extensively been used to constrain the variation of fundamental constants at a redshift $z\simeq 10^3$. Earlier analyses relied on the data from BOOMERanG and MAXIMA~\citep{cmb-kap,cmb-avelino00,cmb-landau01} and on the \WMAP\ data combined with other cosmological probes~\citep{cmb-rocha,cmb-rocha1,cmb-martins,cmb2-2006,cmb3-2007,scoccola,wmap-alpha,menegoni1,menegoni2}, first focusing on the effect of a variation of the fine structure constant only~\citep{cmb-rocha,cmb-rocha1,cmb3-2007,wmap-alpha,menegoni1,naka00,scoccola3} and then also including a possible variation of the electron mass~\citep{cmb2-2006,scoccola,scoccola2,naka00,scoccola3,scoccola2013}. Such studies typically indicate that, on cosmological scales, these two parameters are constant at the percent level. The previous constraints and the set of cosmological parameters used to perform the analyses are summarized in Table~\ref{tab-cmb}.


\begin{table*}[htbp]
\begingroup
\newdimen\tblskip \tblskip=5pt
\caption{Summary of the $68\,\%$ confidence limit (CL) bounds on the variation
of fundamental constants obtained from previous analyses of cosmological data and
more particularly of CMB data.  All assume $\Omega_K=0$, i.e., no spatial
curvature.  Here $\Omega_{\rm b}$, $\Omega_{\rm c}$, and $\Omega_\Lambda$
refer to the density parameters for the baryons, cold dark matter, and
cosmological constant, respectively.  In addition $h$ is the reduced Hubble
constant, $\tau$ the reionization optical depth, $\Theta_\ast$ the angular
size of the sound horizon at last scattering, and the spectrum of the initial
curvature perturbation is characterized by an amplitude $\As$, a spectral index
$\ns$ and potentially a running, $dn_{\rm s}/d\ln k$.  For comparison, our new
analysis, based on the \Planck\ data, uses
$(\Omega_{\rm c} h^2,\Omega_{\rm b} h^2,H_0,\tau, \ns,\As,\alpha,m_{\rm e})$.}
\label{tab-cmb}
\vskip -3mm
\footnotesize
\setbox\tablebox=\vbox{
 \newdimen\digitwidth
 \setbox0=\hbox{\rm 0}
 \digitwidth=\wd0
 \catcode`*=\active
 \def*{\kern\digitwidth}
 \newdimen\signwidth
 \setbox0=\hbox{+}
 \signwidth=\wd0
 \catcode`!=\active
 \def!{\kern\signwidth}
 \halign{\tabskip=0pt#\hfil\tabskip=2em&
 #\hfil\tabskip=2em&
 #\hfil\tabskip=2em&
 #\hfil\tabskip 0pt\cr
\noalign{\doubleline}
Constraint&  Data& Other parameters& Reference\cr
($\aem\times10^2$)& & &\cr
\noalign{\vskip 4pt\hrule\vskip 6pt}
$[-9,2]$& {\it COBE\/}-BOOMERanG-DASI + BBN& BBN with $\aem$ only& \citep{cmb-avelino01}\cr
 & & ($\Omega_{\rm c},\Omega_{\rm b},h, \ns$)&\cr
$[-1.4,2]$& {\it COBE\/}-BOOMERanG-MAXIMA& ($\Omega_{\rm c},\Omega_{\rm b},h, \ns$)& \citep{cmb-landau01}\cr
$[-5,2]$& \WMAP-1& ($\Omega_{\rm c} h^2,\Omega_{\rm b} h^2,\Omega_\Lambda h^2,\tau, \ns,dn_{\rm s}/d\ln k$)& \citep{cmb-rocha}\cr
$[-6,1]$& \WMAP-1& Same as above + $dn_{\rm s}/d\ln k=0$&  \citep{cmb-rocha}\cr
$[-9.7,3.4]$& \WMAP-1& ($\Omega_{\rm c},\Omega_{\rm b},h, \ns,\tau,m_{\rm e}$)& \citep{cmb2-2006}\cr
$[-4.2,2.6]$& \WMAP-1 + \HST&  ($\Omega_{\rm c},\Omega_{\rm b},h, \ns,\tau,m_{\rm e}$)&  \citep{cmb2-2006}\cr
$[-3.9,1.0]$& \WMAP-3 (\TT,\TE,\EE) + \HST& ($\Omega_{\rm c},\Omega_{\rm b},h, \ns,z_{\rm re},\As$)& \citep{cmb3-2007}\cr
$[-1.2,1.8]$& \WMAP-5 + ACBAR + CBI + 2df& ($\Omega_{\rm c} h^2,\Omega_{\rm b} h^2,\Theta_\ast,\tau, \ns,\As,m_{\rm e}$)& \citep{scoccola}\cr
$[-1.9,1.7]$& \WMAP-5 + ACBAR + CBI + 2df& ($\Omega_{\rm c} h^2,\Omega_{\rm b} h^2,\Theta_\ast,\tau, \ns,\As,m_{\rm e}$)& \citep{scoccola2}\cr
$[-5.0,4.2]$& \WMAP-5 + \HST&  ($\Omega_{\rm c} h^2,\Omega_{\rm b} h^2,h,\tau, \ns,\As$)& \citep{wmap-alpha}\cr
$[-4.3,3.8]$& \WMAP-5 + ACBAR + QUAD + BICEP& ($\Omega_{\rm c} h^2,\Omega_{\rm b} h^2,h,\tau, \ns$)& \citep{menegoni1}\cr
$[-1.3,1.5]$& \WMAP-5 + ACBAR + QUAD + BICEP + \HST&  ($\Omega_{\rm c} h^2,\Omega_{\rm b} h^2,h,\tau, \ns$)& \citep{menegoni1}\cr
$[-0.83,0.18]$& \WMAP-5 (\TT,\TE,\EE)& ($\Omega_{\rm c} h^2,\Omega_{\rm b} h^2,h,\tau, \ns,\As,m_{\rm e},\mu$)& \citep{naka00}\cr
$[-2.5,-0.3]$& \WMAP-7 + $H_0$ + SDSS&  ($\Omega_{\rm c} h^2,\Omega_{\rm b} h^2,\Theta_\ast,\tau, \ns,\As,m_{\rm e})$& \citep{scoccola3}\cr
\noalign{\vskip 4pt\hrule\vskip 6pt}
}}
\endPlancktablewide
\endgroup
\end{table*}


From a physical point of view, the effects of a variation of the fundamental constants on the CMB are mostly caused by the modifications to the recombination process. As first discussed in~\cite{cmb-han}, a variation of the fine structure constant can be implemented in the \RECFAST\ code \citep{recfast}, but one should also include the variation of the mass of the electron~\citep{cmb-bat},
or equivalently the electron-to-proton mass ratio.  In this new study, we implement a possible variation of the fundamental constants in a modified version of \RECFAST, as used in earlier works~\citep{cmb-avelino01,cmb-rocha,cmb-rocha1,scoccola}. 

The shape of the CMB power spectrum depends on the cosmological parameters and, as we shall see in Sect.~\ref{sec2}, there exist degeneracies between the fundamental constants and these cosmological parameters. Previous studies (see e.g., Table~\ref{tab-cmb} for concrete examples) have constrained at the percent level
the variation of the fundamental constants between the time of recombination and today.  As we shall show, the resolution of the \Planck\ data allows us to break some of the degeneracies and to improve these constraints. They can be further enhanced by combining CMB data with other cosmological probes (see Table~\ref{tab-cmb}).

The goal of the present study is to constrain the variation of the fundamental constants using the recent \Planck\  \footnote{\Planck\
(http://www.esa.int/\Planck) is a project of the European Space Agency (ESA) with instruments provided by two scientific consortia funded by ESA member states (in particular the lead countries, France and Italy) with contributions from NASA (USA), and telescope reflectors provided in a collaboration between ESA and a scientific consortium led and funded by Denmark.} data \citep{planck2013-p01}. We will start by considering only a time variation of the fine structure constant. This analysis extends the one presented in \cite{planck2013-p11} by showing the effect of combining the \planck\ data with a number of additional data sets. We then discuss the time variation of the mass of the electron and the simultaneous time variation of these two constants. Furthermore, we also address the possibility of a spatial dependence of the fine structure constant.

The article is organized as follows. We start by discussing the phenomenological implementation of the physics that depends on the fundamental constants in Sect.~\ref{sec0}. Section~\ref{sec2} focuses on a pure (spatially homogeneous) time variation of the fine structure constant or of the mass of the electron, assuming that all other constants remain strictly constant. Section~\ref{sec2b} considers the case in which both constants are allowed to vary.  Section~\ref{sec3} investigates the possibility of a spatial variation, focusing on a dipolar modulation of the fine structure constant on the sky. It is first shown that such a variation induces mode coupling between the $a_{\ell m}$, which can then be constrained. We summarize the constraints in Sect.~\ref{sec4} and also discuss extensions and limitations of our approach. 
Technical details of the effects on the recombination process
are summarized in Appendix~\ref{appA}, while Appendix~\ref{effects_alphame} provides an in-depth description of the difference between a variation of $\alpha$ and a variation of $m_{\rm e}$.

It is worth making some preparatory statements about notation.
We generically represent the set of constants by $c_p$. We denote the CMB
temperature anisotropy observed in the direction $\hatn$ by either
$\bar\Theta(\hatn)$ or $\Theta(\hatn;c_p)$, the former assuming that it is
evaluated using the current values of the fundamental constants as determined
by laboratory experiments, $\bar\Theta(\hatn)\equiv\Theta(\hatn;c_{p,0})$.
These can both be expanded in terms of spherical harmonics as
\begin{equation}\label{e.sph-dec}
 \bar\Theta(\hatn) = \sum_{\ell m} \bar a _{\ell m} Y_{\ell m}(\hatn),
\end{equation} 
and similarly for $\Theta$, with multipolar coefficients $a_{\ell m}$. We define the angular power spectrum as usual by
\begin{equation}
 (2\ell+1)\bar C_\ell \equiv \sum_{m} \langle \bar a_{\ell m}\bar a^*_{\ell m}\rangle.
\end{equation}
Statistical homogeneity and isotropy imply that $\langle \bar a_{\ell m}\bar a_{\ell' m'}\rangle=\bar C_\ell \delta_{\ell\ell'}\delta_{mm'}$.  However, this property has no general reason to hold, in particular when we consider a possible spatial variation of the fundamental constants (see Sect.~\ref{sec3}).

The physics of recombination is mostly determined by the atomic physics of hydrogen and helium, and in principle depends on the fine structure constant
\begin{equation}
\alpha\equiv\frac{e^2}{4\pi\varepsilon_0\hbar c}, 
\end{equation}
the masses of the electron and of the proton, $m_{\rm e}$ and $m_{\rm p}$, the proton gyromagnetic factor, $g_{\rm p}$, and the gravitational constant, $G$, which enters through the expansion dynamics and the Friedmann equation. As explained in Sect.~\ref{subsec0a}, the effect of a variation of $m_{\rm p}$ on recombination is subdominant compared to a variation of $m_{\rm e}$, while $g_{\rm p}$ enters only into the hyperfine structure, which is not relevant for the recombination process. In this paper, we assume that $G$ is kept fixed, and so one could consider that a variation of $m_{\rm e}$ actually corresponds to a variation of the dimensionless quantity $Gm_{\rm e}^2/\hbar c$ (however, see the discussion in Sect.~\ref{subsec0b}).

\section{Implementation of the variation of fundamental constants}\label{sec0}

The main effect of a variation of the fundamental constants is to induce a modification of the recombination rates that are mostly dependent on the fine structure constant and on the electron mass. These two constants are thus our primary focus and we summarize their impact on the recombination process in Sect.~\ref{subsec0a}. However, assuming that a mass may be dynamical requires us to specify the model further, since we can imagine, e.g., that all the masses are varying while keeping their ratios constant, or that the mass ratios are varying. We briefly discuss these possibilities in Sect.~\ref{subsec0b}.

\subsection{Modification of the recombination dynamics} \label{subsec0a}

CMB angular power spectra strongly depend on the time and width of last scattering, i.e., on when and how photons decoupled from electrons. This information is encoded in the visibility function, which quantifies the probability distribution that a photon last scatters at a certain conformal time $\eta$, and is defined as
\begin{equation}
g(\eta)d\eta=\dot{\tau} e^{-\tau}d\eta,
\end{equation}
where $\tau=\int \dot\tau d\eta$ is the optical depth and $\dot{\tau}$ is the Thomson scattering rate:
\begin{equation}
 \dot\tau = n_\elec c \sigma_{\rm T} a.
\end{equation}
Here $n_\elec$ is the number density of free electrons, $c$ the speed of light, $a$ the scale factor, and $\sigma_{\rm T}$ the Thomson scattering cross-section.  A change in the fundamental constants modifies the Thompson scattering rate $\dot\tau$ (and thus the visibility function) in a number of different ways. 
First, the Thomson scattering cross-section is given by
\begin{equation}
 \sigma_{\rm T} = \frac{8\pi}{3}\frac{\hbar^2}{m_\elec^2c^2}\alpha^2.
 \label{thomson}
\end{equation}
A variation of the fundamental constants implies that
\begin{equation}
 \sigma_{\rm T}=\sigma_{\rm T0}\left(\frac{\alpha}{\alpha_0}\right)^2 \left(\frac{m_\elec}{m_{\elec0}}\right)^{-2}.
 \label{sigmat} 
\end{equation}

Second, a change in fundamental constants modifies the equations determining the evolution with time of the free electron fraction $x_\elec$. The free electron fraction can be determined as the sum of the free proton fraction $x_{\rm p}$ and of the singly ionized helium fraction $x_{\rm He}$, whose evolution can be calculated by solving the system of differential equations summarized in Appendix~\ref{appA}, together with the evolution of the matter temperature $T_{\rm M}$ (see, e.g., \citealt{recfast}). These equations involve a series of quantities that have to be modified when assuming a variation of the fundamental constants. We will describe the most important dependences below.
 
Since the energy levels scale as $B_{\rm i} \propto \alpha^2 \me$, it follows that
the transition frequencies behave as
\begin{equation}
  \nu_{\rm i}= \nu_{\rm i0}\left(\frac{\alpha}{\alpha_0}\right)^2 \left(\frac{m_\elec}{m_{\elec0}}\right),
\label{eq:energylevels}
\end{equation}
where any quantity with a subscript ``$0$'' has to be understood to be evaluated with its standard value, as known experimentally (see, e.g.,~\citealt{codatalist}). Note that it is in fact the reduced mass $m_{\rm r}=m_\elec M/(m_\elec + M)$, $M$ being the mass of the nucleus, which should appear in the transition frequencies, but clearly at lowest order $\delta m_{\rm r}/m_{\rm r} = \delta m_\elec/m_\elec(1+m_\elec/M)\simeq\delta m_\elec/m_\elec$. This explains why the dependency is in $m_\elec$ and not $m_{\rm r}$. Note also that the gyromagnetic factor does not enter in the dipolar transitions or in the energies, so that it does not enter the discussion at all.

The photoionization cross-sections all scale as $\alpha^{-1}m_\elec^{-2}$, so that
\begin{equation}
 \sigma_{n}=\sigma_{n0}\left(\frac{\alpha}{\alpha_0}\right)^{-1} \left(\frac{m_\elec}{m_{\elec0}}\right)^{-2}.
\end{equation}
The recombination coefficient ${\widetilde \alpha}_{\rm i}$ scales as\footnote{\label{foot:scaling}
The scaling of the effective recombination coefficient with the fine structure
constant is not clearly established; throughout the paper we assume it to be
cubic, following the scaling for the rates between individual atomic levels.
However, there have been other suggested scalings published,
e.g., \cite{ali} adopted $\alpha^5$, while \cite{cmb-kap} suggested a scaling
motivated by the dependence of the effective recombination rate on temperature,
yielding $\alpha^{3.4}$.
We checked however that this difference in scaling has a subdominant impact on the spectra and that the results are essentially unmodified,
whichever scaling we choose.} $\alpha^3m_\elec^{-3/2}$, so that
\begin{equation}
 {\widetilde \alpha}_{\rm i}={\widetilde \alpha}_{\rm i0}\left(\frac{\alpha}{\alpha_0}\right)^{3} \left(\frac{m_\elec}{m_{\elec0}}\right)^{-3/2}.
\end{equation}
The ionization coefficient can then be defined as $\beta_{\rm i}={\widetilde \alpha}_{\rm i}(2\pi m_\elec k T_{\rm M}/h^2)^{3/2}\exp(-h\nu_{\rm i}/kT_{\rm M})$, so that it scales as
\begin{equation}
  \beta_{\rm i}=\beta_{\rm i0}\left(\frac{\alpha}{\alpha_0}\right)^{3} \exp(-h\nu_{\rm i}/kT_{\rm M}),
\end{equation}
where the frequency $\nu_{\rm i}$ depends on the constants as in Eq. \ref{eq:energylevels}.
The ``$K$-factors'' (see Appendix \ref{appA} for definition) account for the cosmological redshifting of the photons, and scale as
\begin{equation}
 \label{eq:redshiftly}
 K_{\rm i}=K_{\rm i0}\left(\frac{\alpha}{\alpha_0}\right)^{-6}\left(\frac{m_\elec}{m_{\elec0}}\right)^{-3}.
 \end{equation}
To finish, the Einstein $A$ coefficient scales as $\alpha^5m_\elec$ and the two-photon decay rates as $\alpha^8m_\elec$, so that
\begin{equation}
 A_{\rm i}=A_{\rm i0}\left(\frac{\alpha}{\alpha_0}\right)^{5} \left(\frac{m_\elec}{m_{\elec0}}\right),
 \quad
  \Lambda_{\rm i}=\Lambda_{\rm i0}\left(\frac{\alpha}{\alpha_0}\right)^{8} \left(\frac{m_\elec}{m_{\elec0}}\right).
  \label{eq:lambda}
\end{equation}
We incorporate all these modifications in the code \RECFAST\ \citep{recfast,recfast_long}, so that the impact of a variation of the fundamental constants on recombination can be inferred. In the following, we impose the condition that the value of the constants can differ from the standard ones only at redshifts higher than around 50,\footnote{\label{expl}It is assumed that the constants do not vary on a time scale of the order of the width of the last scattering surface. As long as matter and radiation are tightly coupled, then after decoupling the equation of evolution of the distribution of the photons is described by a Liouville equation, instead of a Boltzmann equation,  so that neither $\alpha$ nor $m_\elec$ enter the discussion. Thus, we actually implicitly assume that the constants relax to their current value before reionization, i.e., before abour $z=10$. The dynamics of this relaxation is indeed model-dependent, but the constraints at lower redshifts suggest that the variation between redshifts around 4 and today has to be smaller than $10^{-5}$. Any relaxation dynamics of the constants from their values at the epoch of recombination to their laboratory ones will not affect the analysis as long as it takes place after recombination and before the reionization era.  In any case, we are essentially assuming that there is one
value at the recombination epoch and another value today.} and we consistently include variations both for hydrogen and helium recombination physics.

\begin{figure}[th]
\centering
\includegraphics[height=7cm]{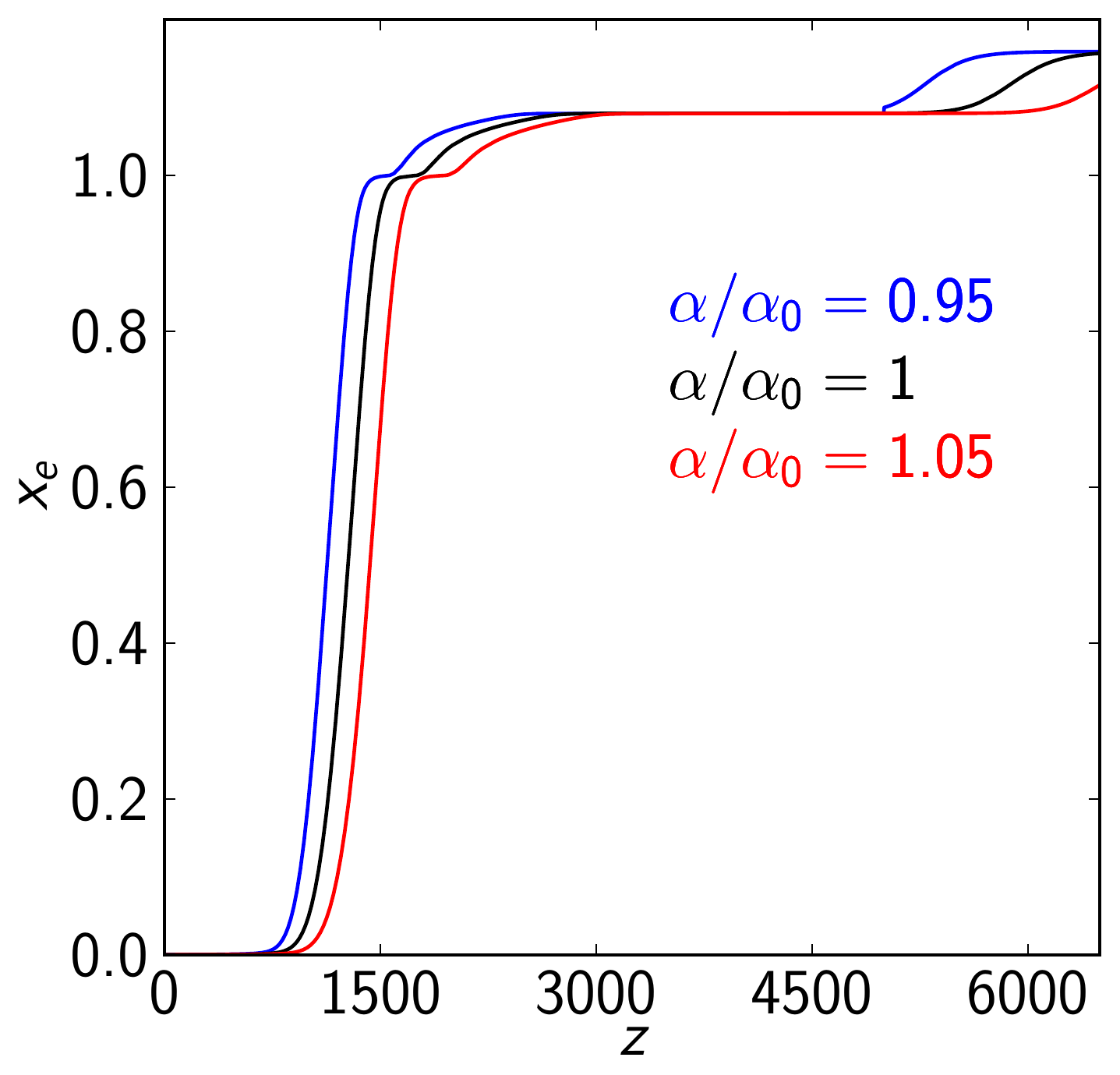}

\includegraphics[height=7cm]{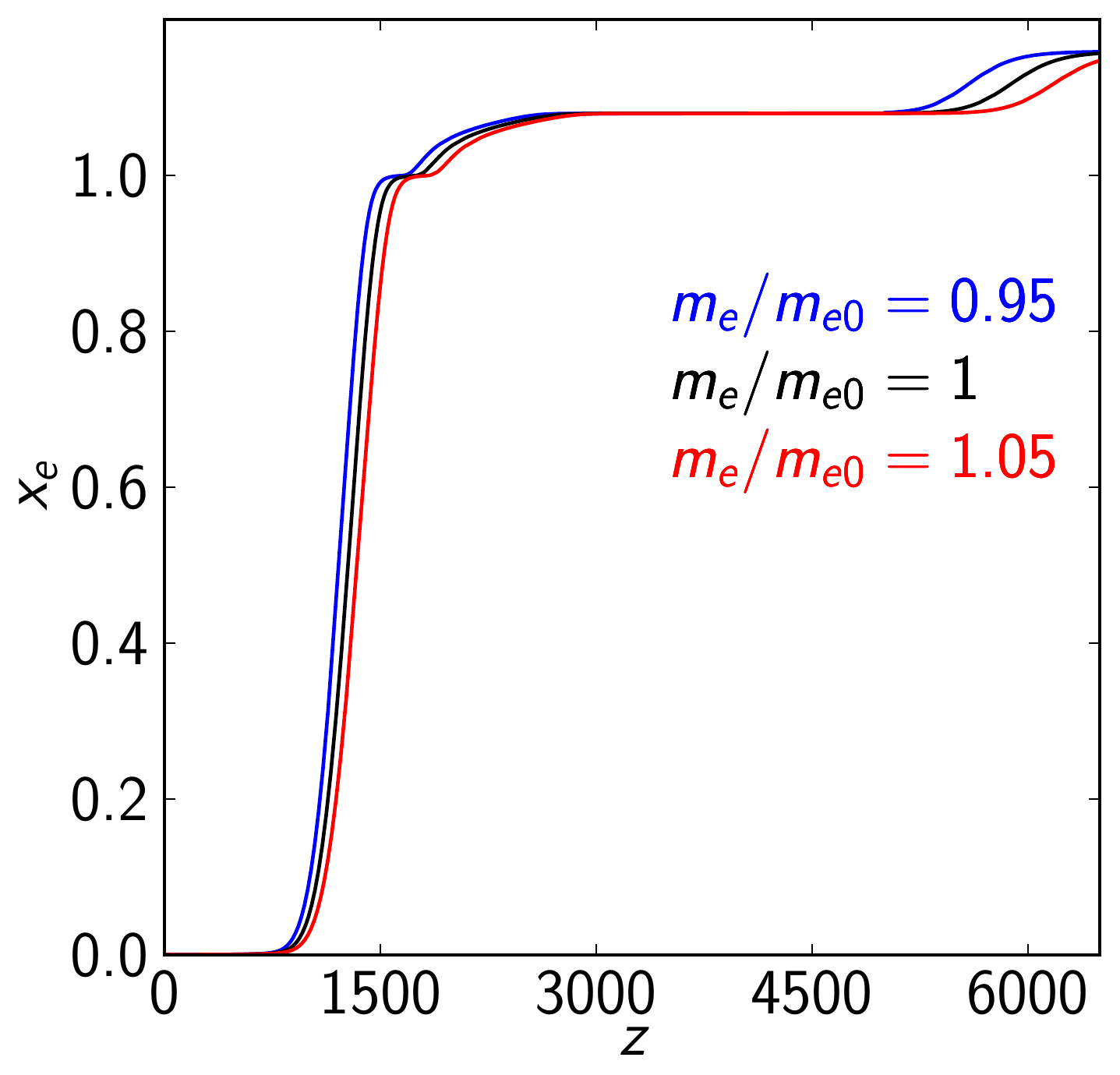}
\caption{Evolution of the free electron fraction $x_{\rm e}$ with redshift $z$ for different values of the fine structure constant $\alpha$ (top) and of the mass of the electron (bottom). The coloured lines refer to a $-5\,\%$ (blue) and $+5\,\%$ (red) variation of the constants, while the black line shows the standard case. The decreases at redshift around 6000 and 2000 correspond to the first and second recombination of helium, while the large decrease at redshift $1300$ is due to the recombination of hydrogen.}
\label{fig1}
\end{figure}
 
Fig.~\ref{fig1} shows the evolution of the free electron fraction $x_\elec(z)$ with redshift under a relative variation of $\pm5\,\%$ of either the fine structure constant, $\alpha$, or of the mass of the electron, $\me$. As can be noticed, a larger value of either of the two constants shifts the recombination epoch to earlier times, i.e., to higher redshifts. The effect is stronger for $\alpha$ than for $\me$, mainly because of the different dependence of the transition frequencies on the two constants (see Eq.~(\ref{eq:energylevels})). A detailed explanation of how the different dependences listed in this section affect the dynamics of recombination is reported in Appendix \ref{effects_alphame}. 

Figure~\ref{fig:clTTalphame} shows the effect of a variation in $\alpha$ or $\me$ on the CMB angular power spectra (temperature, $EE$ polarization and $TE$ cross-correlation). For a larger value of the constants, recombination happens earlier, resulting in three main effects on the power spectra.

The first effect is that the sound horizon at recombination is smaller, and the angular diameter distance to the last scattering surface is larger. As a consequence,  the positions of the acoustic peaks shift to higher multipoles, in a way that can be degenerate with other cosmological parameters, e.g., the Hubble constant. However, these degeneracies can be broken, since a change in the value of the constants also affects the amplitude of the peaks, as described below. 
 
The second effect of an earlier recombination is an increase of the amplitude of the peaks at small scales, due to a decrease of the Silk damping \citep{silk68}. This effect is, however, more relevant for a change in the fine structure constant rather than for a change in the mass of the electron. This is because the Silk damping length can be defined as \citep{zaldarriaga95,hu95,kaiser83}

\begin{equation}
\lambda_{\rm D}^{2}=\frac{1}{6}\int_0^{\eta_{\rm dec}}\frac{d\eta}{\sigma_{\rm T} n_\elec a}\left[\frac{R^2+\frac{16}{15}(1+R)}{(1+R)^2}\right],
\end{equation}
where $R=3\rho_{\rm b}/(4\rho_{\rm r})=3\omega_{\rm b}/(4\omega_{\rm \gamma})\, a$, is the baryon-radiation ratio and $\eta_{\rm dec}$ the conformal time of decoupling. An earlier and faster recombination process, due to a larger value of the constants, results in a smaller $\lambda_{\rm D}$ and thus in damping affecting only smaller scales. However, $\lambda_{\rm D}^2$ is also inversely proportional to the Thomson scattering cross-section, which depends on $\alpha$ and $\me$ as (see also Eq. (\ref{sigmat}))
$$
\lambda_{\rm D}^2\propto \frac{1}{\sigma_{\rm T}} \propto \frac{1}{\alpha^2 \me^{-2}}.
$$
Thus, an increase of $\alpha$ decreases the Silk damping scale, not only by precipitating recombination, but also by increasing the Thomson scattering cross-section, as both these effects decrease $\lambda_{\rm D}$. On the other hand, an increase of $\me$ decreases the Thomson scattering cross-section, thus partially compensating for the decrease of $\lambda_{\rm D}$ due to the earlier recombination. For this reason, $\alpha$ has a larger impact on the damping tail than $\me$.
Finally, the third effect of an earlier recombination due to a larger value of the constants is an increase of the amplitude of the spectra at large scales. 
This is due to the shorter time interval between the redshift of matter-radiation equality and the redshift of decoupling, which has two main consequences.
On the one hand, the early integrated Sachs-Wolfe effect \citep{hu95} is enhanced in the \TT\ power spectrum, increasing the amplitude of the first peak. 
On the other hand, the baryon-radiation ratio $R$  is smaller at the epoch of recombination. This impacts the amplitude of all the spectra \citep{hu95}, 
since it decreases the shift of the equilibrium point in the photo-baryonic oscillations due to the baryon load, resulting in a smaller enhancement of the amplitude of 
odd peaks compared to even ones in the \TT\ power spectrum.

In conclusion, the overall amplitude of the peaks is less affected by a change in $\me$ than by a change in $\alpha$, due to the different effect on the damping tail. This is the reason why high resolution data on the damping tail, as provided by \Planck, allow one to break the degeneracy between $\alpha$ and $H_0$, while one can hardly do this for $\me$ and $H_0$. Further details of these effects can be found in Appendix~\ref{effects_alphame}.

\begin{figure*}[ht]
\centering
\vskip1cm
\includegraphics[height=5.4cm]{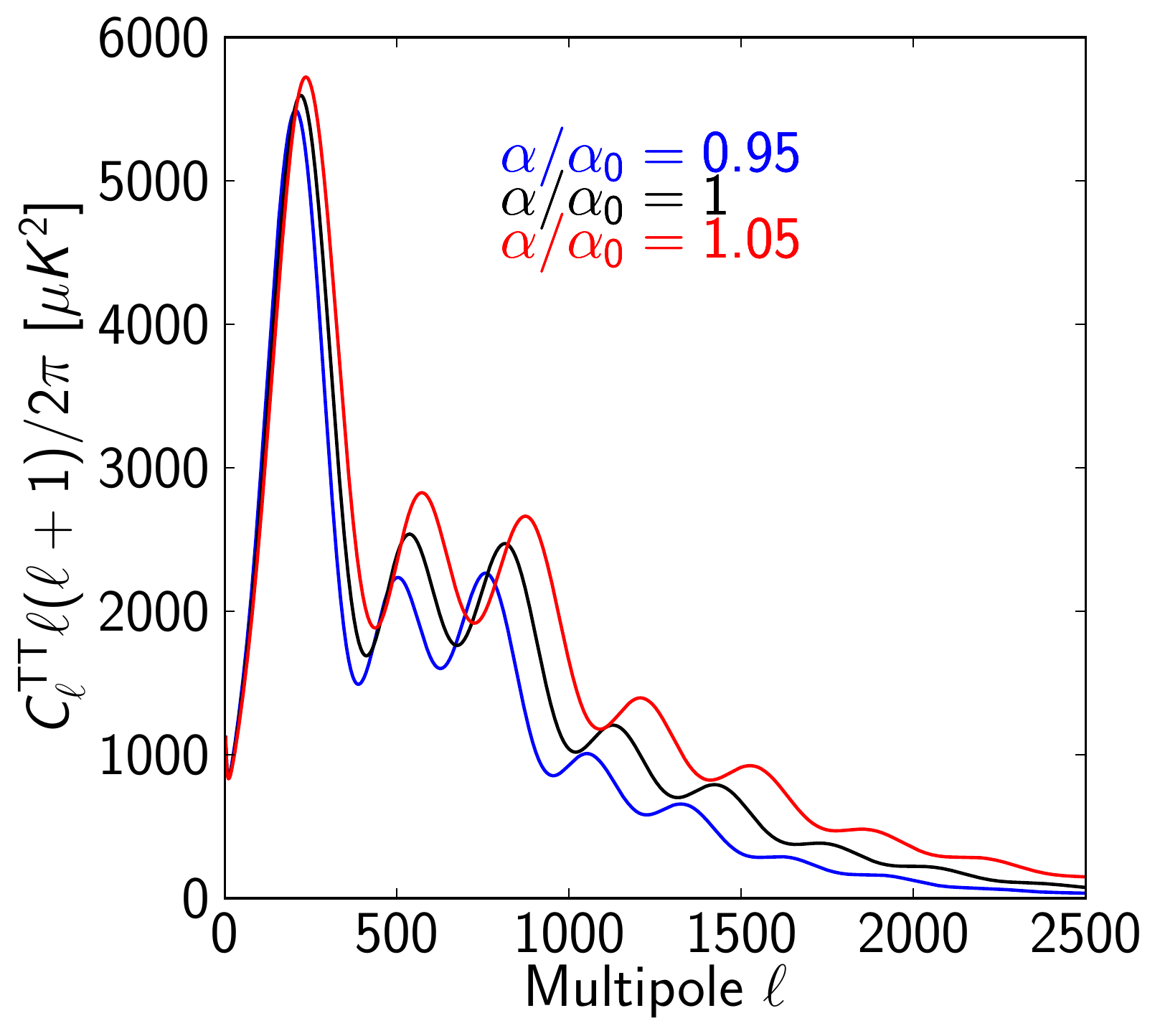}
\includegraphics[height=5.4cm]{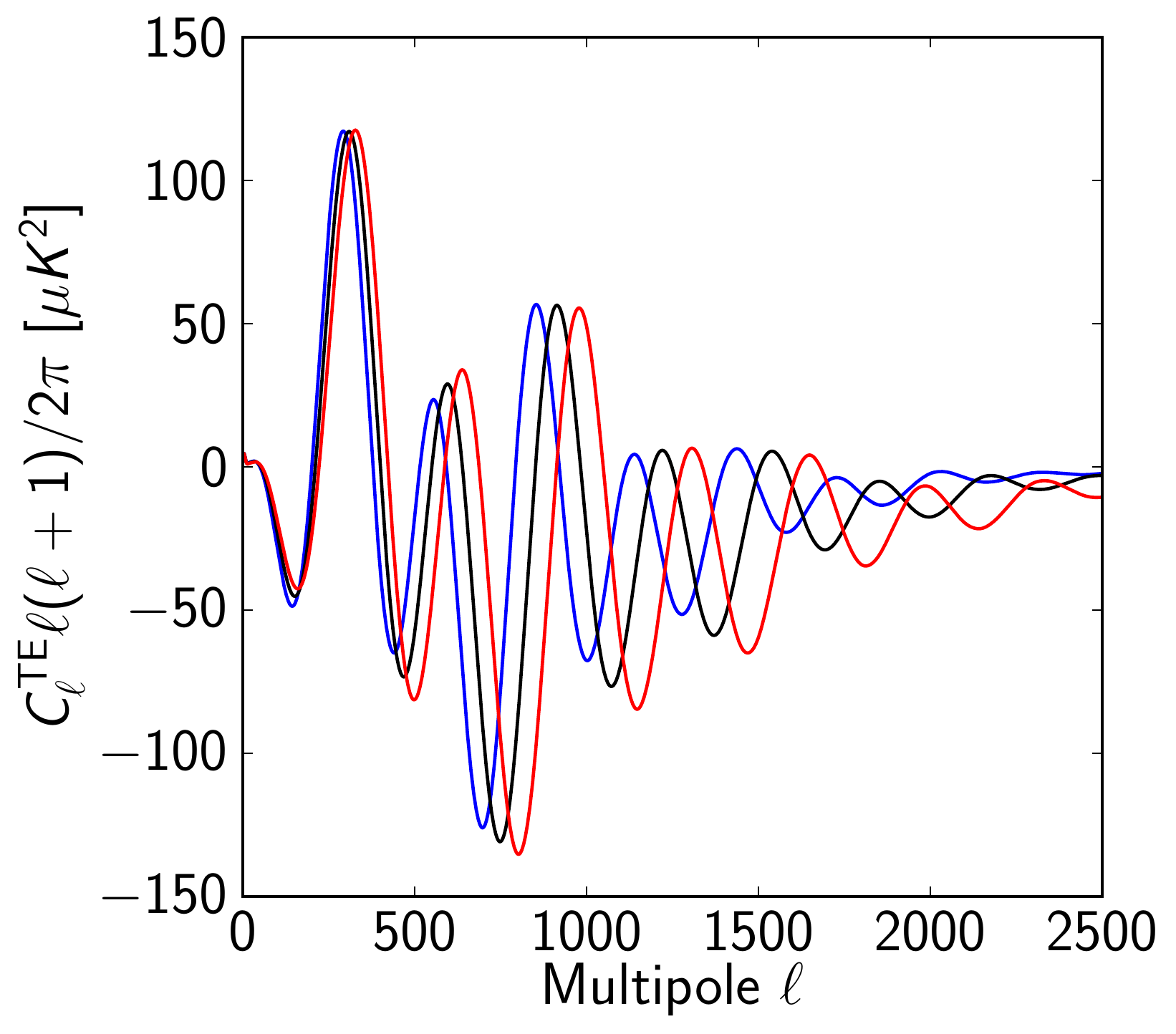}
\includegraphics[height=5.4cm]{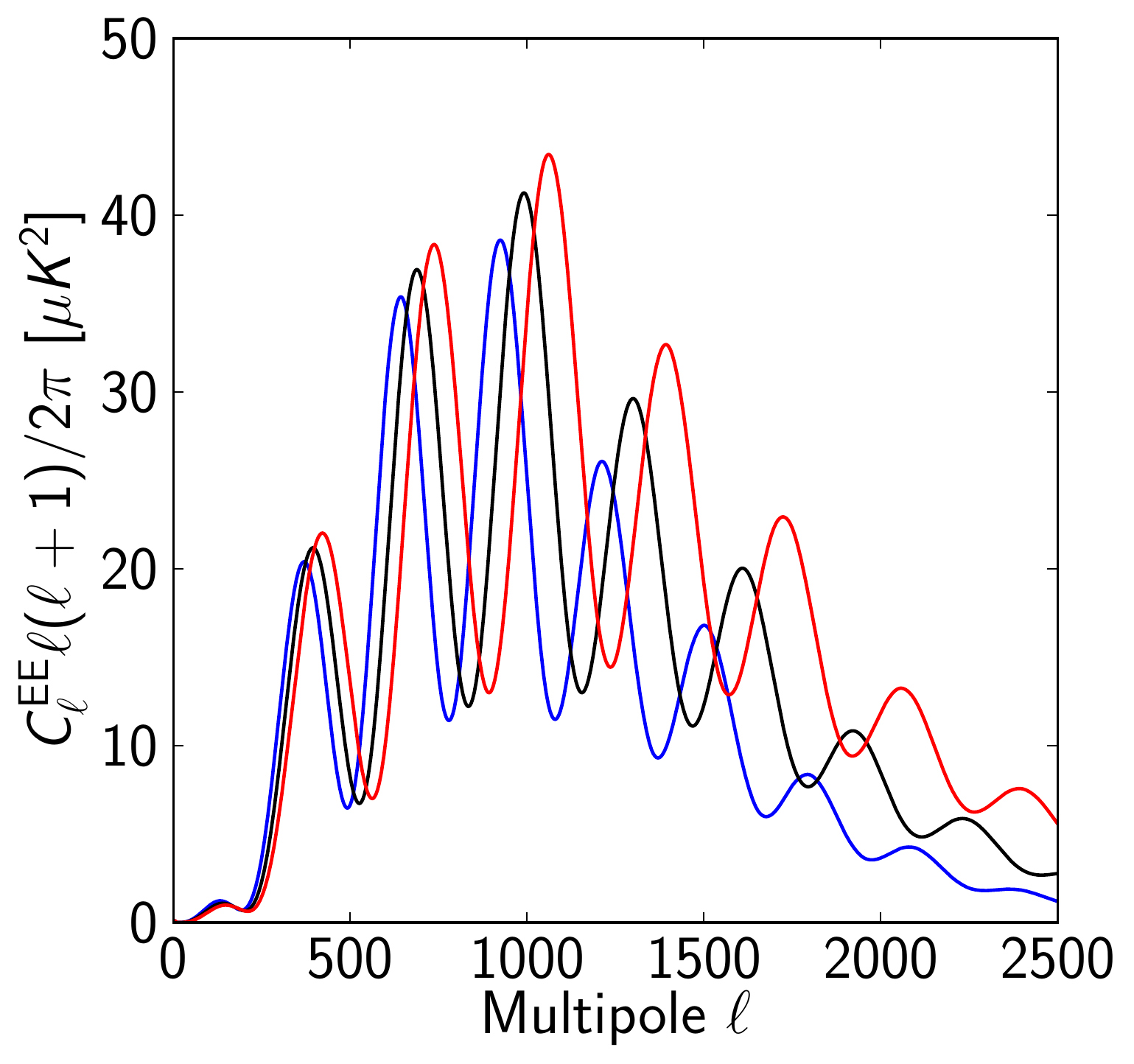}

\includegraphics[height=5.4cm]{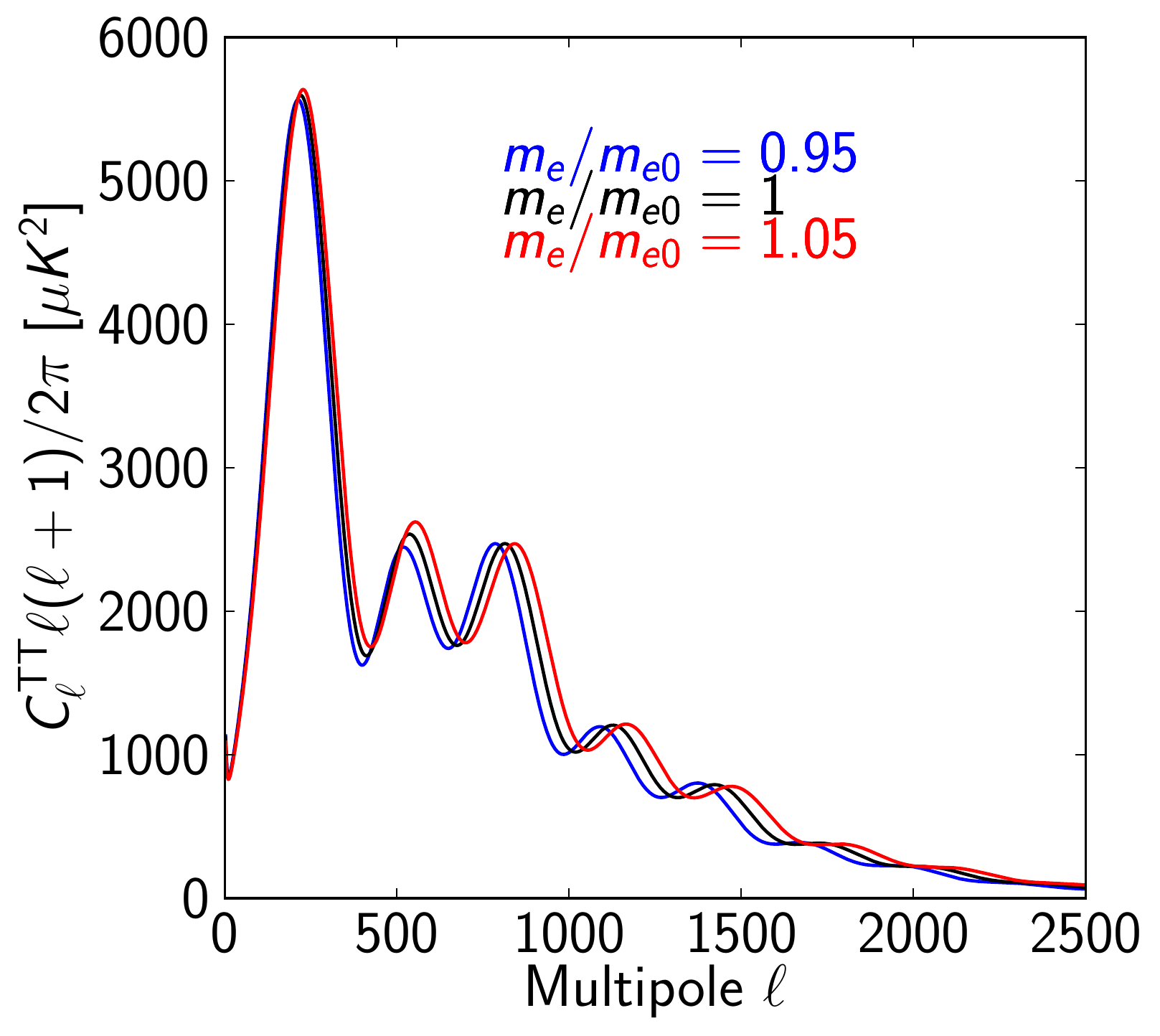}
\includegraphics[height=5.4cm]{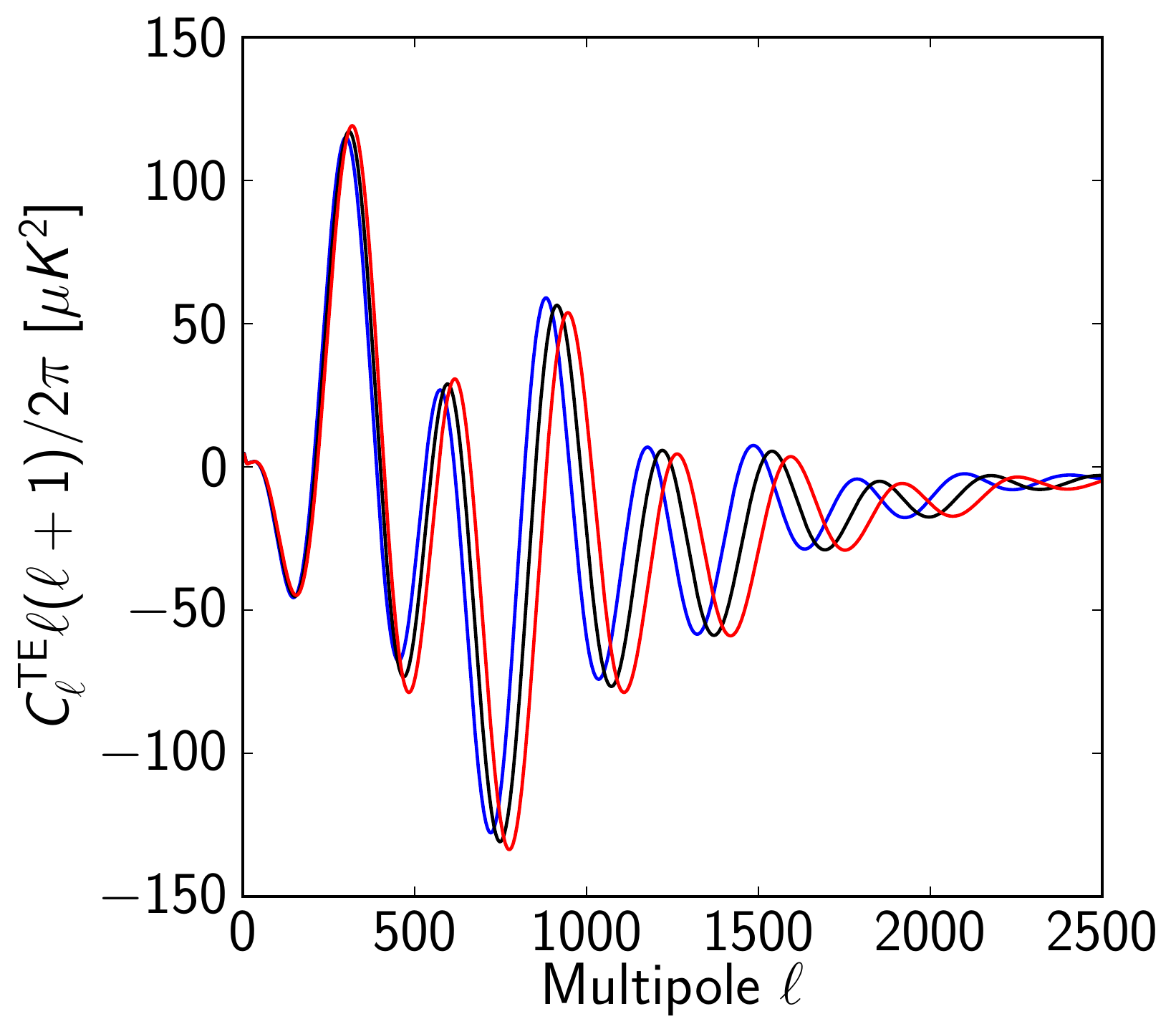}
\includegraphics[height=5.4cm]{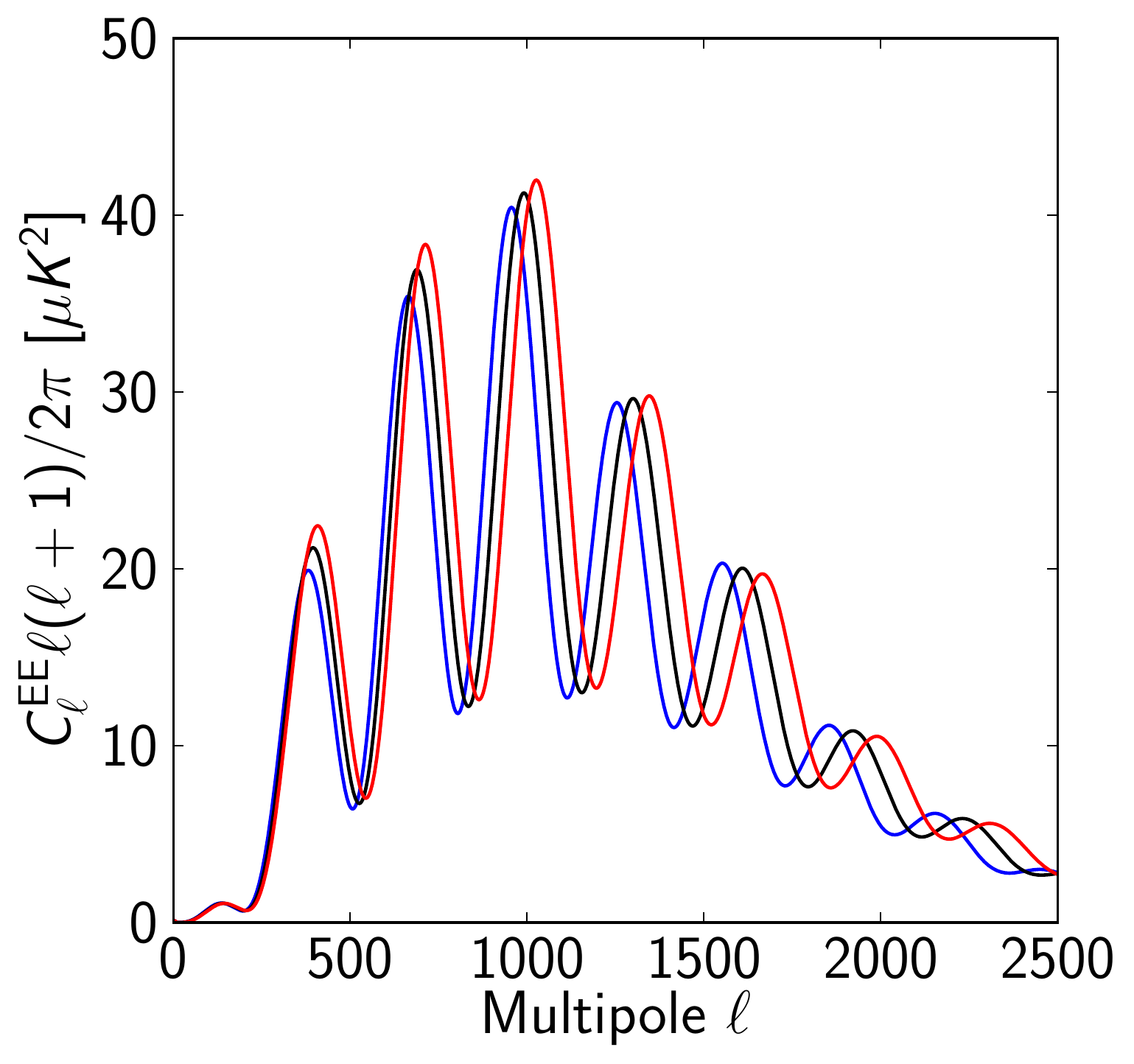}
\caption{CMB \TT, \TE\ and \EE\ angular power spectra for different values of $\alpha$ (upper plots) and $\me$ (lower plots). The lines refer to variations of $-5\,\%$ (blue) and $+5\,\%$ (red), while the standard case is shown in black.}
\label{fig:clTTalphame}
\end{figure*}

\subsection{Beyond ($\alpha$, $m_{\rm e}$)}\label{subsec0b}

As explained above, only the fine structure constant and the electron mass have a direct impact on the recombination history. This work, as in most previous analyses, assumes that the gravitational sector is not modified, so that the dimensionless Friedmann equation takes its standard form:
\begin{equation}\label{e.fl1}
 E^2(z)=\Omega_{\rm m}(1+z)^3 + \Omega_{\rm r}(1+z)^4 + \Omega_K(1+z)^2 +\Omega_\Lambda,
\end{equation}
with $\Omega_{\rm m}=\Omega_{\rm c}+\Omega_{\rm b}$ and $H(z)=H_0E(z)$. However, in full generality, and to be self-consistent, the variation of any constant would represent a violation of the Einstein equivalence principle and hence invalidates the use of General Relativity to describe gravitation, and thus cosmology. The Einstein equations, and thus the Friedmann equations, should be modified due to the existence of new degrees of freedom that: (i) couple to the standard matter fields and are responsible for a long range interaction; and (ii) have their own energy densities, which should be included in the Friedmann equation \citep{uellis}.  

A large number of theoretical extensions to standard physics have been proposed and it is not our purpose here to list and investigate all these ideas. It is, however, useful to mention that we can roughly consider two classes of models. First, one can assume that all masses vary identically, i.e., that $m_a/m_b$ remains constant for all types of field. Such a theory can be rewritten as a varying $G$ theory, i.e., as a scalar-tensor theory \citep{will81}. Cosmological signatures of these theories have been discussed in, e.g.,~\cite{bbn-ST1} and~\cite{bbn-ST2} for big-bang nucleosynthesis (BBN); a complete and consistent description of their signatures on the CMB can be found in~\cite{cmb-G1}, to be compared to the phenomenological approaches adopted in \cite{zahn} and \cite{gallig}. It is however important to realize that as soon as one assumes that $\alpha$ is varying, one expects the masses of all nucleons to vary as well, since the field responsible for the variation of $\alpha$ couples radiatively to nucleons. We thus expect the proton and neutron masses to be time dependent~\citep{xiN}, even if one assumes that the masses of all fundamental particles remain constant, in which case the model cannot be rephrased as a varying $G$ theory.

Another aspect of the variation fundamental constants, which has been much
discussed in the literature \citep[see e.g.,][]{dicke62,duff02,jpu-revue,ali},
is that only dimensionless combinations of constants can really be measured.
Because of this, many previous studies have focussed on the parameter
$\mu\equiv m_{\rm e}/m_{\rm p}$.  We have checked that
for the physics of recombination our consideration of $m_{\rm e}$ is
entirely equivalent to variation of $\mu$.  Hence our study of constraints
on $(\alpha,m_{\rm e})$ is consistent with arguments that the only
meaningful variations are dimensionless ones.  However, the situation would
be more complicated if we were to consider additional constants, and in
particular $G$.  As already stated, $G$ enters the Friedmann equation, and
so even if one considers a dimensionless ratio, such as
$G m_{\rm p}^2/\hbar c$, there are still complications over whether the
cosmological framework is even self-consistent, in addition to whether the
cosmological perturbations might evolve differently.  This can only be done
within the context of specified theories of modified gravity.

This short discussion emphasizes the difficulty in implementing in a self-consistent way the variation of several constants without fully specifying a theory, thus becoming model-dependent. Let us also point out that in many specific constructions the variations of the different constants are correlated \citep[see, e.g.,][]{cnouv,luo} and will enter both the equations governing the dynamics of the background spacetime, i.e., new terms in Eq.~(\ref{e.fl1}), and the recombination process. In our work, we focus on the two constants that have a dominant effect on the recombination history, and we have neglected their effect on the background expansion, which is a good approximation for small variations. However, it is important to keep in mind that this is an approximation, and that one should in principle rely on a completely defined theoretical model (see, e.g., ~\citealt{bbn-ST1} and~\citealt{bbn-ST2} for cases in which such an approximation is not accurate).


\section{Time variation of a single constant}\label{sec2}

\subsection{Standard \Planck\ data analysis}
\label{sec:datamethods}

In this section we determine the \planck\ constraints on the value of the fine-structure constant $\alpha$ and on the mass of the electron $m_{\rm e}$,  assuming that only one constant can vary at a time.

The cosmological picture we assume is a flat $\Lambda$CDM\ model with one additional varying constant ($\alpha$ or $\me$), purely adiabatic initial conditions with an almost scale invariant power spectrum, and no primordial gravity waves. It is thus described by a seven-dimensional parameter space that includes the baryon and cold dark matter densities $\omega_{\rm b}=\Omega_{\rm b}h^2$ and $\omega_{\rm c}=\Omega_{\rm c}h^2$, the Hubble constant $H_0$, the optical depth at reionization $\tau$, the scalar spectral index $\ns$, the overall normalization of the spectrum $\As$ (defined here at $k=0.05\Mpc^{-1}$) and a parameter for the varying constant:
$$
 \lbrace\omega_{\rm b},\omega_{\rm c},H_0,\tau,\ns,\As, \alpha\,\,\hbox{or}\,\,\me\rbrace.
$$

Unless otherwise stated, we derive the value of the primordial helium abundance as in \cite{Hamann:2011ge}. This uses interpolated results from the {\tt PArthENoPE} BBN code \citep{Pisanti:2007hk}, which calculates the value of the helium abundance given the number of relativistic species $N_{\rm eff}$ and the physical baryon density $\omega_{\rm b}$, assuming standard BBN. Furthermore, unless explicitly varied, we fix the number of relativistic species to $N_{\rm eff}=3.046$ and the sum of the neutrino masses to $\mnu=0.06$, as in \cite{planck2013-p11}.

In order to determine the constraints, we use \Planck\ temperature anisotropy data~\citep{planck2013-p08} in combination with several other data sets, and use the likelihood code publicly released by the \Planck\ collaboration \VerbatimFootnotes \footnote{\verb+http://pla.esac.esa.int/pla/aio/planckProducts.html+}.
We use the same assumptions and data sets used in~\cite{planck2013-p11} to determine the cosmological parameters, and we refer the reader to this paper for further details. The data sets we use are: 

\renewcommand{\labelitemi}{$\bullet$}
\begin{itemize}
\item \planck\
The \Planck\ CMB power spectra are analysed using two different likelihood codes.
In the multipole range $2\leq\ell\leq 49$, the likelihood is based on a component-separation approach over 91\,\% of the sky \citep{planck2013-p06,planck2013-p08}, while at higher multipoles
it is constructed from cross-spectra over the frequency range
100--217\,GHz, as discussed in ~\cite{planck2013-p08}. In the latter case, the amplitude of unresolved foregrounds, as well as beam and calibration uncertainties, are explored, along with the cosmological parameters. In the following, we will refer to the combination of these two likelihoods simply as \planck.

\item \textsl{WP}
In combination with the \planck\ temperature data, we include polarization data from the \WMAP\ satellite \citep{Bennett:2012fp} in the multipole range $2 < \ell < 23$, using the likelihood code provided by the \Planck\ collaboration. We will refer to this data set as \textsl{WP}.

\item \textsl{Lensing}
We include information from the lensing potential power spectrum $C_\ell^{\phi\phi}$, as determined from the trispectrum computed on \Planck's maps. We specifically use the data and likelihood provided in ~\cite{planck2013-p12}.

\item \emph{highL}
We combine \planck\ data with CMB \TT\ information coming from ground-based high resolution experiments. We use the $2013$ data release of the Atacama Cosmology Telescope (ACT) as described in ~\cite{das13}, in particular, the ACT $148\times148\,$GHz power spectrum at $\ell > 1000$, and the ACT $148\times218\,$GHz and $218\times218\,$GHz power spectra at $\ell > 1500$. Furthermore, we use the $2012$ data release of the South Pole Telescope (SPT), as described in~\cite{Reichardt:12}, including data at $\ell > 2000$.

\item \emph{BAO}
We also present results using Baryon Acoustic Oscillation (BAO) data from the following redshift
surveys: the SDSS DR7 measurement at $z = 0.35$
 \citep{Padmanabhan:2012hf};
the BOSS DR9 measurement at $z=0.57$ 
\citep{Anderson:2012sa}; and the 6dF Galaxy Survey measurement at $z=0.1$ \citep{Beutler:11}.

\item \emph{HST}
We include a Gaussian prior on the Hubble constant $H_0$
as determined by \cite{Riess:2011yx},
\begin{equation}
H_0 =  (73.8 \pm 2.4)  \; {\rm km}{\,\rm s}^{-1}{\rm Mpc}^{-1},  \label{H0}
\end{equation}
using cepheids and type Ia supernovae (SNe Ia). This value is determined using {\it Hubble Space Telescope\/} ({\it HST\/})
observations of cepheid variables in the host galaxies of eight type~Ia SNe to calibrate the supernova magnitude-redshift relation.

\end{itemize}

In the following, we also present results using the \WMAP-9 data release (temperature and polarization), utilizing the likelihood code provided by the \WMAP\ team \citep{bennett2012}.

We use a modified version of the \RECFAST\ (v.~1.5.2)\footnote{Available at
\url{http://www.astro.ubc.ca/people/scott/recfast.html}.} recombination code~\citep{recfast, cmb-han, cmb-rocha1, cmb-rocha, menegoni1} to include the variation of constants, as described in Sect.~\ref{subsec0a}. 
 To calculate constraints, we use the publicly available Markov Chain Monte Carlo package \texttt{cosmomc} ~\citep{Lewis:2002ah}. 
The MCMC convergence diagnostic tests are performed on four chains using the
Gelman and Rubin ``variance of chain mean''$/$``mean of chain variances'' $R-1$
statistic for each parameter \citep{gelman}. Our constraints are obtained
after marginalization over the remaining ``nuisance'' parameters, again using
the programs included in the \texttt{cosmomc} package.
We use a cosmic age top-hat prior of 10\,Gyr $ \le t_0 \le$ 20\,Gyr, which is wide enough to have no affect on our results.
We sample the seven-dimensional set of cosmological parameters, adopting flat priors on them, together with foreground, beam and calibration parameters when using the \planck\ data ($14$ additional parameters) and the \highL\ data ($17$ additional parameters).
For these ``non-cosmological'' parameters we use the same assumptions and priors as in table~4 of ~\cite{planck2013-p11}.

\subsection{Fine structure constant ($\alpha$)}

\begin{table*}[tmb]                 
\caption{Constraints on cosmological parameters for a standard $\Lambda$CDM\ model with the addition
of a varying fine-structure constant. We quote errors at the $68\,\%$ confidence level. The Hubble constant $H_0$ is in units of $\rm km\,s^{-1}\,Mpc^{-1}$.}                        
\label{table:alpha-tab1}                            
\input table_alpha
\end{table*} 

\begin{figure}[tb]
\centering
\vskip1cm
\includegraphics[height=7cm]{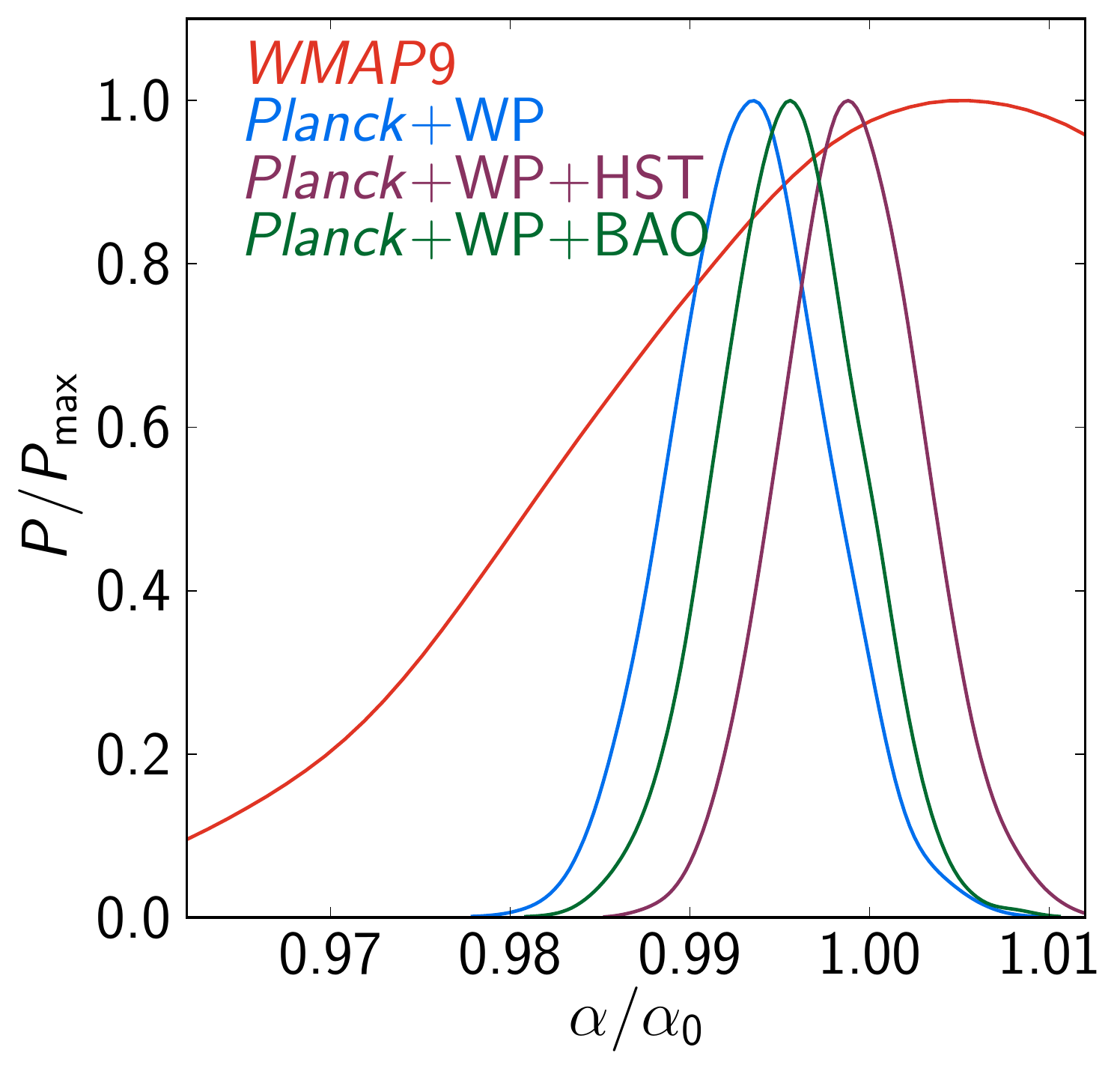}
\caption{ Marginalized posterior distributions of $\alpha/\alpha_0$ for different data combinations, as described in the text.}
\label{alpha1D}
\end{figure}

\begin{figure*}[htb]
\centering
\vskip1cm
\includegraphics[width=\textwidth]{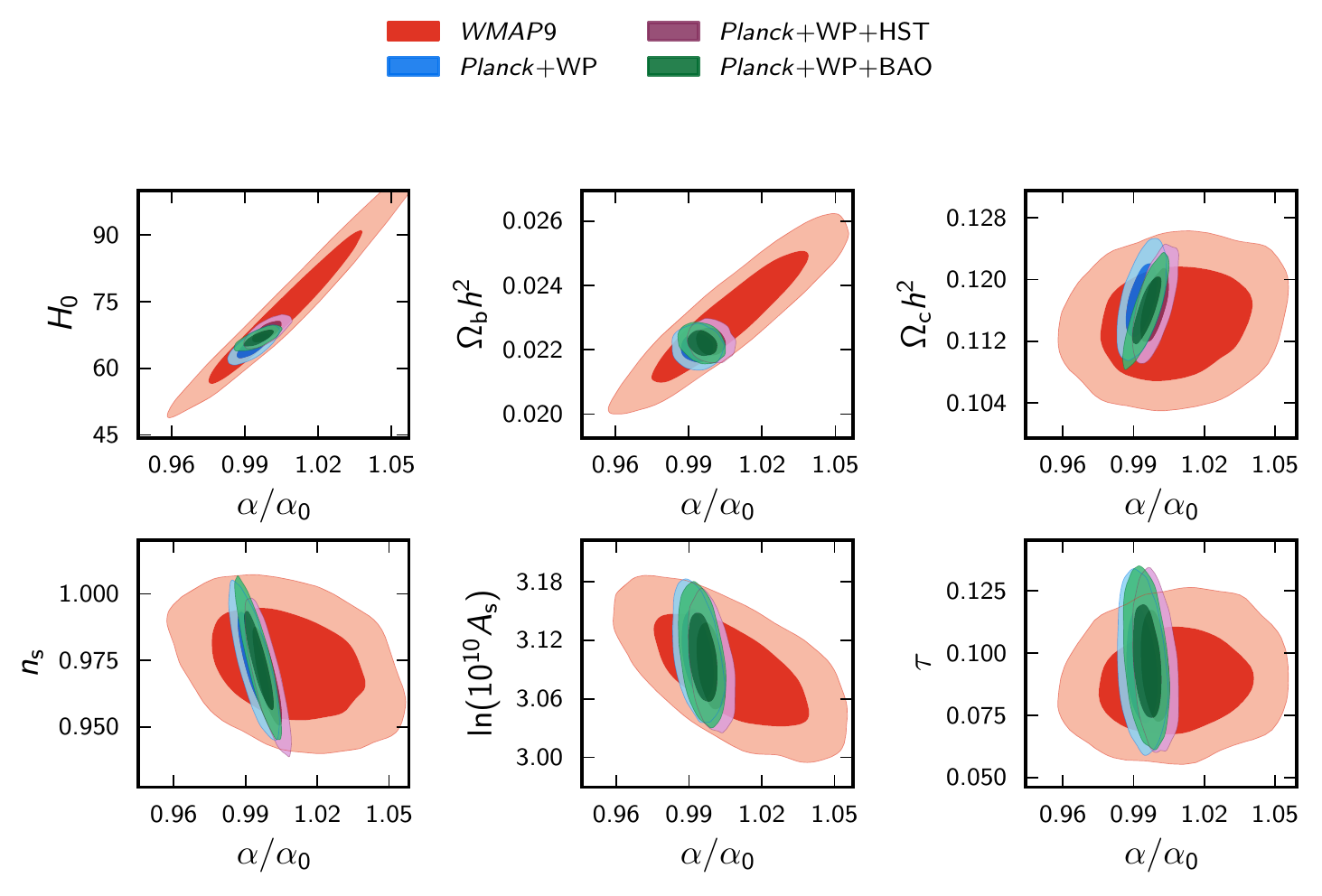}
\caption{Two-dimensional likelihood contours (68\,\% and 95\,\%) for $\alpha/\alpha_0,$  versus other cosmological parameters for the \WMAP-9 (red), \Planck+\WP\ (blue), \Planck+\WP+\HST\ (purple), and \Planck+\WP+BAO (green) data combinations.}
\label{fig3b}
\end{figure*}

The results of the analysis on the variation of the fine structure constant are summarized in Table~\ref{table:alpha-tab1}, which compares parameter constraints from \Planck\ (combined with different data sets) and those obtained with \WMAP-9 data. 
Fig.~\ref{alpha1D} compares the one-dimensional likelihood profiles for $\alpha$ obtained with different data sets, while Fig.~\ref{fig3b} shows the degeneracies between $\alpha$ and the six other cosmological parameters used in our analysis. 

From CMB data alone, \Planck\ improves the constraints from a $2\,\%$ variation on $\alpha$ (from \WMAP-9) to about $0.4\,\%$ (in agreement with what was found
in table~11 of \citealt{planck2013-p11}). \Planck\ thus improves the limit by about a factor of five, in good agreement from what was expected from previous forecasts (see e.g., \citealt{cmb-rocha, galli}). The improvement is mainly due to the fact that \planck\ is able to break the strong degeneracy between $\alpha$ and $H_0$ by observing the damping tail, as already pointed out in Sect.~\ref{subsec0a}. This constraint is also better (although of comparable magnitude) than the one obtained by combining \WMAP\ data with small-scale experiments (see, e.g., \cite{sievers} for a $0.5\,\%$ constraint).

We also observe that the constraints on the parameters of the reference $\Lambda$CDM\ model change very little with the addition of a varying $\alpha$, exceptions being for $\ns$ and $H_0$. 
In fact, the constraint on $\ns$ for \planck+\WP\ is $0.9603\pm 0.0073$ assuming a $\Lambda$CDM model, while it is $0.974\pm 0.012$ for a $\Lambda$CDM+$\alpha$ model; the mean value of $\ns$ shifts about  $1\,\sigma$ to higher values and the uncertainty increases by a factor of $1.6$. On the other hand, the constraint on $H_0$ in a $\Lambda$CDM model is $(67.3\pm1.2)$ $\rm km\, s^{-1}Mpc^{-1}$, while in a $\Lambda$CDM+$\alpha$ model it is $(65.1\pm 1.8)$ $\rm km\, s^{-1}Mpc^{-1}$. Thus, the value of the Hubble constant shifts by about $1.2\,\sigma$ to even lower values; this exacerbates the tension with the value of the Hubble constant found by \cite{Riess:2011yx} and reported in Eq.~(\ref{H0}).

Finally, we find that $\alpha$ is weakly degenerate with foreground, beam and calibration parameters, as shown in Fig.~\ref{fig:alpha_fg}. 

\subsubsection{The low- versus high-$\ell$ tension and $\alpha$}

The value of $\alpha/\alpha_0$ found using \Planck+\WP\ data and reported in Table~\ref{table:alpha-tab1} is $1.6\,\sigma$ lower than unity. Although this is statistically not very significant, it is interesting to emphasize that this deviation is mainly caused by the apparent ``tension" between the low and high multipoles in the \Planck\ data; as already found in \cite{planck2013-p08} and \cite{planck2013-p11}, this tension is partially responsible for small hints of anomalies found by the \Planck\ collaboration when exploring extensions of the $\Lambda$CDM model. In order to assess the importance of this effect for the fine structure constant, we remove the \Planck\ low-$\ell$ likelihood in temperature and the \WMAP\ low-$\ell$ likelihood in polarization, and add a Gaussian prior to constrain the reionization optical depth, $\tau=0.09\pm0.013$; this value approximates the constraint obtained by \cite{hinshaw2012} using \WMAP\ polarization data.
The results of this exercise are shown in Table~\ref{table:alpha-tau}. As expected, for the \Planck($-$low $\ell$)+$\tau$ prior\  case, the constraint for $\alpha/\alpha_0$ is $0.9972\pm 0.0052$, in agreement with unity within $1\,\sigma$.

\subsubsection{Combination with other data sets}

We can combine \Planck\ data with a number of different data sets, as already described in Sect.~\ref{sec:datamethods}. We find that adding \highL\ data improves the constraint on $\alpha$ by a small amount, as most of the meaningful information on the damping tail is already provided by \Planck. Similarly, adding BAO data does not improve the error bars significantly.

Given the apparent tension between the reference $\Lambda$CDM\ parameters from \planck\  and direct measurements of $H_0$ (see e.g., \citealt{planck2013-p11}), we also investigate the effect of including \HST\ data, although the value of the Gaussian prior on $H_0$ in Eq. (\ref{H0}) is roughly $4\,\sigma$
higher than the value of $H_0$ obtained with the \Planck+\WP\ data in Table~\ref{table:alpha-tab1}.
Adding the prior on $H_0$ has the effect of increasing the mean value of $\alpha$ to be closer to $\alpha_0$, due to the positive correlation between the two parameters. However, the uncertainty on $\alpha/\alpha_0$ does not improve significantly.

\subsection{Fine structure constant, number of relativistic species and helium abundance}

\begin{figure}[htb]
\centering
\includegraphics[height=7cm]{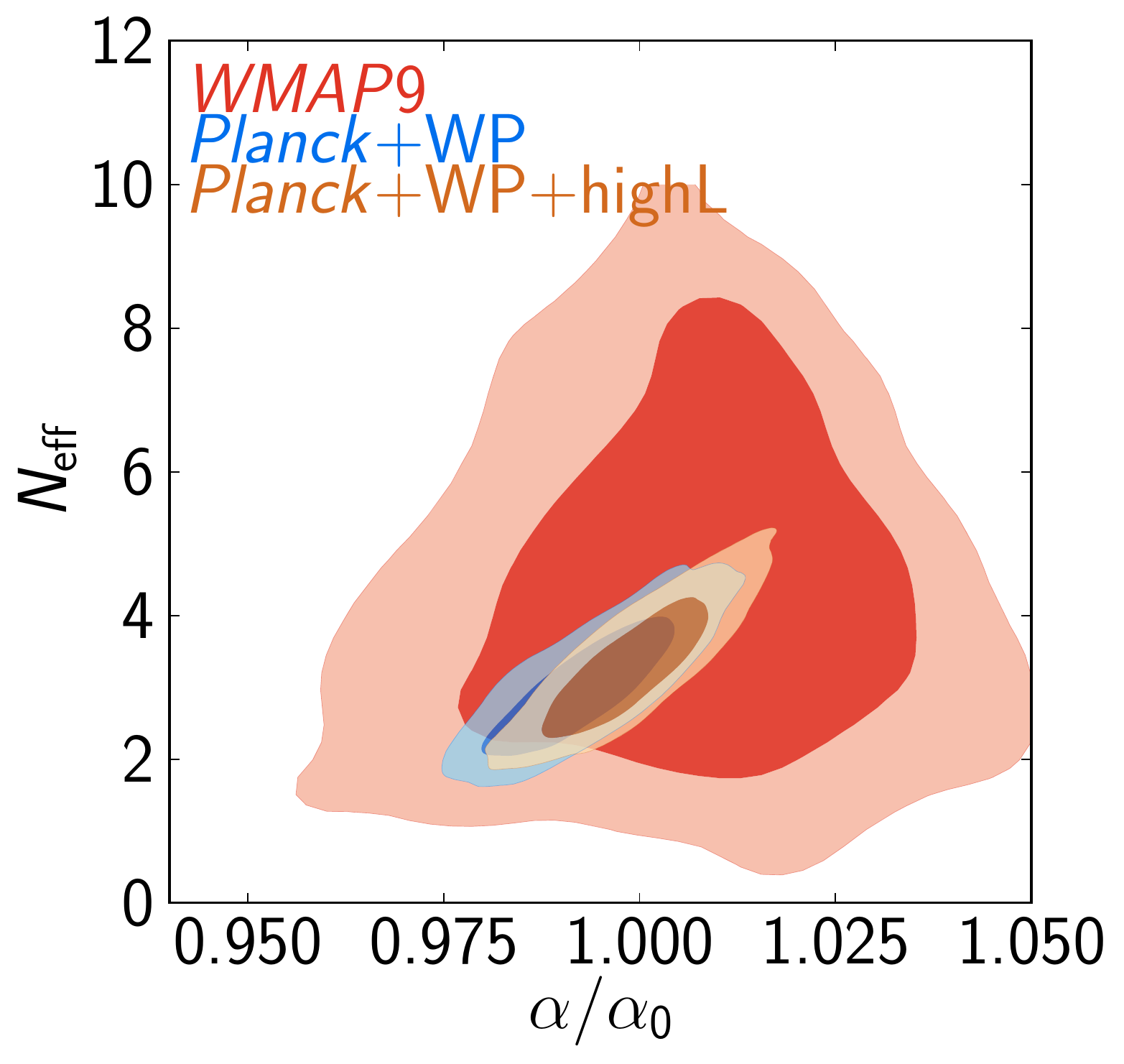}
\caption{Two-dimensional likelihood contours (68\,\% and 95\,\%) in the ($\alpha/\alpha_0,N_\nu$) plane for \WMAP\ (red), \Planck+\WP\ (blue), and \Planck+\WP+\highL\ (orange) data combinations.}
\label{fig_alphannu}
\end{figure}

In order to assess the robustness of our constraints to the chosen cosmological model, we explore how much the constraint on $\alpha$ is weakened when the number of relativistic species ($N_{\rm eff}$) or the helium abundance ($\yhe$) are allowed to vary as well.
We thus explore in this case an eight-dimensional parameter space that includes the six $\Lambda$CDM parameters, together with $\alpha$ and $N_{\rm eff}$ or $\yhe$.
A degeneracy between these parameters can be expected (see e.g., \citealt{Menegoni:2012tq}) since they change the position and the amplitude of the peaks in similar ways (see e.g., \citealt{Hou:2011ec,hinshaw2012}).
In fact, $N_{\rm eff}$ changes the angular scale of Silk damping with respect to the angular scale of the peaks, while varying $\yhe$ changes the recombination history, and in particular the recombination time and the Thomson scattering rate before recombination. Since these effects could also come about through a change in the fine structure constant, we could expect degeneracies between these parameters.
 Table~\ref{table:alpha-nnu} shows the constraints on parameters for the $\Lambda$CDM+$\alpha$+$N_{\rm eff}$ model, while Fig.~\ref{fig_alphannu} shows the important degeneracy present between $\alpha$ and $N_{\rm eff}$. We find that including a variable number of relativistic species increases the uncertainties on the value of $\alpha$, from $\alpha/\alpha_0=0.9934\pm0.0042$ to $\alpha/\alpha_0=0.9933^{+0.0071}_{-0.0045}$ when $N_{\rm eff}$ is also allowed to vary. Furthermore, we find that the constraint on the number of relativistic species in this case, $N_{\rm eff}=3.04^{+0.54}_{-0.73}$, remains in perfect agreement with the standard value $N_{\rm eff}=3.046$.

\begin{figure}[htb]
\centering
\includegraphics[height=7cm]{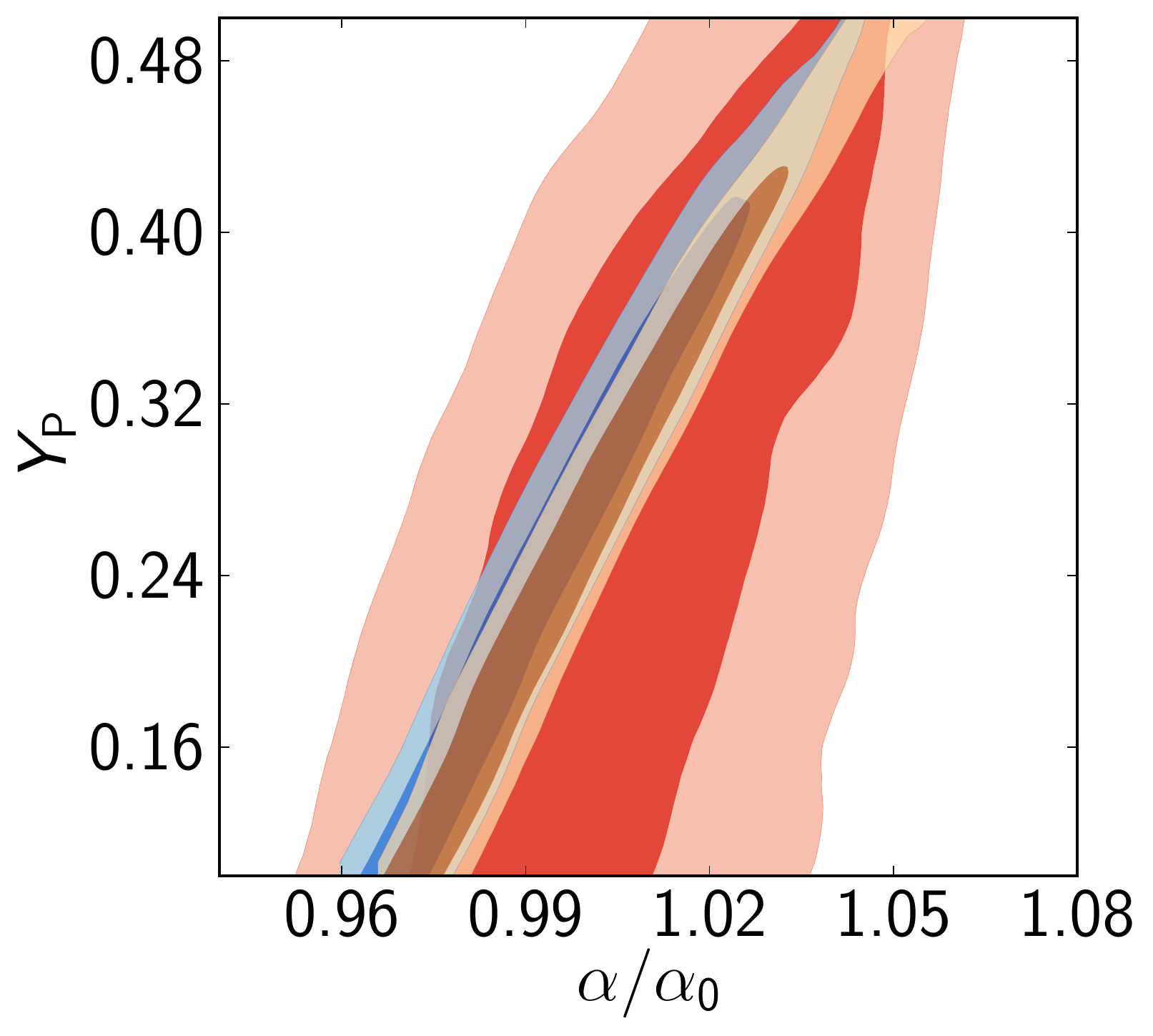}
\caption{Two-dimensional likelihood contours (68\,\% and 95\,\%) in the ($\alpha/\alpha_0,\yhe$) plane for \WMAP\ (red), \Planck+\WP\ (blue) and \Planck+\WP+\highL\ (orange) data combinations.}
\label{fig_alphayhe}
\end{figure}

On the other hand, we observe a much stronger degeneracy between $\alpha$ and $\yhe$, as shown in Table~\ref{table:alpha-yhe} and Fig.~\ref{fig_alphayhe}. 
We find that leaving the helium abundance completely free to vary in the range $0.08< \yhe <0.55$ leads to a constraint on $\alpha$  at the $1\,\%$ level, while the value of $\yhe$ is almost unconstrained. \footnote{We checked that the flat prior we impose on the age of the Universe, described in Sect.~\ref{sec:datamethods}, does not affect these constraints.} 
We emphasize that the chosen variation range for $Y_{\rm p}$ is unphysically large, considering the latest spectroscopic constraints on primordial helium abundance yield $Y_{\rm p}=0.2465\pm0.0097$ \citep{aver2013}.


\subsection{Mass of the electron ($m_{\rm e}$)}
\begin{figure}[tb]
\centering
\includegraphics[height=7cm]{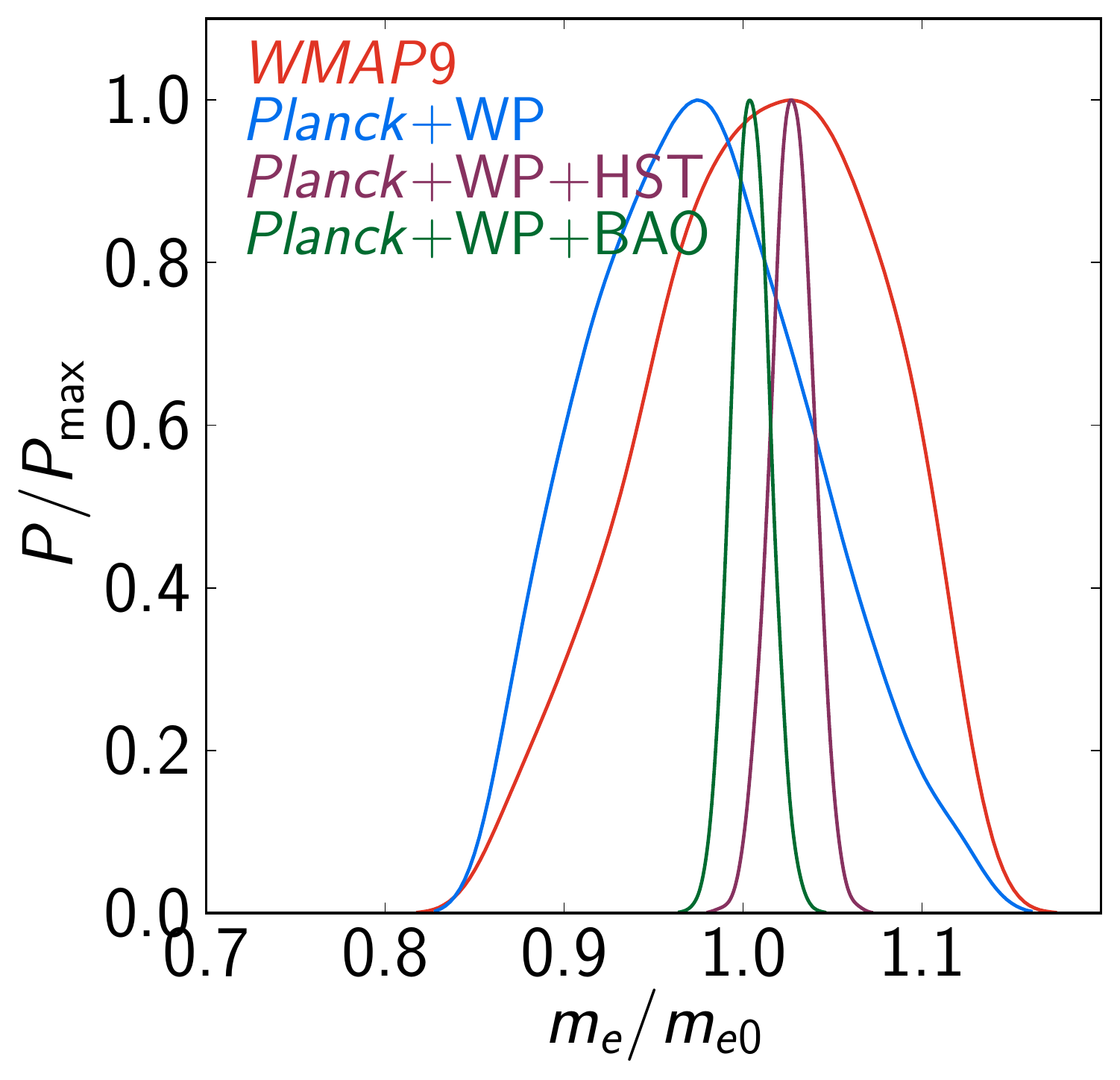}
\caption{Marginalized posterior distributions of $m_{\rm e}/m_{\rm e0}$ for the \WMAP-9 (red), \Planck+\WP (blue), \Planck+\WP+\HST\ (purple), and \Planck+\WP+BAO (green) data combinations.}
\label{fig:me}
\end{figure}

\begin{figure*}[htb]
\centering
\vskip1cm
\includegraphics[width=\textwidth]{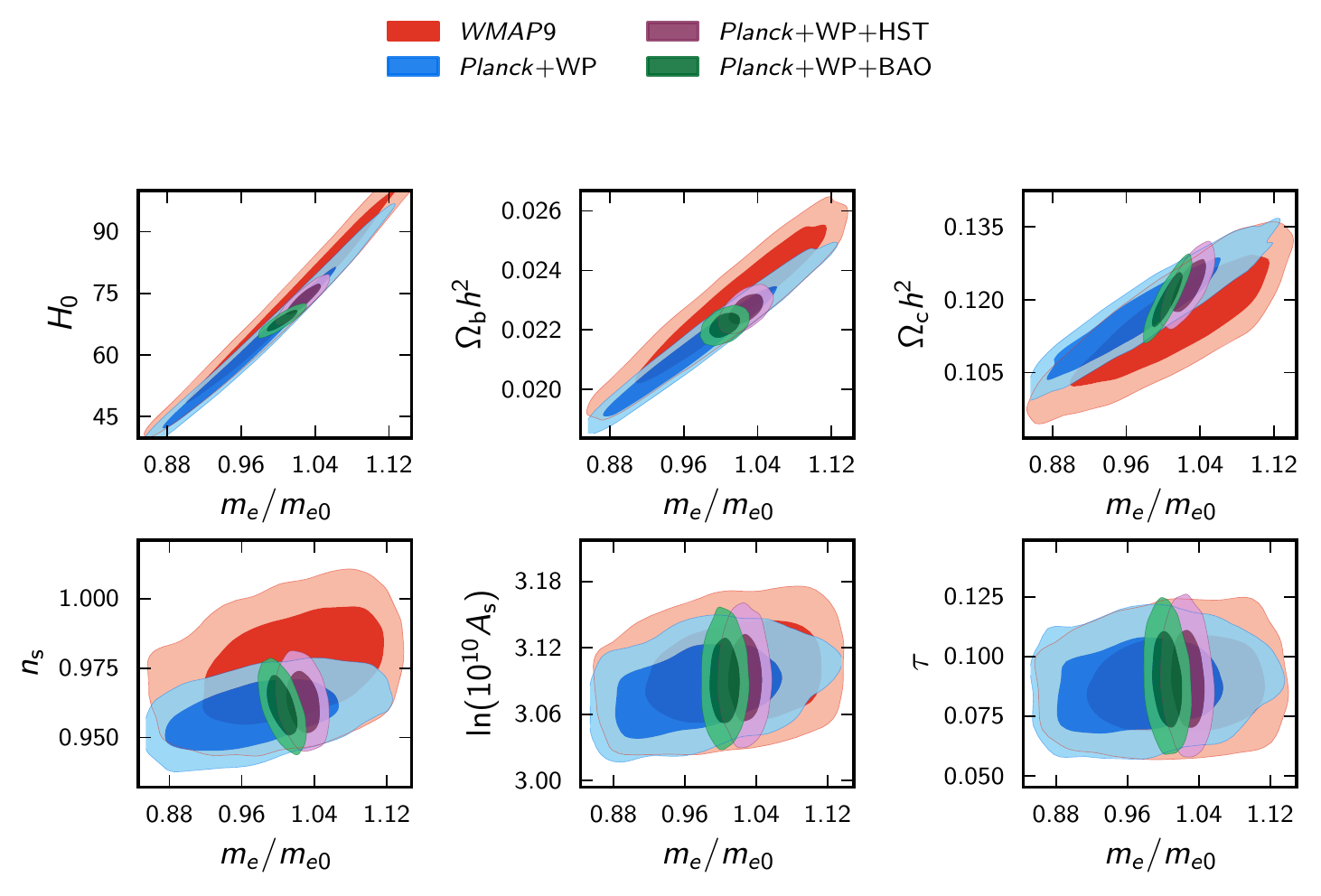}
\caption{Two-dimensional likelihood contours (68\,\% and 95\,\%) for $\me/\mezero$  versus other cosmological parameters, for the \WMAP-9 (red), \Planck+\WP (blue), \Planck+\WP+\HST\ (purple), and \Planck+\WP+BAO (green) data combinations.}
\label{fig:2Dme}
\end{figure*}

Now we present the results of our analysis on the variation of the electron mass, adopting the same procedure followed for the variation of $\alpha$ that was described in the previous sections. Here we assume the fine structure constant to be fixed to its standard value and consider variations of $m_{\rm e}$.
The results of the analysis are presented in Table~\ref{table:me-tab1}, while Fig.~\ref{fig:me} depicts the one-dimensional likelihood profiles for $\me$ determined with different combinations of data. Fig.~\ref{fig:2Dme} shows the two-dimensional contour plots between $\me$ and the other cosmological parameters.  The constraint obtained from \Planck+\WP\ is $\me/\mezero=0.977^{+0.055}_{-0.070}$, 68\,\% CL, to be compared to the one obtained with \WMAP-9 data, $\me/\mezero=1.011^{+0.077}_{-0.057}$.
It is interesting to note that in the case of $m_{\rm e}$, \Planck\, is not able to efficiently break the degeneracy with the Hubble constant, contrary to what happens for the fine structure constant. This is due to the fact that $m_{\rm e}$ does not strongly change the amplitude of the damping tail, as detailed in Sect.~\ref{subsec0a} and Appendix \ref{effects_alphame}.

We also find that, similarly to the case of the fine structure constant, $\me$ is not strongly correlated with foreground, beam and calibration parameters. This is shown in Fig.~\ref{fig:me_fg}. In Table~\ref{table:me-tau} we also show the effect of removing the low-$\ell$ multipoles and placing a Gaussian prior on the optical depth. We find that for the electron mass (just as we found for
$\alpha$), the effect of suppressing the low multipoles is to increase the mean value of $m_{\rm e}/m_{\rm e0}$  to values closer to unity, by typically $0.5 \,\sigma$.

\begin{table*}[tmb]                 
\caption{Constraints on the cosmological parameters of the base $\Lambda$CDM\ model with the addition
of a varying electron mass. We quote $\pm 1\,\sigma$ errors here.
Note that for \WMAP\ there is a strong degeneracy
between $H_0$ and $\me$, which is why the uncertainty on $\me/\mezero$ is 
much larger than for
\Planck.}                          
\label{table:me-tab1}                            
\input table_me
\end{table*}                        

Contrary to the case of the fine structure constant, the constraint on $m_{\rm e}/m_{\rm e0}$ is not dramatically improved by \Planck\ data compared to the \WMAP\ constraint, due to the strong degeneracy with $H_0$. However, we find that adding external data sets can dramatically break this degeneracy. As shown in Table~\ref{table:me-tab1}, including BAO data decreases the uncertainty on $m_{\rm e}/m_{\rm e0}$ by a factor of around 5, from $m_{\rm e}/m_{\rm e0}=0.977^{+0.055}_{-0.070}$ to $m_{\rm e}/m_{\rm e0}=1.004\pm 0.011$.

Similarly, adding \HST\ data provides a tight constraint at roughly the 1\,\% level, $m_{\rm e}/m_{\rm e0}=1.027\pm 0.012$; however, in this case, the mean value of $m_{\rm e}/m_{\rm e0}$ is $2.3\,\sigma$ higher than unity. This is expected from the positive correlation between $m_{\rm e}/m_{\rm e0}$ and $H_0$ and from the ``high" (compared to the \Planck\ determination) \HST\ value of $H_0$. 

\section{Simultaneous variation of 
$\boldsymbol\alpha$ and $\boldsymbol m_\elec$}\label{sec2b}

\begin{figure}[htb]
\centering
\includegraphics[width=7cm]{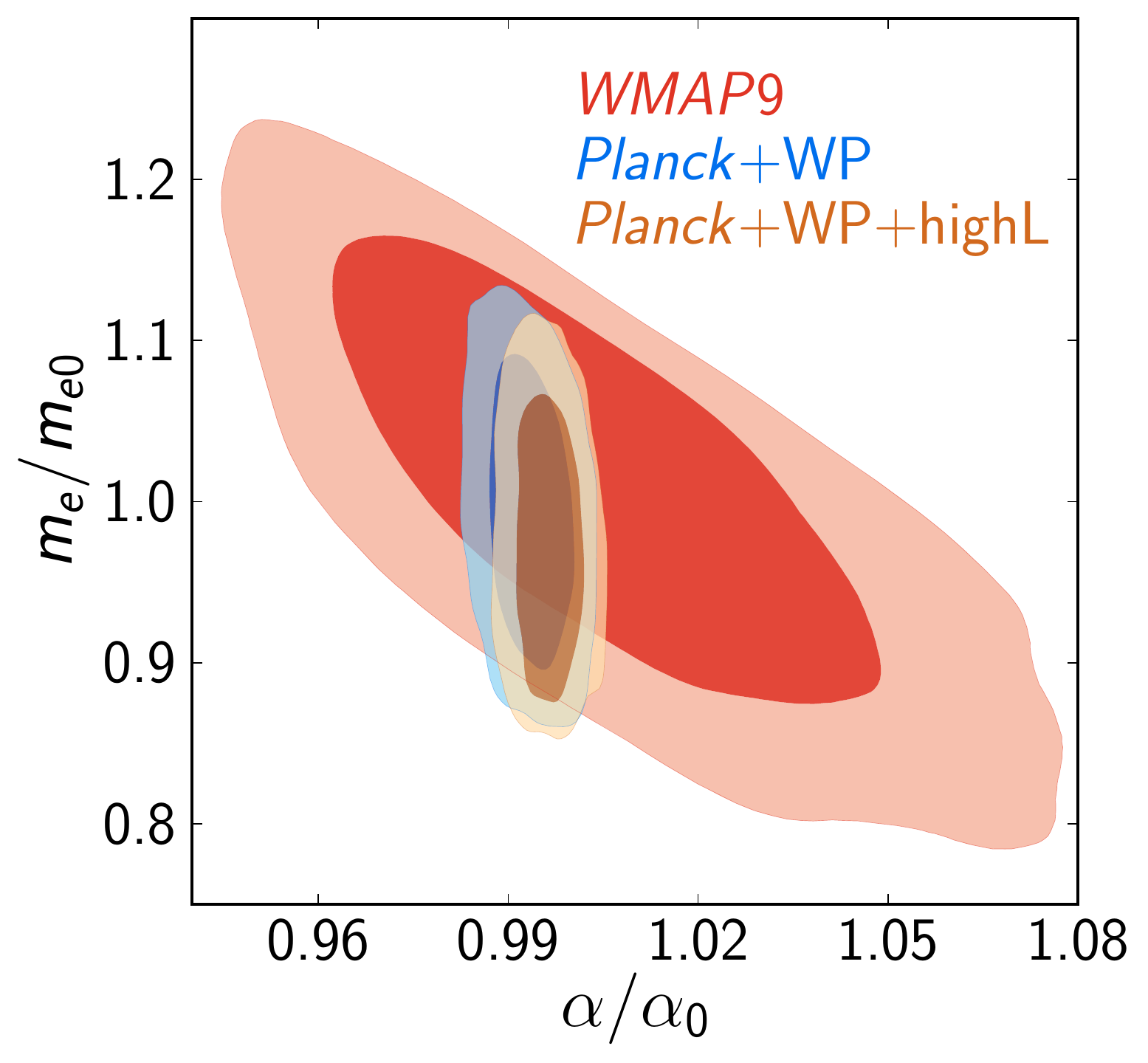}
\caption{Two-dimensional likelihood contours (68\,\% and 95\,\%) in the ($\alpha/\alpha_0,m_{\rm e}/m_{\rm e0}$) plane for \Planck+\WP\ (blue) and \Planck+\WP+\highL\ (yellow) data combinations. We also show the results using \WMAP\ data in red.}
\label{alpha_me}
\end{figure}

\begin{table*}[htmb]                 
\caption{Constraints on the cosmological parameters for the base $\Lambda$CDM\ model with the simultaneous addition
of a varying fine-structure constant and mass of the electron. We quote $68\,\%$ CL uncertainties.
}                        
\label{table:alpha-me}                            
\input table_alpha_me
\end{table*} 

The previous analysis can easily be generalized to include simultaneous variations of the electron mass $m_{\rm e}$ and $\alpha$.
We follow exactly the same procedure as in the previous section but now with an 8-dimensional parameter space. The constraint in the plane $(\alpha,\me)$ is depicted in Fig.~\ref{alpha_me}, while constraints on single parameters are presented in Table~\ref{table:alpha-me}. It is interesting to note that the constraints on $\alpha$ are not substantially changed by varying $\me$ at the same time, as we obtain $\alpha/\alpha_0=0.993\pm0.0045$ and $\me/\mezero=0.994\pm0.059$. This is due to the fact that the effect of the fine structure constant on the damping tail is different enough to break the degeneracy both with $\me$ and $H_0$, as described in Appendix \ref{effects_alphame}.

\section{Spatial variation of $\boldsymbol\alpha$}\label{sec3}

Recent analyses of quasar absorption spectra data have led to a claim that there might exist a dipole in the fine structure constant~\citep{dipole1,dipole2,dipole3}. Combining the observations of 154 absorption systems from VLT and 161 absorption systems from the Keck telescope, it was concluded \citep{dipole3} that the variation of the fine structure constant is well-represented by an angular dipole pointing in the direction ${\rm RA} = 17.3~h \pm 1.0~h$, ${\rm Dec} = -61^\circ \pm 10^\circ$, with amplitude $\Delta\alpha/\alpha=0.97^{+0.22}_{-0.20}\times10^{-5}$ (68\,\% CL). This measured value thus appears to be discrepant with zero at the $4\,\sigma$ level. However, this claim has been met with some scepticism. Webb et al.\ themselves admitted the presence of an unexplained error term in the quasar data set, and a compilation of potential issues in their analysis can be found in~\cite{cameron2012evidence} and \cite{cameron2013evidence}. In these studies, it is argued that the close alignment between the equator of the dipole and the North-South divide between the typical sight-lines of the two telescopes (VLT and Keck) used to collect the data might play a role in the detection of a dipole modulation; systematic errors of opposite signs could be responsible for the apparent signal. Furthermore, all estimates of statistical significance in \cite{dipole1} assumed unbiased Gaussian errors, but this might not be correct in the case considered. Finally~\cite{cameron2012evidence} performed a parametric Bayesian model selection analysis of the very same data set as~\cite{dipole1}, pointing out an incomplete understanding of the observational errors and a lack of theoretical expectation for a spatial variation of $\alpha$.
All these issues motivate us to test for dipole modulation of $\alpha$ using different methods and alternative kinds of data. 

From a theoretical point of view,  a dipolar modulation could be realized in some models~\citep{wall,peloso} and extend to high redshift.  Moreover, it was pointed out that a specific signature on the CMB anisotropy is expected~\citep{tilted} if we simply consider a gradient in $\alpha$ across our Hubble volume.  Such a modulation is different from a randomly fluctuating fine structure constant, which may appear if $\alpha$ depends on a light scalar field that has developped super-Hubble correlations during inflation.
A spatial fluctuation of the fine structure constant also induces non-Gaussian temperature and polarization correlations in the CMB, as well as $B$-modes in the polarization~\citep{Sigurdson,Pitrou08}. The amplitude of such a stochastic fluctuation has recently been constrained to be $\delta \alpha/\alpha_0= (1.34\pm 5.82) \times 10^{-2}$ at the 95\,\% confidence level~\citep{bryan} on scales larger than $10^\circ$ using \Planck\ data.

Here we will search specifically for a dipole signature on the last-scattering
surface.
We first recall in Sect.~\ref{subsec4.1} the effect of a spatial variation of fundamental constants on the CMB, in order to show that it implies mode couplings between the $a_{\ell m}$. We describe a statistical estimator based on~\cite{HansonLewis} in Sect.~\ref{subsec4.3}. This estimator is then used in \Planck-like simulations without (Sect.~\ref{subsec4.4}) and with (Sect.~\ref{subsec4.5}) a dipole modulation in $\alpha$ in order to study the effect of the masking, which induces additional bias in this estimator. We calibrate the estimator in Sect.~\ref{subsec4.6}, and we finally apply the method to the \Planck\ data in Sect.~\ref{resultsspace}.

\subsection{Dipolar modulation in the fundamental constants}\label{subsec4.1}

Following the claim of~\cite{dipole1}, we assume that some constants $c_p$ have a spatial variation that can be described by a dipolar modulation, i.e.,
\begin{equation}
 c_p(\hatn,z) = c_{p}(z) + \sum_{i=-1}^1\delta c_p^{(i)}(z)Y_{1i}(\hatn).
\end{equation}
Here the monopole $c_{p}(z)=c_{p,0}$ is assumed to be independent of time (otherwise we are back to the analysis of the previous sections), so that it reduces to the value measured locally today. The quantities $\delta c_p^{(i)}$ are three parameters that characterize the amplitude and direction of the modulation, the amplitude $\delta c_{p}$ being defined as: 
\begin{equation}\label{e:ampli_dipole}
\delta c_{p}\equiv\frac{1}{2}\sqrt{\frac{3}{\pi}}\sqrt{|\delta c_{p}^{(0)}|^{2}+2|\delta c_{p}^{(1)}|^2}.
\end{equation}

Since $c_p(\hatn,z)$ has to be real, they are related by
\begin{equation}
 \delta c_p^{(0)}\in\mathbb{R}, 
 \qquad
  \delta c_p^{(+1)} = -  \left[\delta c_p^{(-1)}\right]^*. 
\end{equation}
As a consequence studying $\{\delta c_p^{(0)},\delta c_p^{(1)},\delta c_p^{(-1)}\}$ is equivalent to studying $\{\delta c_p^{(0)},\mathrm{Re}(\delta c_p^{(1)}),\mathrm{Im}(\delta c_p^{(1)})\}$. For simplicity, we assume that $\delta c_p^{(i)}$ does not depend on redshift in the range probed
by primary CMB anisotropies (i.e., $600<z<{\rm few}\ \times 10^6$).
This is a good approximation as long as: (i) we do not include the variation of the gravitational constant or other physics in the early Universe;
and (ii) there is no high frequency variation compared to the time scale fixed by the Hubble time at recombination (see Footnote~\ref{expl}).  This means
effectively that we are assuming there is a well-defined value of $\alpha$ in
every direction on the last-scattering surface, but that the value could depend
on direction, with a $\cos\theta$ dependence.  The monopole is the value of the
constant at last-scattering -- which for the purposes of this section we
assume to be the same as the value today.

As previously explained, any change in the constants induces a change in the dynamics of the recombination process. It follows that the CMB temperature anisotropy will be modulated as
\begin{eqnarray}
 \Theta(\hatn) &=&  \Theta[\hatn,c_p(\hatn)]\\
                           &=&  \Theta\left[\hatn,c_{p,0} + \sum_{i=-1}^1\delta c_p^{(i)}Y_{1i}(\hatn)\right] \nonumber\\
                           &\simeq&  \bar \Theta(\hatn) +\sum_p \sum_{i=-1}^{+1} \frac{\partial \bar\Theta(\hatn)}{\partial c_{p}}
                           \delta c_p^{(i)}Y_{1i}(\hatn),
                		\label{e.taylor}
\end{eqnarray}
where, again, $\bar\Theta$ refers to the usual temperature fluctuation field, which is computed assuming the standard values of the constants and which is statistically homogeneous and isotropic. In the following, we assume that $ \Theta$ depends only on $c_{p}$, but not on its space-time derivative. On the one hand, this is a good approximation when dealing with a dipole fluctuation of weak amplitude because higher orders will be negligible.  However, at a theoretical level the dependence of $\bar\Theta$ versus $\hatn$ might be more complicated,
depending on the model.

Decomposing both $ \Theta(\hatn)$ and $\bar \Theta(\hatn)$ in spherical harmonics, as in Eq.~(\ref{e.sph-dec}), it can be shown that (see \citealt{prunet2005} for details):
\begin{eqnarray}\label{eq7}
 a_{\ell m} &=&\bar a_{\ell m}+
     \sqrt{\frac{3}{4\pi}}\sum_p\sum_i \delta c_p^{(i)} (-1)^m 
       \sum_{LM}\,  \frac{\partial \bar a_{LM}}{\partial c_{p}}\\
   &\times&\sqrt{(2\ell+1)(2L+1)}
     \prodtroisj{\ell}{L}{1}{-m}{M}{i}.\nonumber
\end{eqnarray}
Because of the triangular inequality, the Wigner $3j$-symbols are non-zero only when $L=\ell\pm1$ and $M=m- i$, so that $a_{\ell m}$ is in fact a sum involving $\bar a_{\ell m}$ and $\bar a_{\ell\pm 1 ~m- i}$. 

It is clear from Eq.~(\ref{eq7}) that such a dipolar modulation will develop $(\ell,\ell+1)$ correlations that can be characterized by the two quantities
\begin{eqnarray}
 D_{\ell m}^{(i)}&\equiv& \left<a_{\ell m} \,a_{\ell+1~m+i}^*\right>\label{eq10}
\end{eqnarray}
for $i=0,1$. The $D_{\ell m}$s will be non-zero only if any of the $\delta c_p^{(i)}$ are non-zero. Using the usual diagonal covariance property of the $\bar a_{\ell m}$, we deduce that
\begin{eqnarray}\label{eq12}
 D_{\ell m}^{(i)} &=& f_i(\ell,m) \sum_p \delta c_p^{(i)}\Gamma^{(p)}_\ell,
\end{eqnarray}
with
\begin{eqnarray}\label{eq13}
 f_0(\ell,m)&=& \sqrt{\frac{3}{4\pi}}\,\frac{\sqrt{(\ell+1)^2 - m^2}}{\sqrt{(2\ell+1)(2\ell+3)}},\\
 f_1(\ell,m) &=&\sqrt{\frac{3}{8\pi}}\, \sqrt{\frac{(\ell+2+m)(\ell+1+m)}{(2\ell+1)(2\ell+3)}},
\end{eqnarray}
and where we have defined the quantity
\begin{equation}
\Gamma^{(p)}_\ell\equiv \frac{1}{2}\left(\frac{\partial\bar C_\ell}{\partial c_p} + \frac{\partial\bar C_{\ell+1}}{\partial c_p} \right).
\end{equation}
A central ingredient here is the quantity $\partial \bar C_\ell/\partial c_p$, the computation of which is detailed in Appendix~\ref{appA} and explicitly given in Fig.~\ref{fig:dclda} for the case $c_p=\alpha$.
Note that essentially the same estimator for dipole modulation, introduced
by \cite{prunet2005}, has been used to discuss power asymmetry
\cite{HansonLewis}, aberration \citep[e.g.,][]{planck2013-pipaberration},
and more general parameter anisotropy \citep{tilted}, through substituting
different $c_p$.

This construction assumes that we are using the full sky, and one needs to keep in mind that any mask will violate isotropy and induce additional correlations, and thus bias any estimator. Let us note that in general the temperature field on a masked sky is given by
\begin{equation}\label{eq20}
  \Theta(\hatn) =\left\lbrace \bar \Theta(\hatn) +\sum_p \sum_{i=-1}^{+1} \frac{\partial \bar\Theta(\hatn)}{\partial c_{p}}
                           \delta c_p^{(i)}Y_{1i}(\hatn)\right\rbrace W(\hatn),
\end{equation}
where $W(\hatn)$ is a window function for the mask, which can be decomposed into spherical harmonics as
\begin{equation}\label{eq21}
 W(\hatn)=\sum_{\ell m}\,w_{\ell m}\,Y_{\ell m}(\hatn).
\end{equation}
Since $W(\hatn)$ is a real-valued function, this implies that $w_{\ell m}^*=(-1)^mw_{\ell~-m}$. We deduce from Eqs.~(\ref{e.sph-dec}) and (\ref{eq20}) that
\begin{equation}\label{eq22}
 a_{\ell m} = a^{\mathrm{masked}}_{\ell m}  +\sum_i\delta c_p^{(i)}A^{(i)}_{\ell m},
\end{equation}
where $a^{\mathrm{masked}}_{\ell m}$ are the coefficients of the masked primordial temperature field $\Theta^{\mathrm{masked}}(\hatn)=\bar\Theta(\hatn)W(\hatn)$,
\begin{eqnarray}\label{eq22b}
 \ a^{\mathrm{masked}}_{\ell m}&=& \sum_{\ell_1m_1}\,
 \bar a_{\ell_1m_1}\sum_{\ell_2m_2}\,w_{\ell_2m_2}\nonumber\\
&&\times \int\dd^2\hatn\, Y_{\ell_1m_1}(\hatn)Y_{\ell_2m_2}(\hatn)
 Y_{\ell m}^*(\hatn),
\end{eqnarray}
and the effects of the modulation are encoded in the correction
\begin{eqnarray}\label{eq22c}
 A^{(i)}_{\ell m} &=& \sum_{\ell_1m_1}\,
 \frac{\partial \bar a_{\ell_1m_1}}{\partial c_p}\sum_{\ell_2m_2}\,w_{\ell_2m_2}\\
 &&\times\int\dd^2\hatn\, Y_{\ell_1m_1}(\hatn)Y_{\ell_2m_2}(\hatn)Y_{1i}(\hatn)\nonumber
 Y_{\ell m}^*(\hatn).
\end{eqnarray}
These results have already been presented in \cite{prunet2005} to search for a dipole signal in \WMAP\ masked maps; the estimator used in \cite{prunet2005} was the precursor of the optimal estimator of~\cite{HansonLewis} that we use in our analysis.

\subsection{Optimal estimator}\label{subsec4.3}

To constrain the effect of a spatial variation of the fundamental constants, we need to adapt the estimator proposed by~\cite{HansonLewis}. For that purpose, we start from Eq.~(\ref{e.taylor}), which reads, in terms of harmonic coefficients,
\begin{eqnarray}
a_{\ell m} &\simeq &a^{\mathrm{masked}}_{\ell m}\\
 &+& \sum_p\sum_{LM}\sum_i {\partial \bar a_{LM} \over \partial c_p}\delta c_p^{(i)} \int \dd^2 \hatn Y^*_{\ell m}(\hatn)Y_{LM}(\hatn)Y_{1i}(\hatn),
  \nonumber
\end{eqnarray}
where, contrary to Eq.~(\ref{eq7}), we do not evaluate the integral over the sky, but over a window function, as we are working on a masked sky.
This allows us to compute the covariance matrix $C_{\ell_1 m_1,\ell_2 m_2} \equiv \langle a^{\vphantom{*}}_{\ell_1 m_1}a^*_{\ell_2 m_2} \rangle$ to first order in $\delta c_p^{(i)}$:
\begin{eqnarray}
C_{\ell_1 m_1,\ell_2 m_2} 
&=& \delta_{\ell_1\ell_2}\delta_{m_1m_2}C_{\ell_1} \nonumber\\
&&+ {1\over 2}\sum_p\sum_i \delta c_p^{(i)}\left[{\partial C_{\ell_1}\over \partial c_p} + {\partial C_{\ell_2}\over \partial c_p}\right] \nonumber\\
&& \times \int \dd^2\hatn Y_{1i}(\hatn)Y^*_{\ell_1m_1}(\hatn)Y_{\ell_2 m_2}(\hatn).
\end{eqnarray}
It follows that the unnormalized quadratic maximum likelihood (QML) estimator proposed by \cite{HansonLewis} takes the form
\begin{eqnarray}
\widetilde{\delta c_p^{(i)}} &=&\sum_p \int \dd^2\hatn Y^*_{1i}(\hatn)\left[\sum_{\ell_1 m_1}\underline{a}_{\ell_1 m_1}Y_{\ell_1 m_1}(\hatn)\right] \nonumber\\
	&& \times \left[\sum_{\ell_2 m_2}{1\over 2}{\partial C_{\ell_2}\over \partial c_p} \underline{a}_{\ell_2 m_2}Y_{\ell_2 m_2}(\hatn)\right],
\end{eqnarray}
where we have introduced the data weighted by the inverse of the covariance,
\begin{equation}
 \underline{a}_{\ell m} = \sum_{\ell' m'} (C_{\rm obs}^{-1})_{\ell m \ell' m'}a_{\ell' m'}\label{eq:inv-cov-weight}.
\end{equation}
Even with a mask, one can use the full-sky approximation in which $C_{\rm obs}^{-1} \simeq 1/(C_\ell b_\ell^2 + N_\ell)$, where $C_\ell$ is the angular power spectrum of the CMB, $b_\ell^2$ is the beam, and $N_\ell$ is the power spectrum of the noise. However, we emphasize here that even in an ideal case (i.e., no noise and no mask), this estimator is biased; we will show below how to eliminate the bias using simulations.

In the following sub-sections, we restrict the study to a single varying constant, the fine structure constant $\alpha$, so that $c_p=\alpha$.  We assume that
this constant has a dipole pattern around our last-scattering surface, but no
monopole.

\subsection{Simulation of maps without a modulated signal}\label{subsec4.4}

In order to calibrate the estimators described above, we need to simulate CMB maps with and without a modulation of the fine structure constant. We first describe the simulations produced without any modulation. 

The spherical harmonic decomposition reads
\begin{equation}\label{map_no_mod}
\bar\Theta(\hatn) = \sum_{\ell m} \bar a_{\ell m} Y_{\ell m}(\hatn) = \sum_{\ell m} \sqrt{C_\ell}\epsilon_{\ell m} Y_{\ell m}(\hatn)
\end{equation} 
for the isotropic map, where $\epsilon_{\ell m}$ is an ${\cal N}(0,1)$ complex Gaussian random variable. 

We generate synthetic maps with a Gaussian beam of full width half maximum $5'$, which is the resolution of the foreground-cleaned \Planck\ CMB maps \citep{planck2013-p06}. The corresponding noise realizations were obtained from realistic simulations of the noise of each individual \Planck\ frequency map (including noise correlations in the timelines, and anisotropic sky coverage), which were propagated through the component separation filters. These should therefore faithfully sample the noise covariance of the CMB foreground-cleaned map. 

\begin{figure*}[h!tpb]

\includegraphics[width=0.50\textwidth]{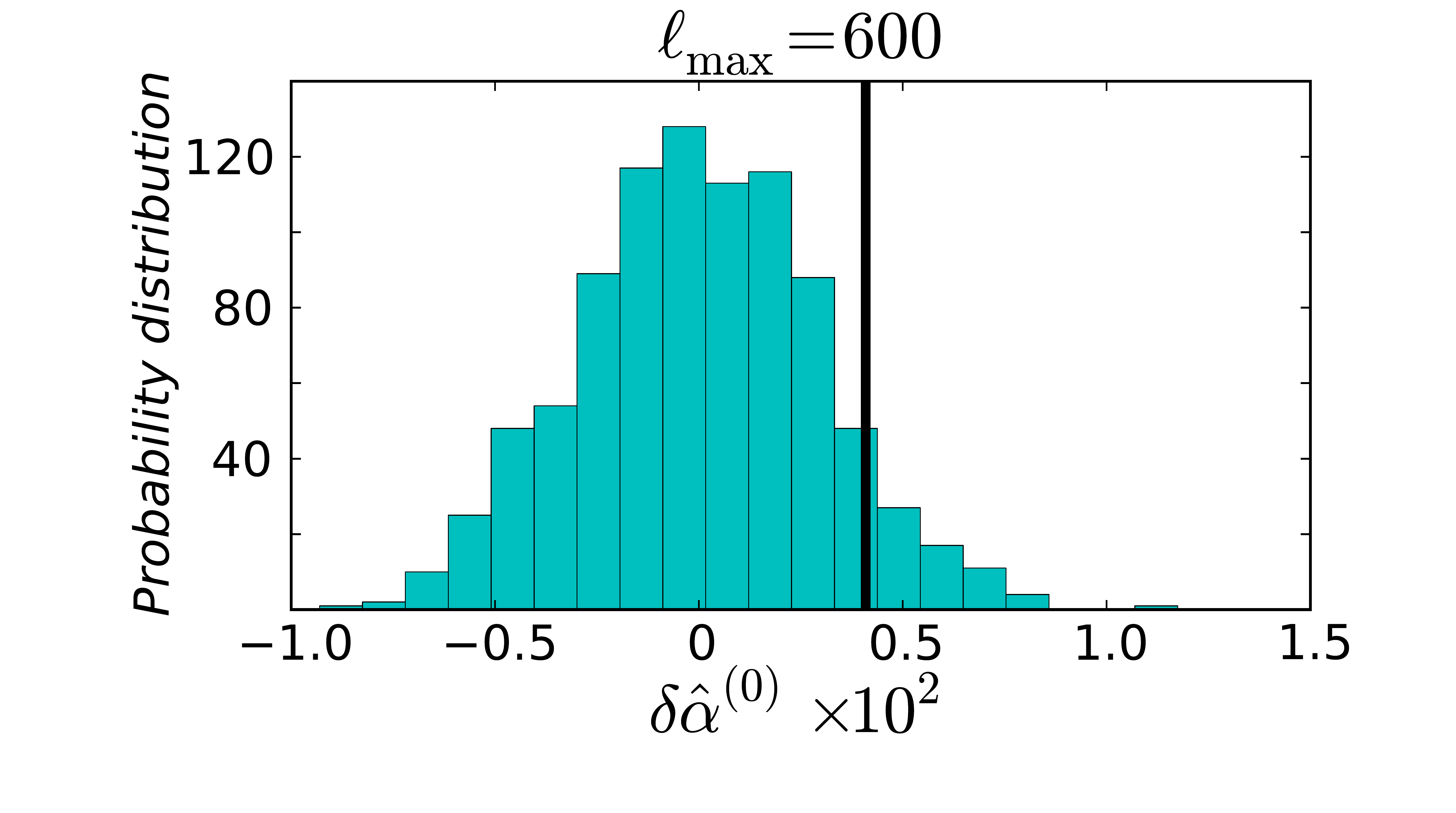}\includegraphics[width=0.50\textwidth]
{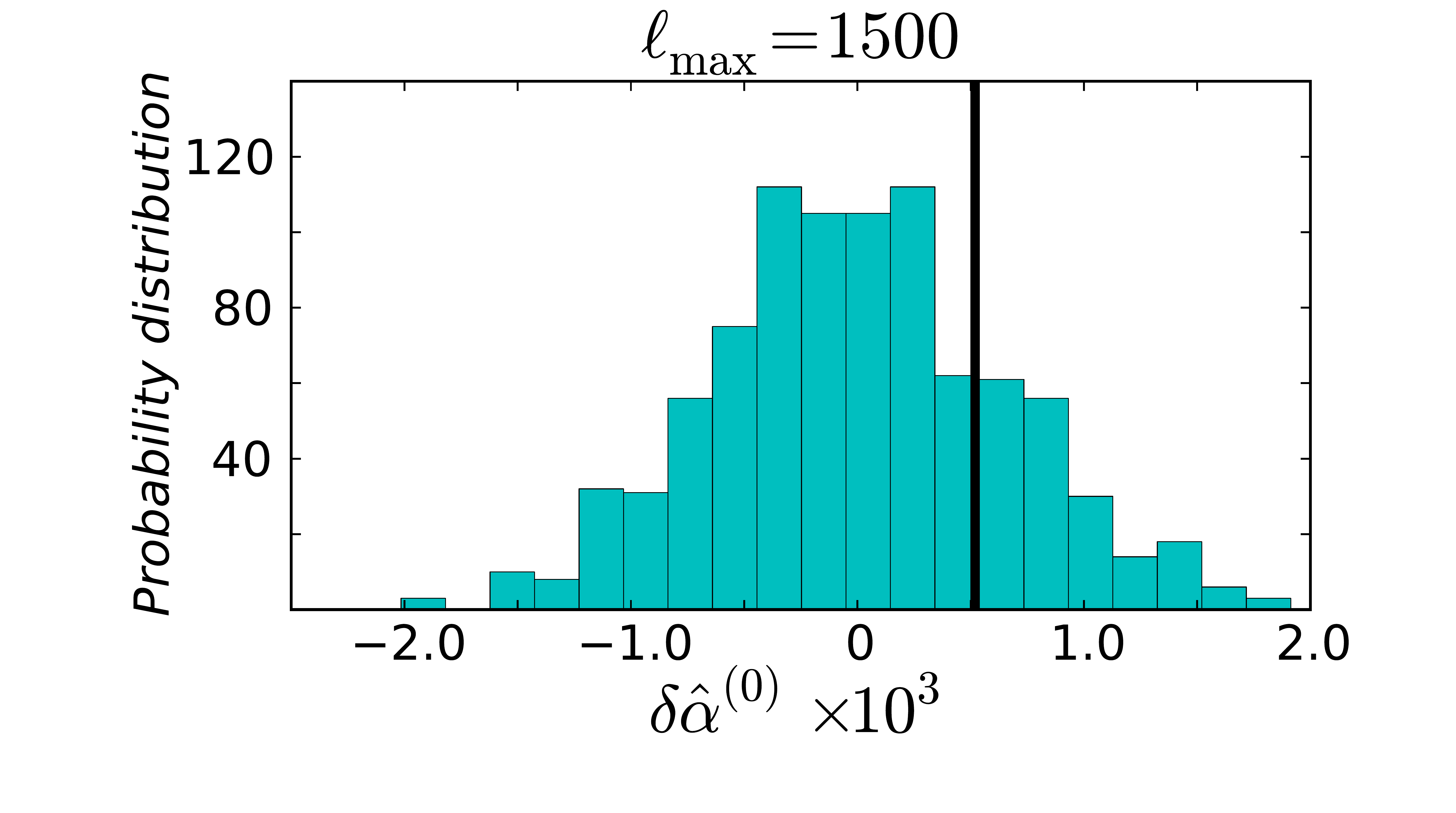}\\
\includegraphics[width=0.50\textwidth]{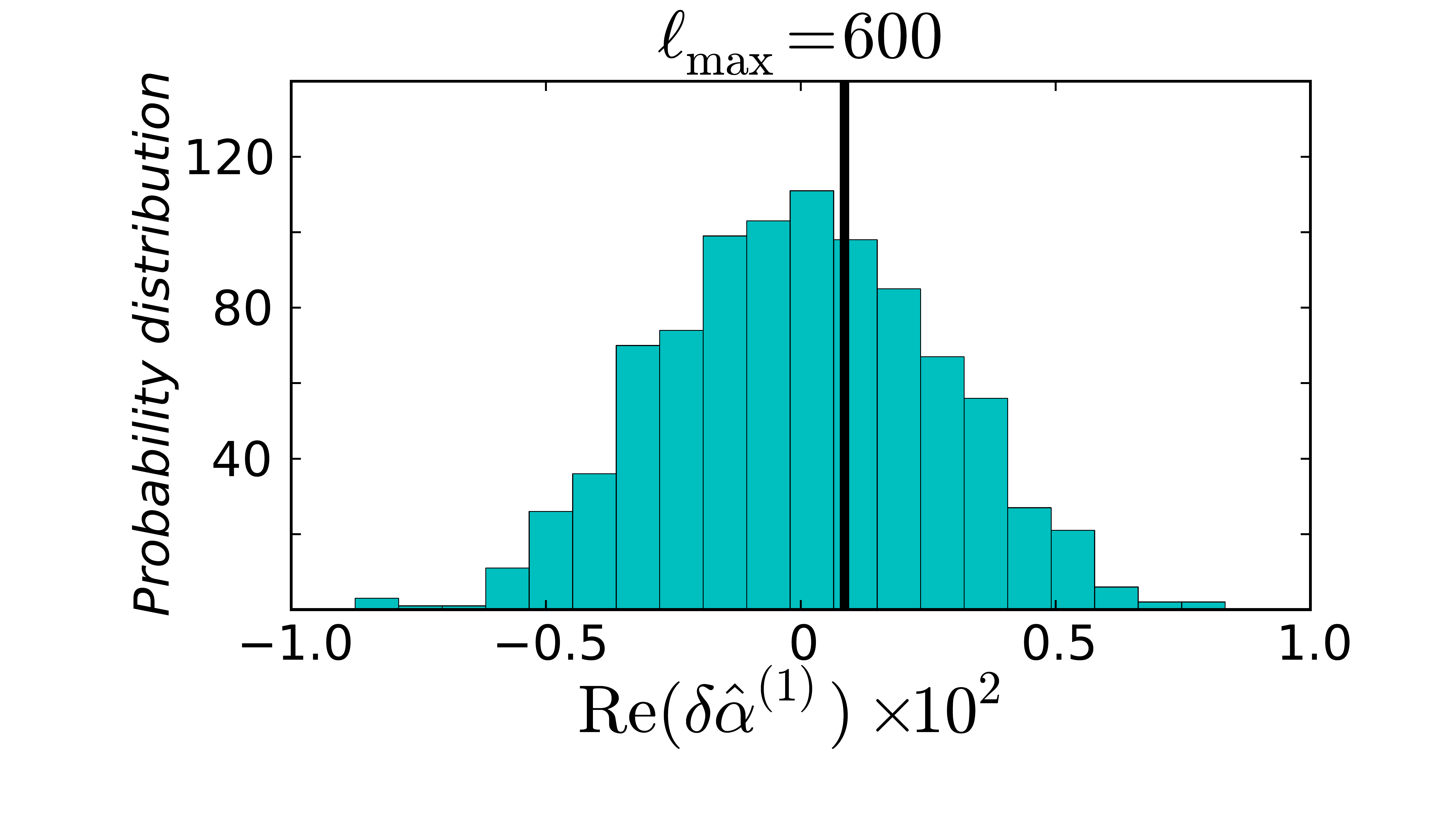}
\includegraphics[width=0.50\textwidth]{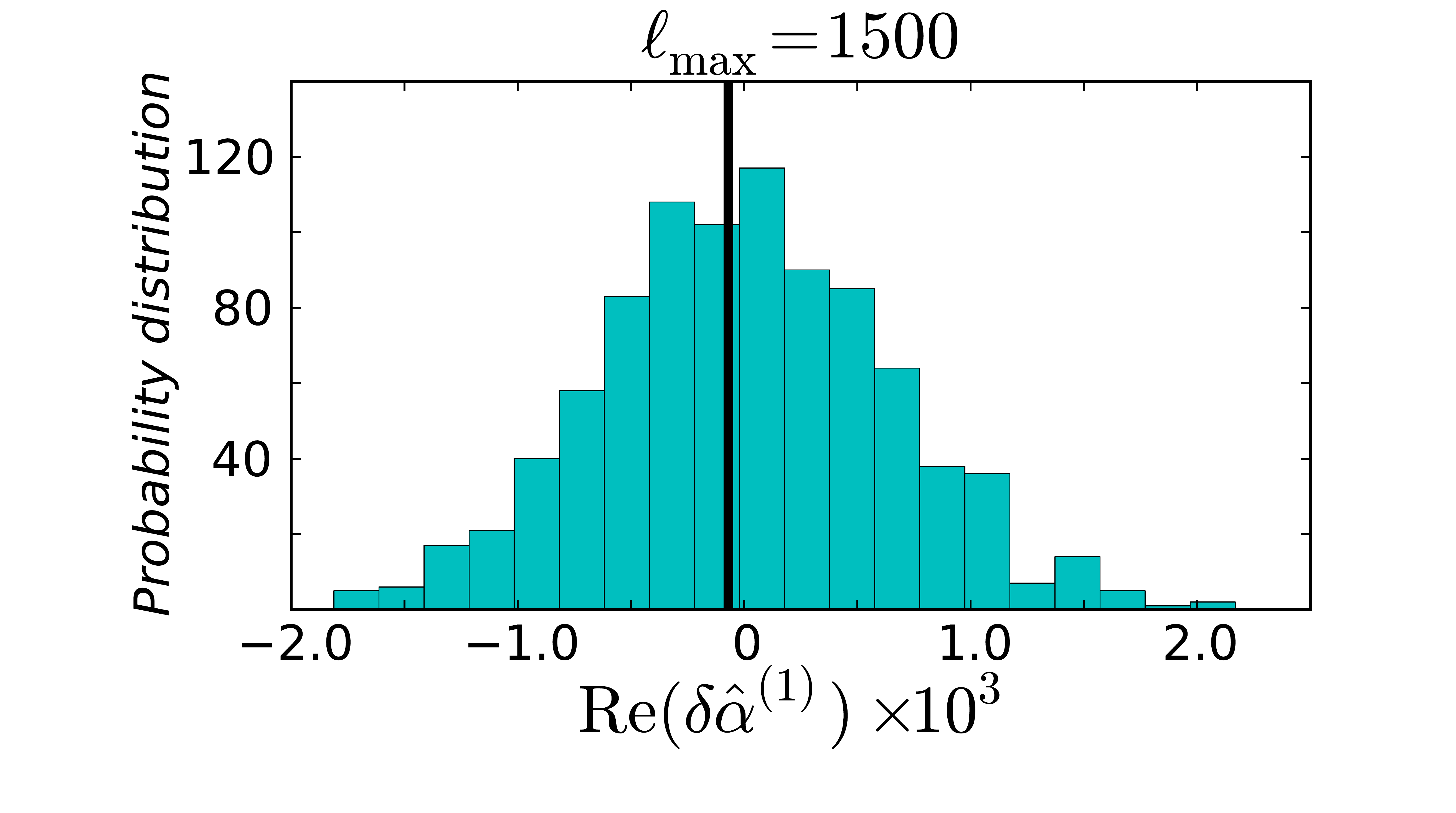}\\
\includegraphics[width=0.50\textwidth]{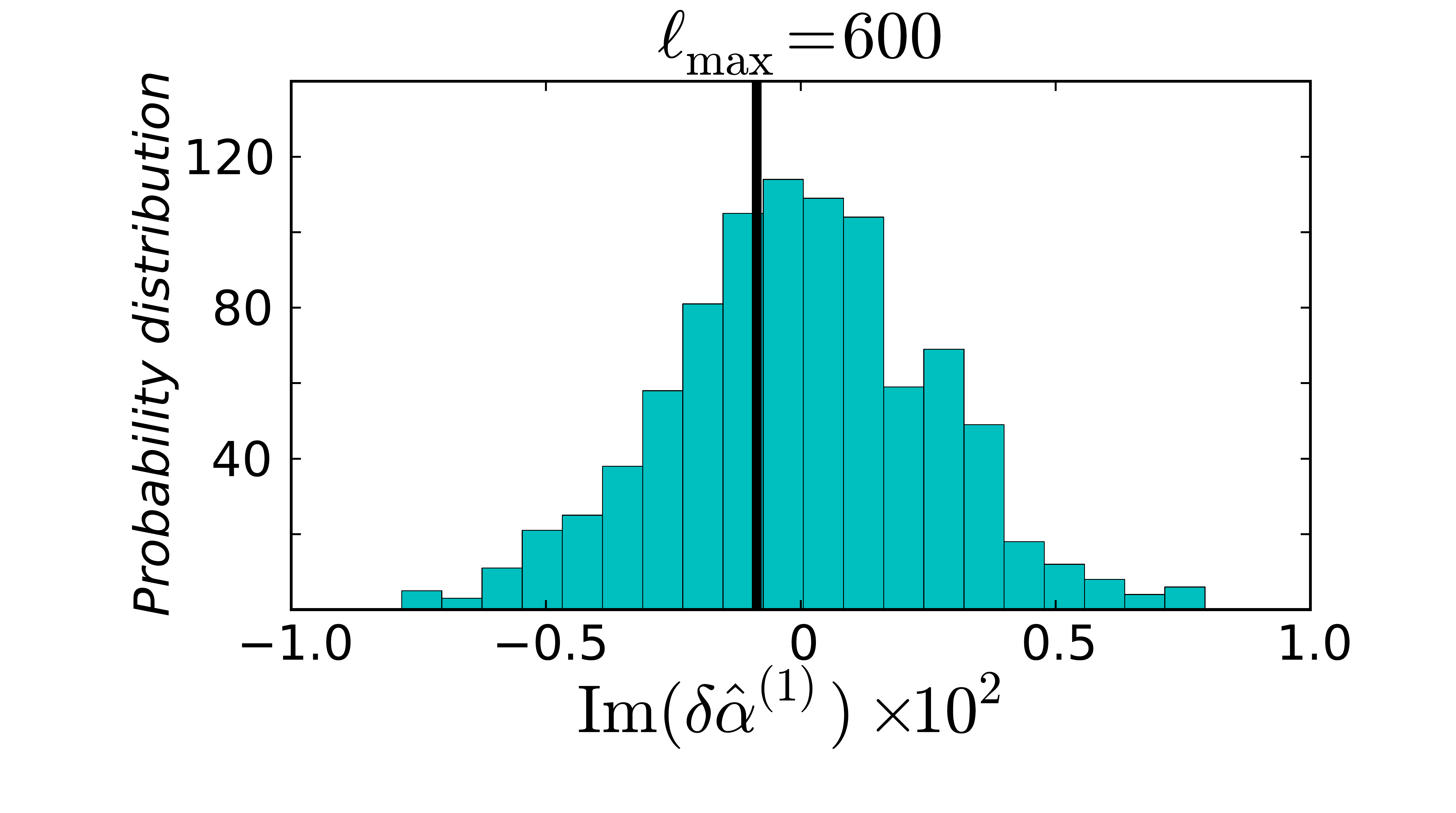}\includegraphics[width=0.50\textwidth]{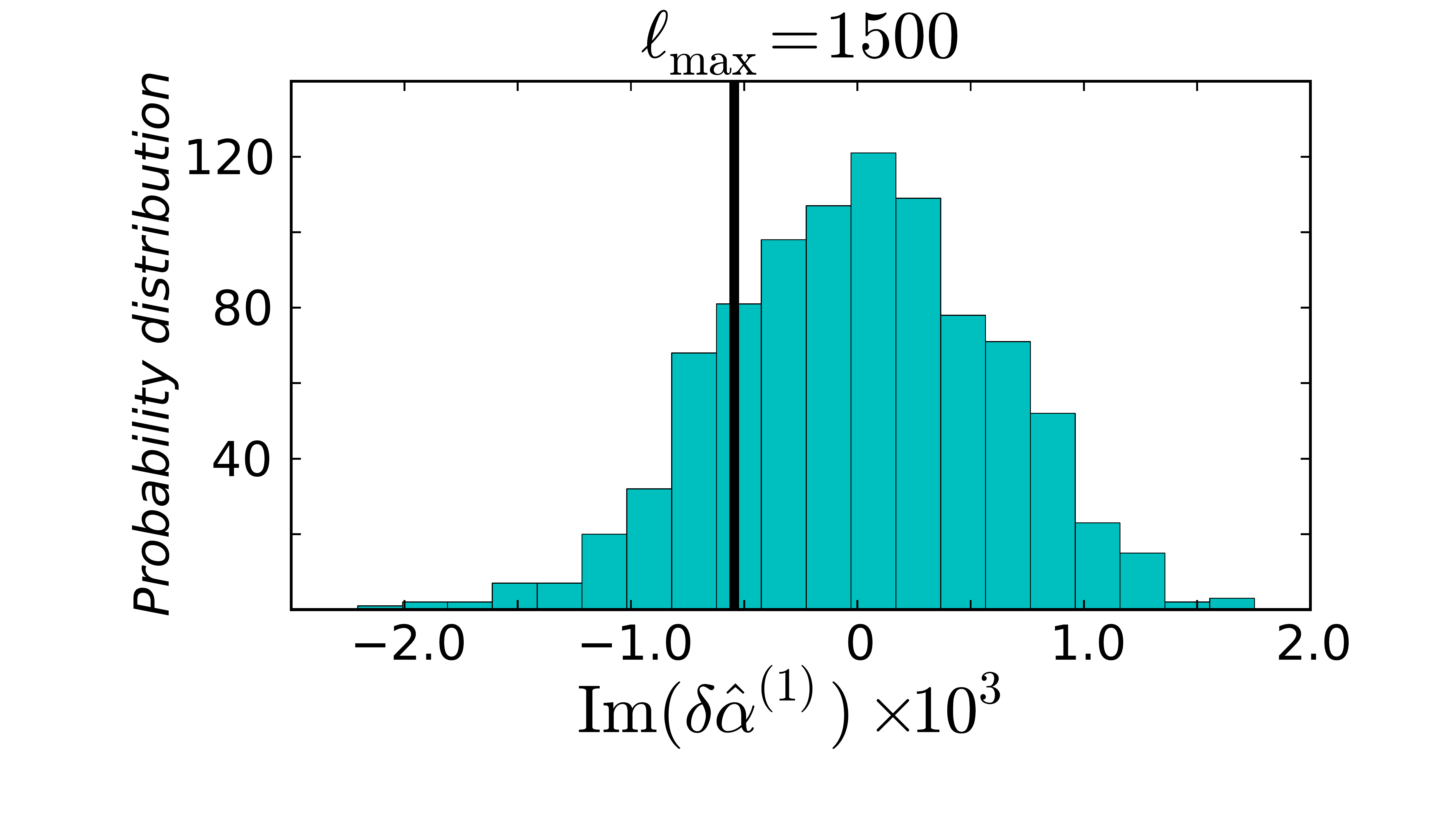} 
\caption{\emph{Left}: histogram of the distribution of the components of the dipole modulation for 900 unmodulated realizations. The distribution is not normalized. From top to bottom we show $\delta\alpha^{(0)}$, ${\rm Re}(\delta\alpha^{(1)})$ and ${\rm Im}(\delta\alpha^{(1)})$. The three histograms correspond to \Planck-like modulation-free simulated data with $\ell_{\rm max}=600$. 
The black line corresponds to the measurements in the actual \Planck\ CMB map. \emph{Right}: the same thing for $\ell_{\mathrm{max}}=1500$.}
\label{fig:space-simu}
\end{figure*}

\begin{table*}[h!tpb]
\begingroup
\newdimen\tblskip \tblskip=5pt
\caption{Mean fields (computed with 500 realizations of modulation-free \Planck-like maps), variances (computed by estimating dipole components on 900 realizations of modulation-free \Planck-like maps) and diagonal elements of $\mathcal{F}$ (computed with 400 realizations of modulated \Planck-like maps) of each component of the modulation for the \cite{HansonLewis} estimator with $\ell_{\mathrm{max}}=600$ and $\ell_{\mathrm{max}}=1500$.}
\label{tab:compo_table}
\vskip -3mm
\footnotesize
\setbox\tablebox=\vbox{
 \newdimen\digitwidth
 \setbox0=\hbox{\rm 0}
 \digitwidth=\wd0
 \catcode`*=\active
 \def*{\kern\digitwidth}
 \newdimen\signwidth
 \setbox0=\hbox{+}
 \signwidth=\wd0
 \catcode`!=\active
 \def!{\kern\signwidth}
 \halign{\tabskip=0pt\hbox to 1.5in{#\leaderfil}\tabskip=2em&
 \hfil#\hfil\tabskip=2em&
 \hfil#\hfil\tabskip=2em&
 \hfil#\hfil\tabskip 0pt\cr
\noalign{\doubleline}
\multispan2\hfil Quantities\hfil& $!\ell_{\rm max}=600$& $\ell_{\rm max}=1500$\cr
\noalign{\vskip 4pt\hrule\vskip 6pt}
Variances& $\sigma_{\delta\alpha^{(0)}}$&            $!2.95\times 10^{-3}$&
 $!6.50\times 10^{-4}$\cr
\omit    & $\sigma_{\text{Re}(\delta\alpha^{(1)})}$& $!2.70\times 10^{-3}$&
 $!6.45\times 10^{-4}$\cr
\omit    & $\sigma_{\text{Im}(\delta\alpha^{(1)})}$& $!2.61\times 10^{-3}$&
 $!5.97\times 10^{-4}$\cr
\noalign{\vskip 4pt\hrule\vskip 6pt}
Mean fields&
        $\mathcal{F}^{-1}\langle\widetilde{\delta\alpha^{(0)}}\rangle$&
 $!5.93\times10^{-3}$&  $!4.57\times10^{-3}$\cr
\omit& $\mathcal{F}^{-1}\langle\text{Re}(\widetilde{\delta\alpha^{(1)}})\rangle$&
 $-1.50\times10^{-2}$& $-7.97\times10^{-3}$\cr
\omit& $\mathcal{F}^{-1}\langle\text{Im}(\widetilde{\delta\alpha^{(1)}})\rangle$&
 $-1.84\times10^{-2}$& $-9.46\times10^{-3}$\cr
\noalign{\vskip 4pt\hrule\vskip 6pt}
Diagonal elements of $\mathcal{F}$&
  $\mathcal{F}_{00}$&   $!3.24\times10^{5}$& $!1.67\times10^{8}$\cr
\omit& $\mathcal{F}_{11}$&   $!1.75\times10^{5}$& $!9.08\times10^{7}$\cr
\omit& $\mathcal{F}_{-1-1}$& $!2.00\times10^{5}$& $!1.04\times10^{8}$\cr
\noalign{\vskip 4pt\hrule\vskip 6pt}
}}
\endPlancktablewide
\endgroup
\end{table*}

\begin{figure*}[h!tpb]
\includegraphics[width=0.5\textwidth]{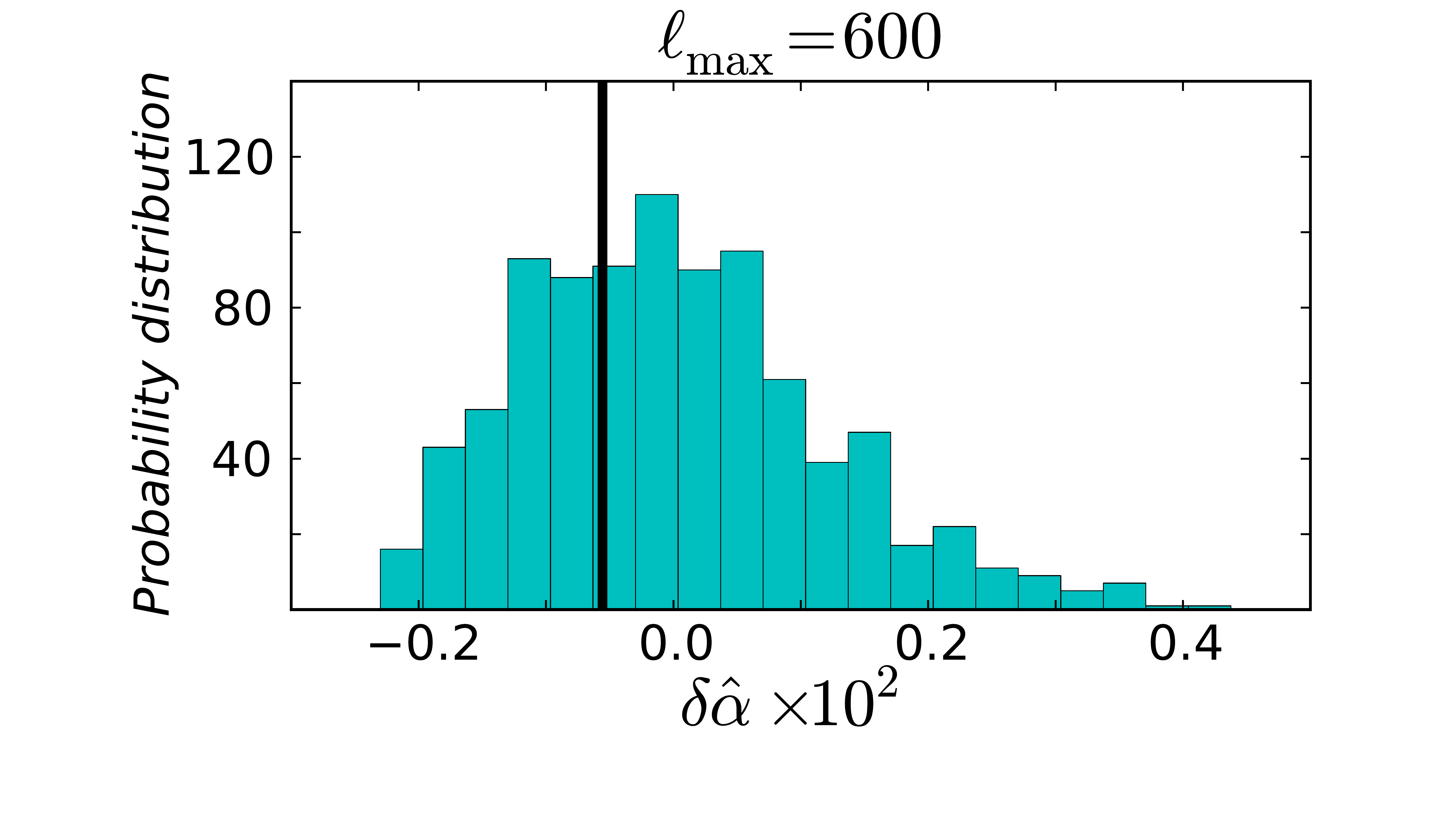}\includegraphics[width=0.5\textwidth]{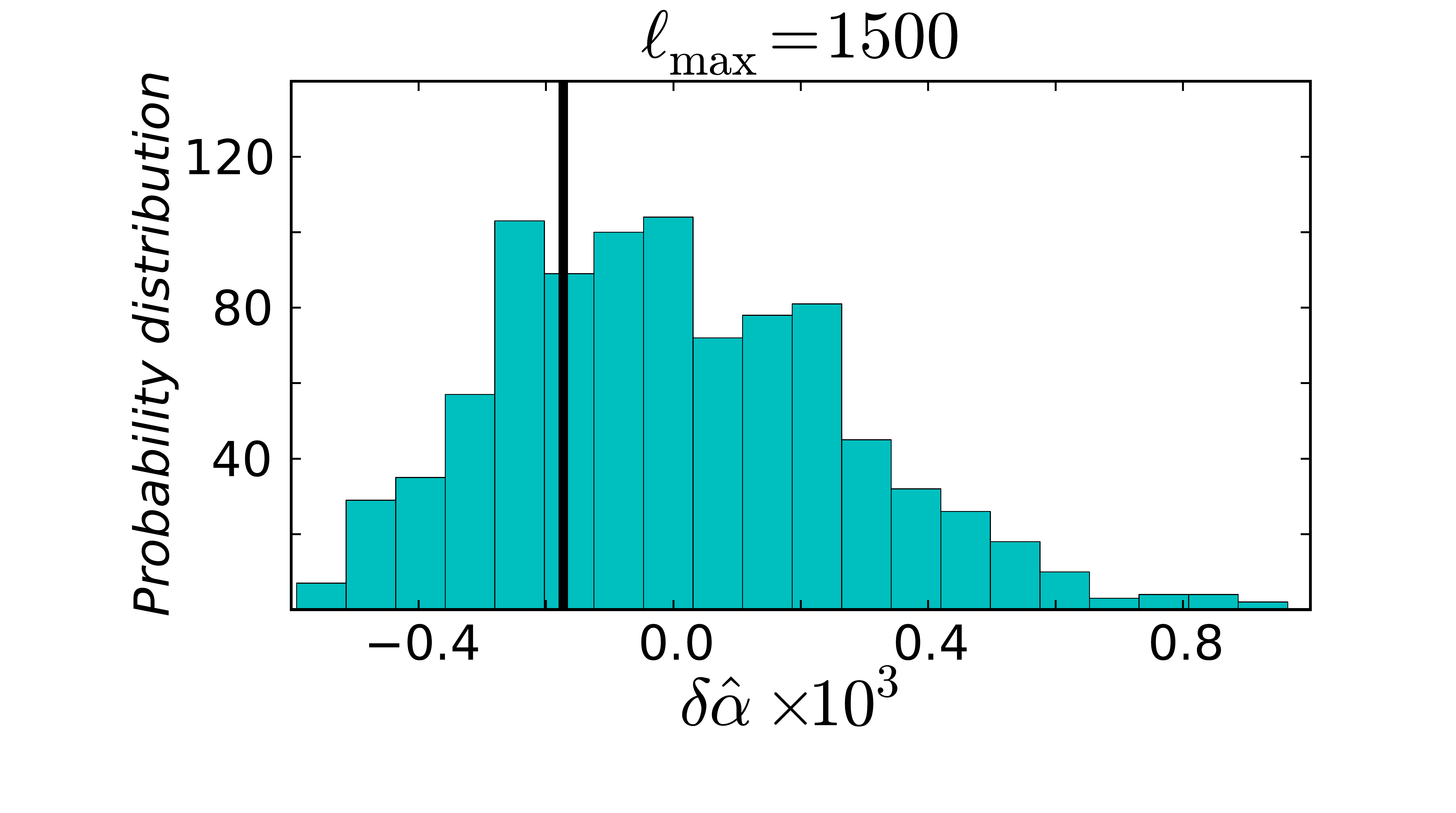}
\caption{Histogram of the distribution of the overall amplitude of the dipole modulation for 900 unmodulated realizations. The distribution is not normalized. 
The histogram on the left corresponds to \Planck-like simulated data without modulation with $\ell_{\mathrm{max}}=600$, while the one on the right os for $\ell_{\mathrm{max}}=1500$ . In each plot, the black line corresponds to the measurements in the actual \Planck\ CMB map. 
}
\label{fig:space-simu-norm}
\end{figure*}

\begin{table*}[h!tpb]
\begingroup
\newdimen\tblskip \tblskip=5pt
\caption{Mean field $\langle\widetilde{\delta \alpha}\rangle$ and variance $\sigma_\alpha^2=\langle\delta\alpha^2\rangle-\langle\delta\alpha\rangle^2$ of the amplitude $\delta\alpha$ of the modulation for 900 \Planck\ realizations of the \cite{HansonLewis} estimator for the specific choices $\ell_{\mathrm{max}}=600$ and $\ell_{\mathrm{max}}=1500$. }
\label{tab:ampli_table}
\vskip -3mm
\footnotesize
\setbox\tablebox=\vbox{
 \newdimen\digitwidth
 \setbox0=\hbox{\rm 0}
 \digitwidth=\wd0
 \catcode`*=\active
 \def*{\kern\digitwidth}
 \newdimen\signwidth
 \setbox0=\hbox{+}
 \signwidth=\wd0
 \catcode`!=\active
 \def!{\kern\signwidth}
 \halign{\tabskip=0pt\hbox to 1.5in{#\leaderfil}\tabskip=2em&
 \hfil#\hfil\tabskip=2em&
 \hfil#\hfil\tabskip 0pt\cr
\noalign{\doubleline}
\omit\hfil Hanson-Lewis estimator\hfil&
 $\ell_{\mathrm{max}}=600$& $\ell_{\mathrm{max}}=1500$\cr
\noalign{\vskip 4pt\hrule\vskip 6pt}
Variances $\sigma_{\delta\alpha}$ $\dotfill$&
 $1.17\times10^{-3}$& $2.71\times10^{-4}$\cr
Mean fields $\langle\widetilde{\delta \alpha}\rangle$ $\dotfill$&
 $2.72\times10^{-3}$& $6.29\times10^{-4}$\cr
\noalign{\vskip 4pt\hrule\vskip 6pt}
}}
\endPlancktablewide
\endgroup
\end{table*}

\begin{table*}[h!tpb]
\begingroup
\newdimen\tblskip \tblskip=5pt
\caption{Summary of the results obtained for the amplitude of the spatial modulation of the fine structure constant with the Hanson \& Lewis \cite{HansonLewis} estimator applied to the \Planck\, data for $\ell_{\mathrm{max}}=600$ and $\ell_{\mathrm{max}}=1500$. We show uncertainties at 68\,\% CL.}
\label{tab:results_tot}
\vskip -3mm
\footnotesize
\setbox\tablebox=\vbox{
 \newdimen\digitwidth
 \setbox0=\hbox{\rm 0}
 \digitwidth=\wd0
 \catcode`*=\active
 \def*{\kern\digitwidth}
 \newdimen\signwidth
 \setbox0=\hbox{+}
 \signwidth=\wd0
 \catcode`!=\active
 \def!{\kern\signwidth}
 \halign{\tabskip=0pt\hbox to 1.5in{#\leaderfil}\tabskip=2em&
 \hfil#\hfil\tabskip=2em&
 \hfil#\hfil\tabskip 0pt\cr
\noalign{\doubleline}
\omit\hfil \Planck\ results\hfil& $\ell_{\mathrm{max}}=600$&
 $\ell_{\mathrm{max}}=1500$\cr
\noalign{\vskip 4pt\hrule\vskip 6pt}
$\widehat{\delta\alpha}$&           $-5.56\times10^{-4}\pm1.17\times10^{-3}$&
 $-1.73\times10^{-4}\pm2.71\times10^{-4}$\cr
\noalign{\vskip 4pt\hrule\vskip 6pt}
$\widehat{\delta\alpha^{(0)}}$&     $!4.09\times10^{-3}\pm2.95\times10^{-3}$&
 $!5.20\times10^{-4}\pm6.50\times10^{-4}$\cr
\noalign{\vskip 4pt\hrule\vskip 6pt}
Re$(\widehat{\delta\alpha^{(1)}})$& $!8.57\times10^{-4}\pm2.70\times10^{-3}$&
 $-6.93\times10^{-5}\pm6.45\times10^{-4}$\cr
\noalign{\vskip 4pt\hrule\vskip 6pt}
Im$(\widehat{\delta\alpha^{(1)}})$& $-8.66\times10^{-4}\pm2.61\times10^{-3}$&
 $-5.44\times10^{-4}\pm5.97\times10^{-4}$\cr
\noalign{\vskip 4pt\hrule\vskip 6pt}
}}
\endPlancktablewide
\endgroup
\end{table*}

In order to minimize the impact of foreground residual emission from the Galaxy, we adopt the CG80 mask (combined Galactic mask with 80\,\% sky coverage) from \cite{planck2013-p06}. To minimize artificial mode coupling induced by the mask, we apodize it with a cosine of $8^{\circ}$ width. We also attempted to include an additional point source mask (with or without a $30'$ apodization). However, in this case, we notice that instabilities rise in the Hanson \& Lewis estimator at large multipoles. Therefore, when building the estimator, we decided to mask the Galaxy, but to ``inpaint'' the \Planck\ CMB map inside the point source mask with constrained Gaussian realizations, as was done in the CMB lensing analysis of \cite{planck2013-p12}. 
In the present study, we have not quantified precisely the influence of the size and type of the mask; this would require considerable additional computations,
which would certainly be important if a signal was to be detected.  But in any
case, there is no reason to believe that the mask effects are substantially
different from related studies \citep[e.g.,][]{planck2013-pipaberration}.

\subsection{Simulation of maps with a modulated signal}\label{subsec4.5}

We now simulate a \Planck-like map with a dipole variation of $\alpha$, starting from the previous modulation-free CMB realizations. For this purpose, we use Eq.~(\ref{e.taylor}), which describes the Taylor expansion of the CMB temperature anisotropies in the presence of a dipolar variation of $\alpha$. Now, let us consider the harmonic decomposition of ${\partial \bar\Theta(\hatn)\over \partial \alpha}$. Given Eq.~(\ref{map_no_mod}), we obtain
\begin{equation}\label{alphamap}
{\partial\bar\Theta(\hatn)\over \partial\alpha} = \sum_{\ell m}{\partial \sqrt{C_\ell}\over \partial \alpha}\epsilon_{\ell m}Y_{\ell m}(\hatn)=\sum_{\ell m}{1 \over 2 C_\ell}{\partial C_\ell \over \partial \alpha}\bar a_{\ell m} Y_{\ell m}(\hatn).
\end{equation}
Using Eq.~(\ref{e.taylor}), Eq.~(\ref{map_no_mod}), and Eq.~(\ref{alphamap}) it is thus straightforward to generate a synthetic CMB map with a dipole modulation in $\alpha$, starting from the unmodulated maps described in Sect.~\ref{subsec4.4}.

\subsection{Calibration of the Hanson-Lewis estimator}\label{subsec4.6}

We have defined an estimator that allows us to constrain a dipolar spatial variation of the fine structure constant. As explained above, this estimator is expected to be biased, especially in the presence of a mask. 

Therefore, in order to eliminate this bias, we need to subtract the mean field\footnote{Here, $\langle\rangle$ indicate the ensemble average in the absence of modulation.}, $\langle\widetilde{\delta\alpha^{(i)}}\rangle$, computed from unmodulated simulations, and to renormalize with the help of the normalization matrix ${\mathcal{F}}$, calculated from modulated simulations, so that
\begin{eqnarray}
\widehat{\delta\alpha^{(i)}} &=&{\mathcal{F}_{ij}}^{-1}\left(\widetilde{\delta\alpha^{(j)}}-\langle\widetilde{\delta\alpha^{(j)}}\rangle\right).
\end{eqnarray}

The matrix ${\mathcal{F}}$ corresponds to the $3\times 3$ Fisher matrix of the $\delta\alpha^{(i)}$ coefficients, at least in the case where the data are precisely inverse covariance weighted (see Eq.~(\ref{eq:inv-cov-weight}). We can understand it as the inverse covariance of the estimator in the ideal case without modulation (${\mathcal{F}^{-1}_{\mathrm{no~mod}}}\simeq\sigma^2$). 

Let us define an orthonormal basis $(\hat{x},\hat{y},\hat{z})$ for our sky maps. Here $\hat{x}$ and $\hat{y}$ are two orthonormal vectors lying in the Galactic equatorial plane and $\hat{z}$ is the vector normal to this plane. The term ${\mathcal{F}_{00}}$ strongly depend on the shape of the mask in the $\hat{z}$ direction, whereas ${\mathcal{F}_{11}}$ and ${\mathcal{F}_{-1-1}}$  depends on the shape of the mask in the $\hat{x}$ and $\hat{y}$ directions. The terms ${\mathcal{F}_{11}}$ and ${\mathcal{F}_{-1-1}}$ are quite similar, but not exactly the same (${\mathcal{F}_{11}} = {\mathcal{F}_{-1-1}}$ only for a simple azimuthal mask). We expect ${\mathcal{F}_{11}}$ and ${\mathcal{F}}_{-1-1}$ to be smaller than ${\mathcal{F}_{00}}$: more information is lost in the $(\hat{x},\hat{y})$ plane because it is aligned with the Galactic plane mask, and it is more difficult to recover these modulation components than in the $\hat{z}$ direction.


In order to determine the mean field and the $\mathcal{F}$ matrix, we start by generating a set of 900 realizations of CMB temperature maps (without dipolar modulation) at a {\tt HEALPix}~\citep{gorski2005healpix} resolution of $N_{\rm side}=2048$. We then add noise and a Galactic mask to these maps, as described in Sect.~\ref{subsec4.4}.

We used 500 of these unmodulated simulations to estimate the mean field term by calculating the biased estimator $\widetilde{\delta\alpha^{(i)}}$ for each of these maps and by performing the ensemble average over these values.

We then used the remaining 400 unmodulated maps to produce different sets of modulated simulations, as explained in Sect.~\ref{subsec4.5}, in order to determine the elements of ${\mathcal{F}}$. 
In particular, to determine the first column ${\mathcal{F}_{i0}}$ of the matrix ${\mathcal{F}}$ we simulate 400 maps with a fiducial modulation $\delta \alpha^{(0)}=0.1$, with all other components $\mathrm{Re}(\delta \alpha^{(1)})$ and $\mathrm{Im}(\delta \alpha^{(1)})$ set to zero.
We estimate the first column $\mathcal{F}_{i0}$ from the ensemble average of Eq.~(\ref{e.hanson}):

\begin{eqnarray}
{\mathcal{F}_{i0}}&=&\frac{1}{\langle\widehat{\delta\alpha^{(0)}}\rangle}{\mathcal{F}}\left( \begin{array}{c}
\langle\widehat{\delta\alpha^{(0)}}\rangle\\
0\\
0
\end{array} \right)\\
&=&\frac{1}{\langle\widehat{\delta\alpha^{(0)}}\rangle}\left( \begin{array}{c}
\langle\widetilde{\delta\alpha^{(0)}}\rangle_{mod}-
\langle\widetilde{\delta\alpha^{(0)}}\rangle\\
\mathrm{Re}\left(\langle\widetilde{\delta\alpha^{(1)}}\rangle_{mod}-
\langle\widetilde{\delta\alpha^{(1)}}\rangle\right)\\
\mathrm{Im}\left(\langle\widetilde{\delta\alpha^{(1)}}\rangle_{mod}-
\langle\widetilde{\delta\alpha^{(1)}}\rangle\right).
\end{array} \right)
\label{e.hanson}
\end{eqnarray}

\noindent Here $\widetilde{\delta\alpha^{(0)}}$ is the biased estimator calculated on each of the modulated simulations, $\langle\widetilde{\delta\alpha^{(0)}}\rangle$ is the previously calculated mean field, and the expected value of the ensemble average of the unbiased estimator $\langle\widehat{\delta\alpha}^{(0)}\rangle$ is assumed to correspond to the input fiducial value of the simulations. We similarly determine ${\mathcal{F}_{i1}}$ and ${\mathcal{F}_{i-1}}$, assuming as a fiducial $\mathrm{Re}(\delta \alpha^{(1)})=0.1$ or $\mathrm{Im}(\delta \alpha^{(1)})=0.1$, respectively, with all other components set to zero. We report only the diagonal elements of the obtained $\mathcal{F}$ matrix in Table~\ref{tab:compo_table}, although we use the full matrix in our analysis .

After that, we use all the 900 unmodulated simulations in order to determine the variance of the estimator. 
For each of these simulations, we estimate $\widehat{\delta\alpha^{(0)}}$ and the real and imaginary parts of $\widehat{\delta\alpha^{(1)}}$. The histogram of these values are shown in Fig.~\ref{fig:space-simu}, while Table~\ref{tab:compo_table} lists the results for the mean field, the ${\mathcal{F}_{ii}}$ elements, and the variance of the estimator for two specific choices of the maximum multipole included in the analysis, $\ell_{\mathrm{max}}=600$ and $\ell_{\mathrm{max}}=1500$. We notice that the high multipoles contribute considerably in decreasing the variance of the different $\delta\alpha^{(i)}$ components, by up to a factor $5$ for $\ell_{\mathrm{max}}=1500$ compared to the $\ell_{\mathrm{max}}=600$ case (as expected, since the number of modes grows like $\ell_{\rm max}^2$).

Finally, let us focus on the amplitude of the modulation $\widetilde{\delta\alpha}=\frac{1}{2}\sqrt{\frac{3}{\pi}}\sqrt{\widehat{|\delta\alpha^{(0)}}|^2+2\left\lvert\widehat{\delta\alpha^{(1)}}\right\rvert^2}$, as already defined in Eq.~(\ref{e:ampli_dipole}). This quantity is interesting as it reveals the presence of a dipole regardless of its direction.
Although calculated from the unbiased $\delta\alpha^{(i)}$ components, this quantity is still biased. To correct for this, we subtract an additional mean field term $\langle\widetilde{\delta\alpha}\rangle$, computed from 500 unmodulated realizations, in order to obtain the unbiased amplitude $\widehat{\delta\alpha}=\widetilde{\delta\alpha}-\langle\widetilde{\delta\alpha}\rangle$.
Table~\ref{tab:ampli_table} summarizes the variance and the mean field of the amplitude. 

As a final cross-check, we now simulate maps, with the appropriate angular resolution, noise content and mask, for different fiducial modulations, and check whether we recover the input value using our unbiased estimator. We consider the following nine fiducial cases: $\delta\alpha^{(0)}\in\{10^{-1},10^{-2},10^{-3}\}$ with all the other coefficients equal to zero, $\mathrm{Re}(\delta\alpha^{(1)})\in\{10^{-1},10^{-2},10^{-3}\}$ with all the other coefficients equal to zero, and $\mathrm{Im}(\delta\alpha^{(1)})\in\{10^{-1},10^{-2},10^{-3}\}$ with all the other coefficients equal to zero. We apply the unbiased estimator on these maps and check that in all nine cases considered we recover the input value of the different components $\delta\alpha^{(i)}$ and the amplitude $\delta\alpha$ within $3\,\sigma$.

We consider two values of $\ell_{\mathrm{max}}$, in order to determine how much the high multipoles contribute to improve the estimates.
 The first, $\ell_{\mathrm{max}}=600$, roughly corresponds to the $\ell_{\mathrm{max}}$ at which a modulation with $|\delta\alpha^{(i)}|\approx 10^{-1}$ (i.e., a dipolar modulation of amplitude $\delta\alpha\approx 10^{-2}$) is well detected at more than $10\,\sigma$ from the unmodulated distribution. The second value, $\ell_{\mathrm{max}}=1500$, still corresponds to the signal-dominated regime of \Planck\ and is roughly the $\ell_{\mathrm{max}}$ at which a modulation with $|\delta\alpha^{(i)}|\approx 10^{-2}$ (i.e., a dipolar modulation of amplitude $\delta\alpha\approx 10^{-3}$) is well detected at more than $10\,\sigma$ from the unmodulated distribution.  However, for $|\delta\alpha^{(i)}|\approx 10^{-3}$, the estimates can no longer be easily distinguished from the unmodulated case. Note that for \planck\ at $\ell_{\mathrm{max}}>1500$, the rising noise cancels the benefit of adding more modes. 

\subsection{Results on data}
\label{resultsspace}
We finally apply the unbiased Hanson-Lewis estimator to the \Planck\ data. More specifically, we use the CMB foreground-cleaned {\tt SMICA} map \citep{planck2013-p06}, which shares the same resolution and noise covariance as the \Planck\ simulations used above. The results are shown in Table~\ref{tab:results_tot}, for $\ell_{\mathrm{max}}=600$ and $\ell_{\mathrm{max}}=1500$.
All the components of the dipolar modulation of the fine structure constant are compatible with zero at the $1\,\sigma$ level, except for $\widehat{\delta\alpha^{(0)}}$ at $\ell_{\mathrm{max}}=600$, which is still compatible with zero within $2\,\sigma$.  At the $1\,\sigma$ level, the overall amplitude of the modulation is also compatible with zero. There is thus no detection of a dipole modulation in the fine structure constant in the \Planck\ data.  Dividing by the standard value of the fine structure constant, $\alpha_0$, the results on the amplitude reported in Table~\ref{tab:results_tot} correspond to $\widehat{\delta\alpha}/\alpha_0=(-8\pm 16)\times 10^{-2}$ for $\ell_{\mathrm{max}}=600$ and
$\widehat{\delta\alpha}/\alpha_0=(-2.4\pm 3.7)\times 10^{-2}$ for $\ell_{\mathrm{max}}=1500$ at 68\,\% CL.



\section{Summary and conclusions}\label{sec4}

We have provided a detailed analysis of the variation of two fundamental constants, the fine structure constant $\alpha$ and the mass of the electron $\me$, on the CMB angular power spectra.  We have presented the constraints that can be derived from the recent \Planck\ data, focussing on these two constants because they are the ones that mostly affect the cosmic recombination process.

As time as variations are concerned, we find that \Planck\ data improve the constraints on $\alpha/\alpha_0$, with respect to those from \WMAP-9, by a factor of about 5. Our analysis of \Planck\ data limits any variation in the fine-structure constant from $z\sim 10^3$ to the present day to be less than approximately $0.4\,\%$, specifically $\alpha/\alpha_0=0.9934\pm0.0042$ (68\,\% CL) from \planck+\WP\ data . We emphasize that \Planck\ allows one to break the degeneracy between $\alpha$ and $H_0$ from the observation of the damping tail. Furthermore, we stress that the $1.6\,\sigma$ deviation of $\alpha/\alpha_0$ from unity when considering the \Planck+\WP\ case is strongly reduced when we remove the low-$\ell$ data, so that this mild deviation is probably coming from the low versus high $\ell$ tension.

 We have also explored how much the constraint on $\alpha$ is weakened by opening up the parameter space to variations of the number of relativistic species or the helium abundance. We find that the constraint on the fine structure constant weakens by about a factor of $1.5$ when $N_{\rm eff}$ is allowed to float, while it weakens by up to a factor of $4$ when the helium abundance is allowed to float.

As the variation of the mass of the electron is concerned, we find that the constraint from \planck\ is comparable to the one obtained from the \WMAP-9\ data, namely $\me/\mezero=0.977^{+0.055}_{-0.070}$.  This is due to the fact that $\me$ only weakly affects the damping tail, contrary to $\alpha$. Therefore, the degeneracy between $\me$ and $H_0$  is not broken by observing the high multipoles, and that limits the constraint on $\me$. This degeneracy can be alleviated by adding other data sets; typically, \Planck\ data combined with BAO provide a constraint on $\me$ at the 1\,\% level.

The \Planck\ data also permit us to set constraints on $\alpha$ and $m_{\rm e}$ when they are both allowed to vary. We find that the constraints on each of the two constants are only slightly weakened, namely $\alpha/\alpha_0=0.993\pm0.0045$ and $\me/\mezero=0.994\pm0.059$.
In Appendix~\ref{effects_alphame} we have presented a detailed analysis of the signature of the variation of each constant in order to explain how the observation of the damping tail allows one to break the degeneracy between them.

Concerning spatial variations, we have constrained a dipolar modulation of the fine structure constant. Such a modulation induces mode couplings, and we have presented an estimator that allows one to constrain this effect. The main difficulty is to circumvent the effect of the masking. We performed $900$ numerical simulations to calibrate our estimators in order to demonstrate that the \planck\ data set a constraint on the amplitude of such a modulation of $\delta\alpha/\alpha_0=(-2.4\times\pm 3.7)\times 10^{-2}$ (68\,\% CL), using multipoles up to $\ell_{\mathrm{max}}=1500$. The conclusion of the analysis is summarized in Table~\ref{tab:results_tot}.

From a theoretical point of view, our analysis relies on a modified version of the \RECFAST\ code. It would be interesting to compare these results with those using more sophisticated (although computationally slower) recombination codes such as {\tt Hyrec} \citep{hyrec} and {\tt Cosmorec} \citep{cosmorec}.  This would enable us to quantify the accuracy of the description of the recombination process and its effect on the constraints; although the major effect on recombination is through the scaling of the energy levels in hydrogen, it will be worth checking for more subtle effects, in particular when the primordial helium fraction is considered as a free parameter.  There is also the possibility of studying the effects of the variations of other constants, such as the mass of the baryons (induced by radiative corrections as soon as $\alpha$ is allowed to vary) or the strength of gravity.  The investigation of these possibilities, which requires consideration of specific self-consistent theoretical models are postponed to future work.  Similarly, the comparison of the limits derived in this work
with the ones obtained at lower redshifts require one to specify a model, since the constraints are very sensitive to the functional form of the time variation of the fundamental constants.

In conclusion, the angular resolution and sensitivity of \Planck\ enables us to reach higher accuracy and lift existing degeneracies.
Further improvement in studies of both temporal ans spatial variation can be expected in the near future by including polarization data from \Planck, as well as other experiments, such as SPTpol \citep{sptpol}, ACTpol \citep{actpol}, and Polarbear \citep{polarbear}.

\begin{acknowledgements}
The development of \Planck\ has been supported by: ESA; CNES and
CNRS/INSU-IN2P3-INP (France); ASI, CNR, and INAF (Italy); NASA and DoE (USA);
STFC and UKSA (UK); CSIC, MICINN, JA, and RES (Spain); Tekes, AoF, and CSC
(Finland); DLR and MPG (Germany); CSA (Canada); DTU Space (Denmark); SER/SSO
(Switzerland); RCN (Norway); SFI (Ireland); FCT/MCTES (Portugal);
and PRACE (EU). A description of the Planck Collaboration and a list of its
members, including the technical or scientific activities in which they have
been involved, can be found at
\url{http://www.sciops.esa.int/index.php?project=planck&page=Planck_Collaboration}.
Some of the results in this paper have been derived using the {\tt HEALPix}
package.  
We thank Alain Coc and Elisabeth Vangioni for discussions and S. Rouberol for running the {\tt horizon} cluster, where some of the computations were performed. Some of this work was carried out at the ILP LABEX (under reference ANR-10-LABX-63) and was supported by French state funds managed by the ANR within the Investissements d'Avenir programme under reference ANR-11-IDEX-0004-02 and by the ANR VACOUL. This work was made possible thanks to the ANR Chaire d'Excellence ANR-10-CEXC-004-01.

\end{acknowledgements}


\bibliographystyle{aa}
\bibliography{PIP_100_Rocha,Planck_bib}

\appendix
\section{Implementation in {\tt RECFAST}}\label{appA}

The recombination equations form a set of three differential equations for the proton fraction $x_{\rm p}=n_{\rm p}/n_{\rm H}$, the singly ionized helium fraction $x_{\rm He II}=n_{\rm HeII}/n_{\rm H}$ and the matter temperature $T_{\rm M}$. The electron fraction is then obtained from electric neutrality as $x_{\rm e}=x_{\rm p} + x_{\rm He II}$. Following \cite{recfast}, \cite{recfast_long}, and
\cite{wong2008}, these are given by
\begin{eqnarray}
\frac{\dd x_{\rm p}}{\dd z} & =& \frac{C_\hydro}{H_0(1+z)E(z)}\left[x_\elec x_\proton n_\hydro {\widetilde \alpha}_\hydro
 \right. \nonumber\\
 &&\qquad\qquad \left.-\beta_\hydro(1-x_\proton)\,\hbox{e}^{-h \nu_{\hydro{\rm 2s}}/kT_{\rm M}}\right]\label{e.a1}\\
\frac{\dd x_\heliumDeux}{\dd z} & =& \frac{C_\heliumUn}{H_0(1+z)E(z)}\left[x_\elec x_\heliumDeux n_\hydro {\widetilde \alpha}_\heliumUn
 \right. \nonumber\\
 &&\qquad\qquad \left.-\beta_\heliumUn(f_\helium-x_\heliumDeux)\,\hbox{e}^{-h \nu_{\heliumUn, 2^1{\rm s}}/kT_{\rm M}}\right]\nonumber\\
 && + \frac{C_\heliumUn^{\rm t}}{H_0(1+z)E(z)}\left[x_\elec x_\heliumDeux n_\hydro {\widetilde \alpha}^{\rm t}_\heliumUn
 \right. \label{e.a2}\\
 &&\qquad\qquad \left.-\frac{g_{\heliumUn,2^3{\rm s}}}{g_{\heliumUn,1^1{\rm s}}}\beta_\heliumUn^{\rm t}(f_\helium-x_\heliumDeux)\,\hbox{e}^{-h \nu_{\heliumUn, 2^3{\rm s}}/kT_{\rm M}}\right]\nonumber,\\
\frac{\dd T_{\rm M}}{\dd z} & =& \frac{8\sigma_{\rm T} a_{\rm R} T^4}{3 H_0 E(z) (1+z) m_{\rm e}c}(T_{\rm M}-T)+ \frac{2T_{\rm M}}{1+z},\label{e.a3}
\end{eqnarray}
where the second term of equation~(\ref{e.a2}) accounts for recombination through the triplets, via the semi-forbidden transition $2^3{\rm p}\rightarrow1^1{\rm s}$. Here $T$ is the radiation temperature that evolves as $T=T_0(1+z)$.

\begin{figure}[htb]
\centering
 \includegraphics[width=\columnwidth]{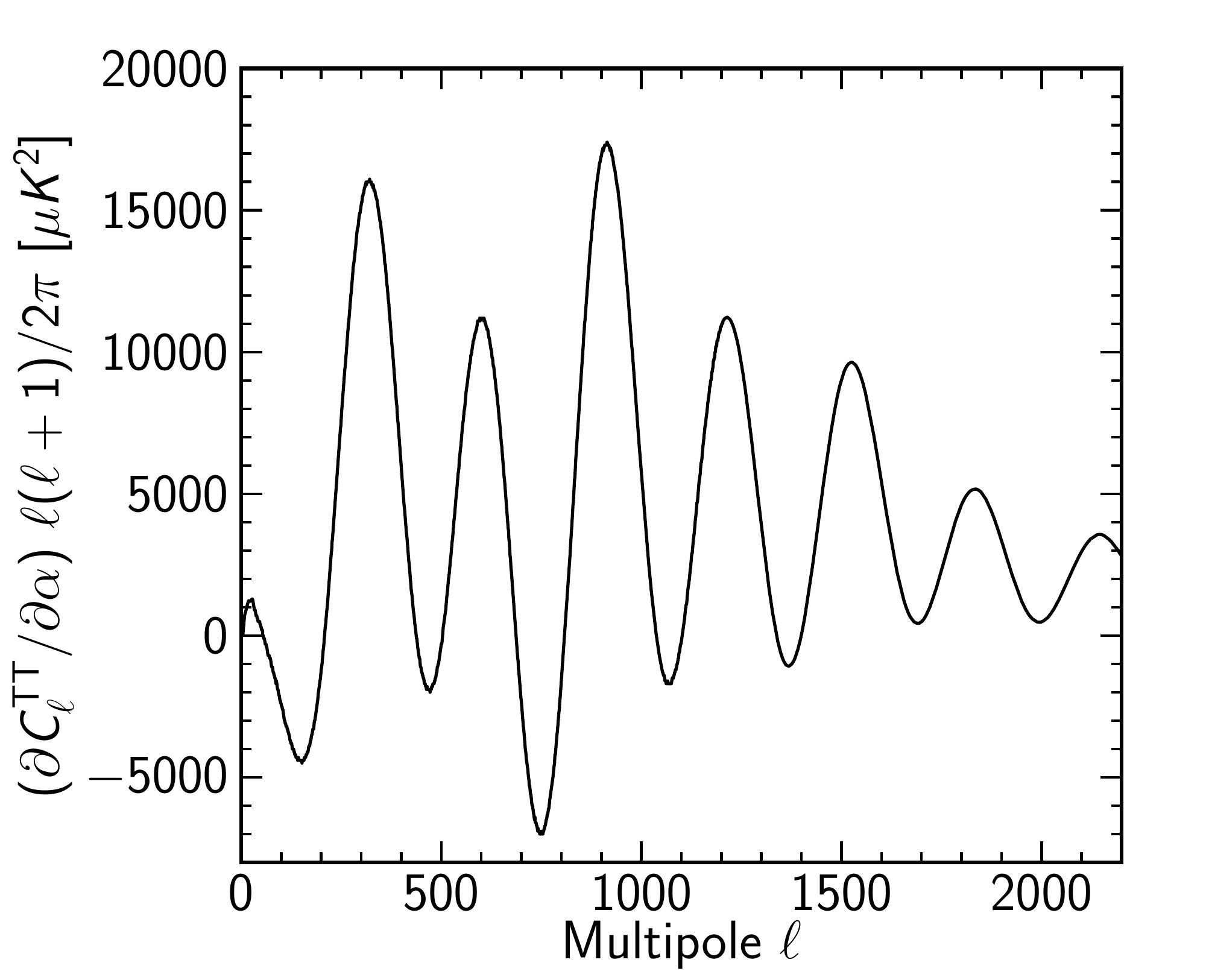}
\caption{The quantity ${\partial C^{TT}_\ell}/{\partial \alpha}$, assuming the \Planck\ best-fit model cosmological parameters.}
\label{fig:dclda}
\end{figure}

Equations (\ref{e.a1}--\ref{e.a3}) involve quantities that remain constant due to our choice of units, such as the speed of light $c$, the \Planck\ constant $h$ and the radiation constant $a_{\rm R}=4\pi^2 k^4/(6\pi c^3\hbar^3)$.
They also involve spectroscopic quantities, such as the hydrogen $2{\rm s}$--$1{\rm s}$ frequency $\nu_{\hydro{\rm 2s}}$, and the helium $2^1{\rm p}$--$1^1{\rm s}$ and $2^3{\rm p}$--$1^1{\rm s}$ frequencies, $\nu_{\heliumUn, 2^1{\rm s}}$ and $\nu_{\heliumUn, 2^3{\rm s}}$. All these frequencies scale as $\alpha^2m_\elec$, as already described in Sect.~\ref{subsec0a}. The importance of these rates on the CMB spectrum has been emphasized recently by~\cite{mukhanov}. The coefficients $C$ are explicitly given by
\begin{eqnarray}
C_\hydro &=& \frac{1+K_\hydro\Lambda_\hydro n_\hydro(1-x_\proton)}{1+K_\hydro(\Lambda_\hydro+\beta_\hydro)n_\hydro(1-x_\proton)}\label{e.a4},\\
C_\heliumUn &=& \frac{1+K_\heliumUn\Lambda_\heliumUn n_\hydro(f_\helium-x_\heliumDeux)\,\hbox{e}^{h\nu_{\rm ps}/kT_{\rm M}}}{1+K_\heliumUn(\Lambda_\helium+\beta_\heliumUn)n_\hydro(f_\helium-x_\heliumDeux)\,\hbox{e}^{h\nu_{\rm ps}/kT_{\rm M}}},\nonumber\\\label{e.a5}\\
C_\heliumUn^{\rm t} &=&\frac{1}{1+K_\heliumUn^{\rm t}\beta_\heliumUn^{\rm t}n_\hydro(f_\helium-x_\heliumDeux)\,\hbox{e}^{h\nu^{\rm t}_{\rm ps}/kT_{\rm M}}}. \label{e.a6}
\end{eqnarray}
These involve the H $2{\rm s}$--$1{\rm s}$ and \ion{He}{i} $2^1{\rm s}$--$1^1{\rm s}$ two-photon decay rates, $\Lambda_\hydro$ and $\Lambda_\heliumUn$, which both scale as $\alpha^8m_\elec$.

We also need the case B recombination coefficient for hydrogen (also by unfortunate convention called $\alpha$), which we shall label ${\widetilde \alpha}_\hydro$, and the two recombination coefficients for helium, ${\widetilde \alpha}_\heliumUn$ (singlet) and ${\widetilde \alpha}_\heliumUn^{\rm t}$ (triplet), which are fitted by the same functional form given in~\cite{pequignot}. These are all assumed to scale as $\alpha^3 m_\elec^{-3/2}$ (see footnote \ref{foot:scaling} for further discussion on this dependence). The photoionization coefficients $\beta_\hydro$ and $\beta_\heliumUn$ are given by $\beta_{\rm i}={\widetilde \alpha}_{\rm i}(2\pi m_\elec k T_{\rm M}/h^2)^{3/2}\exp(-h\nu_{\rm i}/kT_{\rm M})$, so that they scale mostly as the recombination coefficients, up to the dependence induced by the frequency in the exponential factor.

The ``$K$-quantities'', $K_\hydro$, $K_\heliumUn$  and $K_\heliumUn^{\rm t}$, are, respectively, the cosmological redshifting of the hydrogen Lyman-$\alpha$ and helium $2^1{\rm p}$--$1^1{\rm s}$ and $2^3{\rm p}$--$1^1{\rm s}$ transition lines. It can be shown that they scale as $\alpha^{-6}\me^{-3}$.

More details on the physical constant dependence of these equations can be found in \cite{cmb-kap}, \cite{cmb-bat}, \cite{scoccolaPHD}, and \cite{ali}.
\section{How $ \boldsymbol \alpha$ and $ \boldsymbol \me$ affect the power spectrum}
\label{effects_alphame}

In this Appendix, we illustrate how $\alpha$ and $m_{\rm e}$ affect the CMB power spectra through the different terms listed in Sect.~\ref{subsec0a}. The dependences described here have already been discussed in earlier papers \citep{cmb-han,cmb-kap,cmb-bat}. However, we are interested in addressing the following two specific questions. 


The first aims to identify the quantities in Sect.~\ref{subsec0a}
whose change, due to a variation of the constants, gives the strongest effect on the CMB power spectra. 

The second aims to understand what are the dependences that determine the different behaviour of the angular power spectra under a variation of $\alpha$ or $m_{\rm e}$. 

In order to investigate these two issues, we perform the following exercise. We add a variation of $\alpha$ (or $m_{\rm e}$) only to one or a few terms at a time, while keeping the value of the constants at the standard values in all other terms.
We consider the following cases:
\begin{enumerate}
\item variation only in the hydrogen binding energy, as in Eq.~(\ref{eq:energylevels});
\item variation only in the energy of the ``Lyman$-\alpha$'' (Ly$\,\alpha$) transition (here improperly defined as the average of the ($2{\rm s}$--continuum) and ($2{\rm p}$--continuum) energy levels);
\item variation only in the previous two terms together;
\item variation of the previous two terms and the Thomson scattering cross-section $\sigma_{\rm T}$, as in Eq.~(\ref{thomson});
\item variation only in the previous three terms and the $2-$photon decay rate $\Lambda_{\rm i}$, as in Eq.~(\ref{eq:lambda});
\item variation of all the terms where the constants appear.
\end{enumerate}

\begin{figure*}[htb]
\centering
\vskip1cm
\includegraphics[width=8cm]{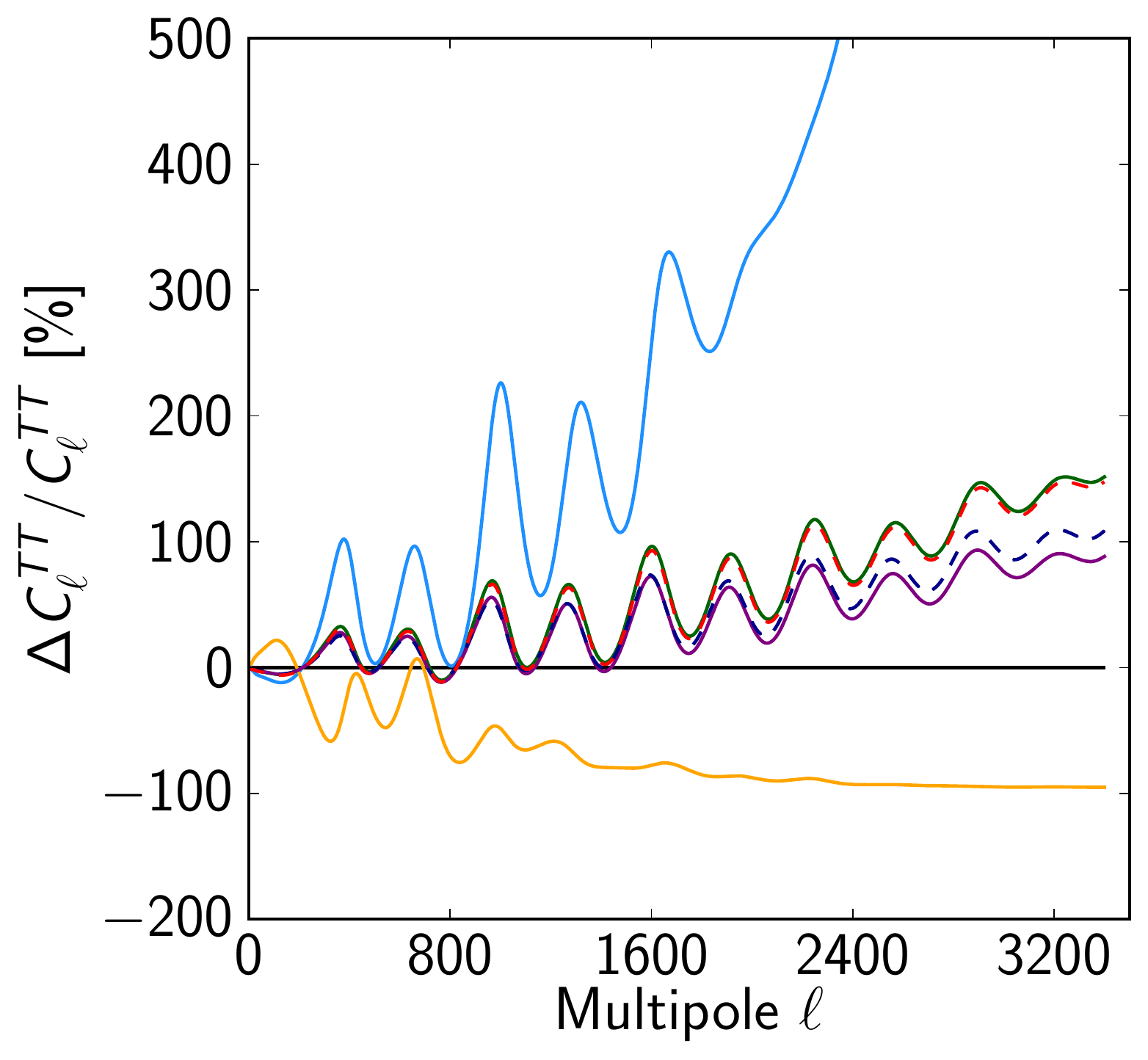}$\qquad$
\includegraphics[width=8cm]{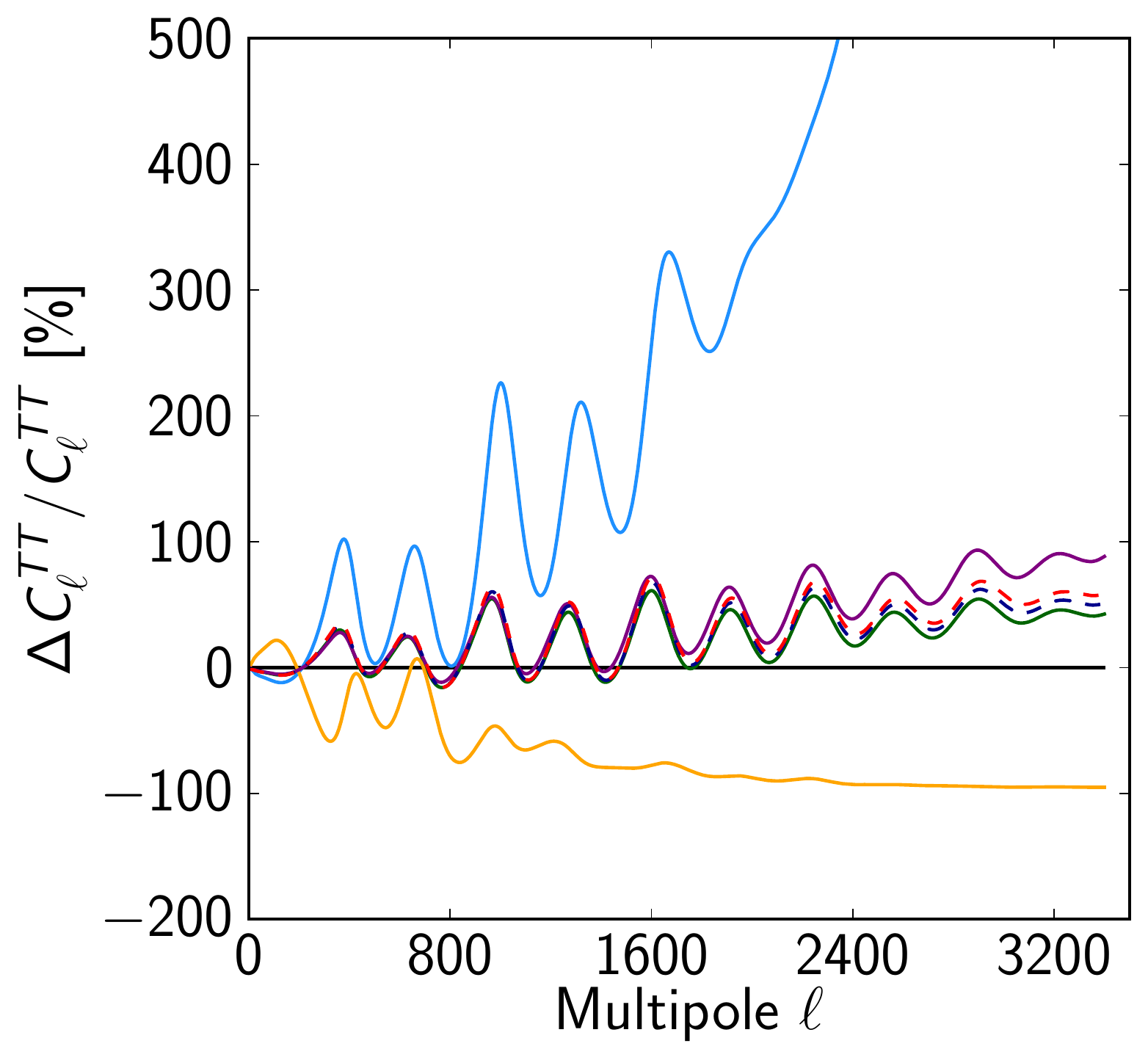}
\caption{{\em Left}: relative difference between the CMB \TT\ power spectrum calculated using a value of $\alpha/\alpha_0$ different from unity (in one, a few or all the terms where it appears), and a power spectrum, $C_{\ell,{\rm st}}$, calculated using a standard value of $\alpha/\alpha_0=1$. We thus plot $\Delta C_\ell/C_\ell=(C_\ell-C_{\ell,{\rm st}})/C_{\ell,{\rm st}}$[\%]. The cases considered are $\alpha$ varying: only in the hydrogen binding energy (solid light blue); only in the Ly$\,\alpha$ energy (solid yellow); in both the previous two terms (solid purple); in both the previous terms and in the Thomson scattering cross-section (dashed dark blue); in the previous three terms and in the $2-$photon decay rate (dashed red); and in all terms (solid green). In each case, we assume that $\alpha$ varies by $+5\,\%$ ($\alpha/\alpha_0=1.05$) only in the terms considered, while it is $\alpha/\alpha_0=1$ in all the others. {\em Right}: same as the cases on the left, but for a variation of $m_{\rm e}$ of $10\,\%$ ($m_{\rm e}/m_{\rm e0}=1.1$). }
\label{fig:effects_alpha_me}
\end{figure*}

We test the effects on the CMB angular power spectra of the sequence of cases listed above,  both for variations of $\alpha$ and $m_{\rm e}$, and show the results in Fig.~\ref{fig:effects_alpha_me}. There we plot the relative difference between: (i) the angular power spectrum obtained assuming that $\alpha$ (or $m_{\rm e}$) is changing only in a few terms as listed above; and (ii) the standard angular power spectrum, with the constants set to their usual values. For $\alpha$ we assume a variation at the $5\,\%$ level, while for $m_{\rm e}$ we assume a variation at the $10\,\%$ level. This latter choice is motivated by the fact that atomic energies scale as $\alpha^2 m_{\rm e}$.  Hence changing $m_{\rm e}$ by twice the $\alpha$ change should result in similar effects on the angular power spectra, making the comparison between the effects on spectra easier. This is what is shown in Fig.~\ref{fig:effects_alpha_me}: the blue line (relative to the change of hydrogen binding energy only, item~1 in the list), the yellow line (relative to a change in the Ly$\,\alpha$ energy level only, item~2) and the purple line (sum of the previous two effects, item~3) are identical for $\alpha$ and for $m_{\rm e}$.

It is evident from the figures that the major contribution to the change in the angular power spectrum induced by a variation of $\alpha$ or $m_{\rm e}$ comes from the change in the hydrogen binding energy (item~1) and Ly$\,\alpha$ energy (item~2). 
The main effect of changing these two energy levels is to modify the hydrogen $2{\rm s}$--$1{\rm s}$ transition energy, $h\nu_{\rm H2s}$, in Eq.~(\ref{e.a1}), since this is, by definition, the difference between the first two mentioned energies. Increasing $h\nu_{\rm H2s}$ through, e.g., an increase of the hydrogen binding energy, weakens the ionization term in Eq.~(\ref{e.a1}) through a decrease of the Boltzmann factor $\hbox{e}^{-h \nu_{\hydro{\rm 2s}}/kT_{\rm M}}$, resulting in earlier recombination.
As a consequence, acoustic peaks move to higher multipoles, early ISW is increased and the Silk damping is decreased, so that the overall amplitude of the peaks is increased, as already described in Sect.~\ref{subsec0a}. On the other hand, increasing the Ly$\,\alpha$ energy has the opposite effect on $h\nu_{\rm H2s}$ and would thus tend to delay recombination. However, this effect is mitigated by the other terms where the Ly$\,\alpha$ energy level appears, such as in the $K$ factors encoding the redshifting of the Ly$\,\alpha$ photons in Eqs. (\ref{eq:redshiftly}) and (\ref{e.a4}). This is why the effects of increasing both the hydrogen binding energy and the Ly$\,\alpha$ energy, through an increase of the value of the constants, do not perfectly cancel, but the first effect dominates over the second.
Furthermore, as already mentioned, these effects are qualitatively the same for $\alpha$ and for $m_{\rm e}$, although of different magnitude. 

A distinction in the effects of $\alpha$ or $\me$ is, however, introduced if we now also consider the impact on the Thomson scattering cross-section $\sigma_{\rm T}$ (item~4). As already described in Sect.~\ref{subsec0a}, $\sigma_{\rm T}\propto \alpha^2/m_{\rm e}^2$, i.e., an increase in the value of $\alpha$ increases $\sigma_{\rm T}$, while an increase in $\me$ decreases $\sigma_{\rm T}$. Consequently, an increase in $\sigma_{\rm T}$ increases Silk damping, while a decrease in $\sigma_{\rm T}$ decreases Silk damping.
 This is the reason why adding the effect of the constants on the Thomson cross-section, shown in the dark-blue dashed lines in Fig.~\ref{fig:effects_alpha_me}, increases the amplitude of the peaks for a larger value of $\alpha$, while it decreases it for a larger value of $\me$. Consequently, this is the reason why $\alpha$ and $\me$ have different effects on the amplitude of the peaks.
 
We now analyse the effect of adding the variation of constants in the 2-photon decay rate (item~5). As shown in Eq. (\ref{eq:lambda}), this ratio depends much more strongly on $\alpha$ than on $\me$, $\Lambda\propto \alpha^8 m_{\rm e}$. The effect of increasing the value of the constants in this term is again to shift recombination to earlier times, but, as expected, the impact is much larger when varying $\alpha$ than when varying $\me$, as shown by the dashed red line in Fig.~\ref{fig:effects_alpha_me}. Finally, adding the variation of the constants in all the remaining terms, including the equations for helium recombination, further adjusts the amplitudes at the few percent level, to finally converge to the green solid line. In particular, we verified that neglecting the effects of the variation of the constants on helium recombination impacts the constraints by less than 5\,\%.

\begin{figure*}[htb]
\centering
\vskip1cm
\includegraphics[width=8cm]{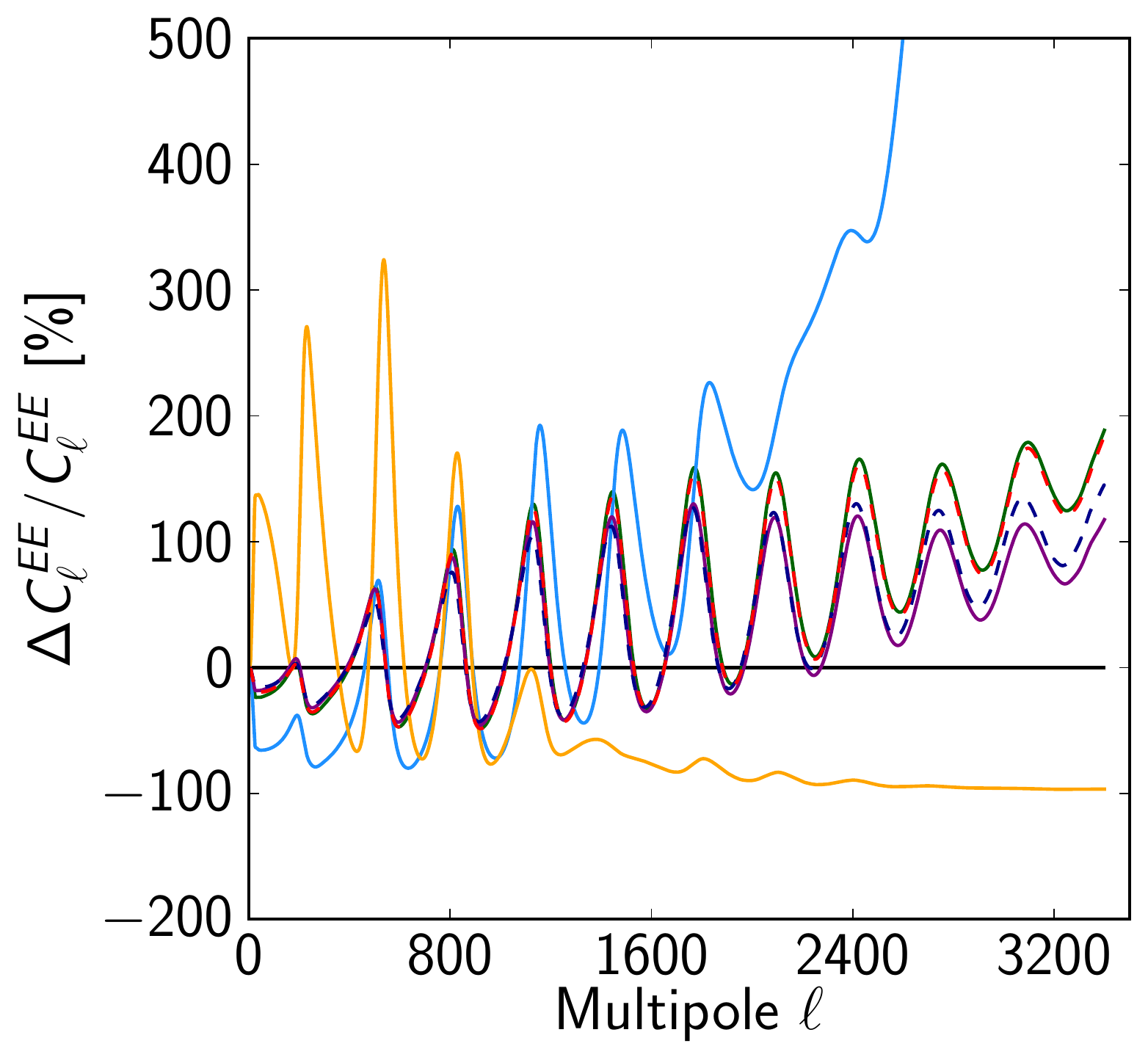}$\qquad$
\includegraphics[width=8cm]{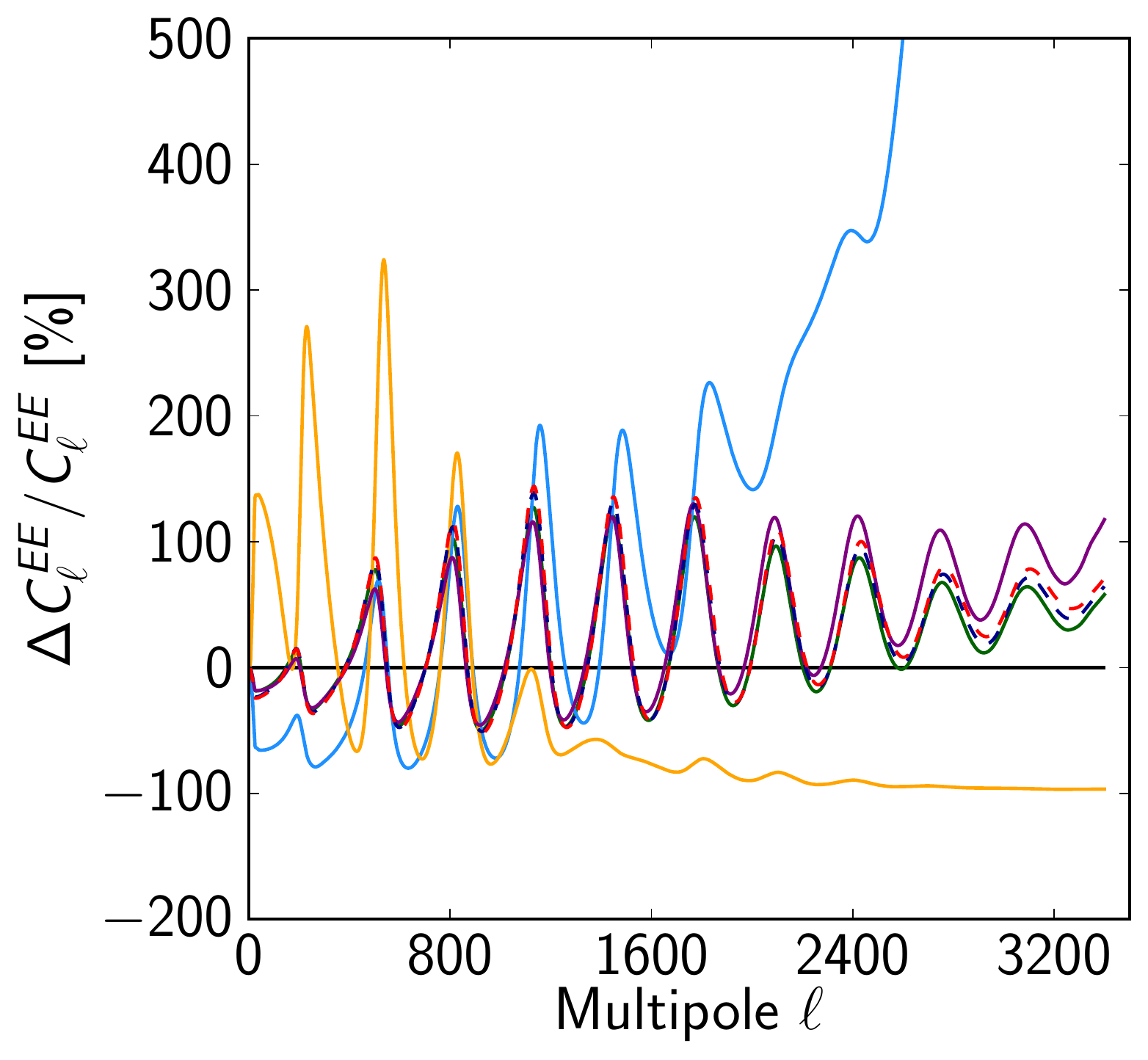}
\caption{Same as Fig.\ref{fig:effects_alpha_me}, but for \EE\ polarization power spectra.}
\label{fig:effects_alpha_me_pol}
\end{figure*}

For the sake of completeness, we show in Fig.~\ref{fig:effects_alpha_me_pol} the effect of varying the constants on the \EE-polarization. The effects are similar to the ones described for temperature, although changes are in this case even larger. High accuracy observations of the polarization power spectra might therefore help in improving the constraints on fundamental constants.

\section{Additional Tables and Figures}
\label{sec:fg}
\begin{figure*}[htmb]
\centering
\vskip1cm
\includegraphics[width=0.8\textwidth]{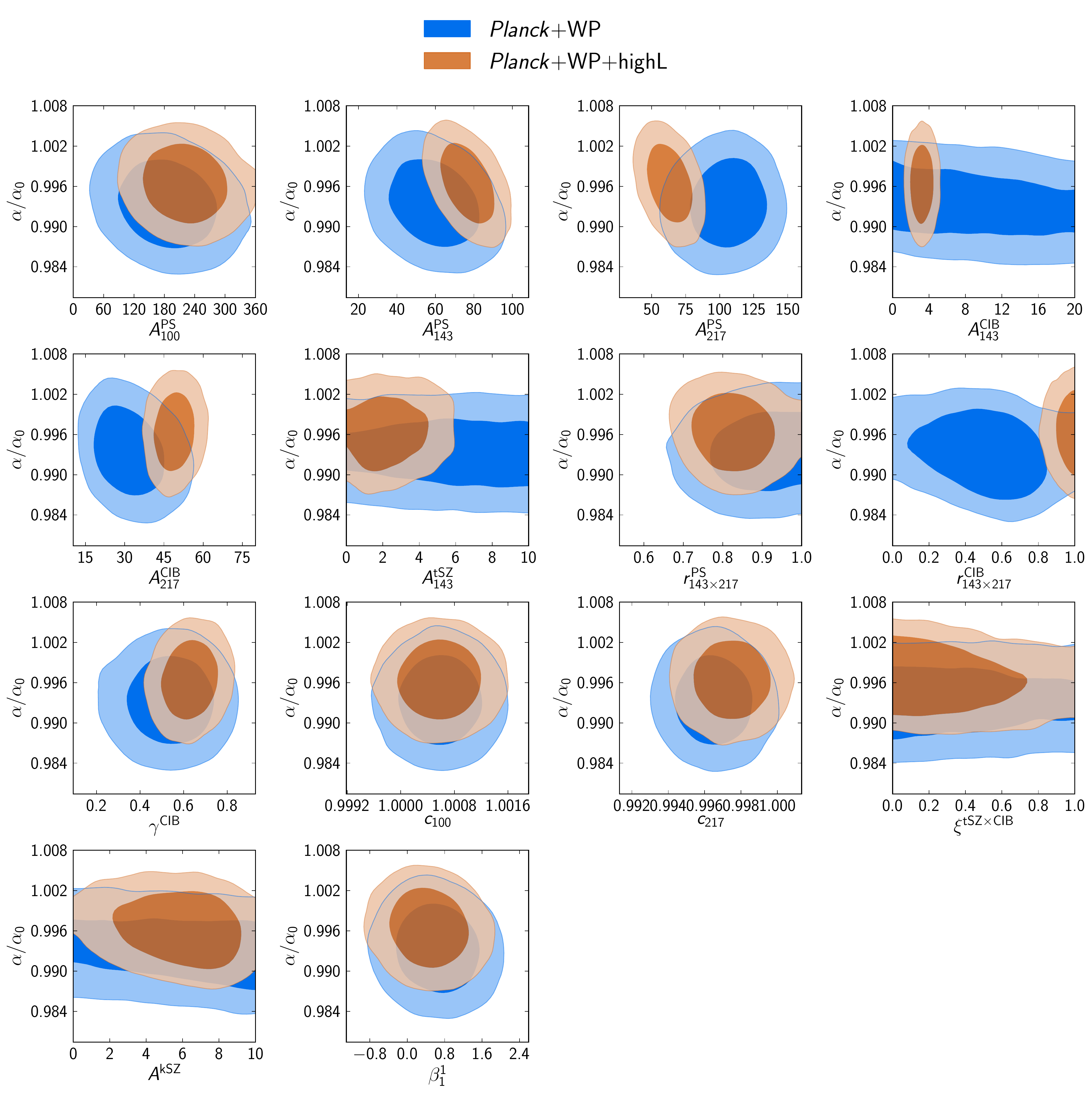}
\caption {Two-dimensional likelihood contours (68\,\% and 95\,\%) for $\alpha/\alpha_0$ versus different foreground/beam/calibration parameters, as defined in table~4 of \cite{planck2013-p11}. We show results for \Planck+\WP\ (blue) and \Planck+\WP+\highL\ (orange) data combinations.  }
\label{fig:alpha_fg}
\end{figure*}

\begin{figure*}[htmb]
\centering
\vskip1cm
\includegraphics[width=0.8\textwidth]{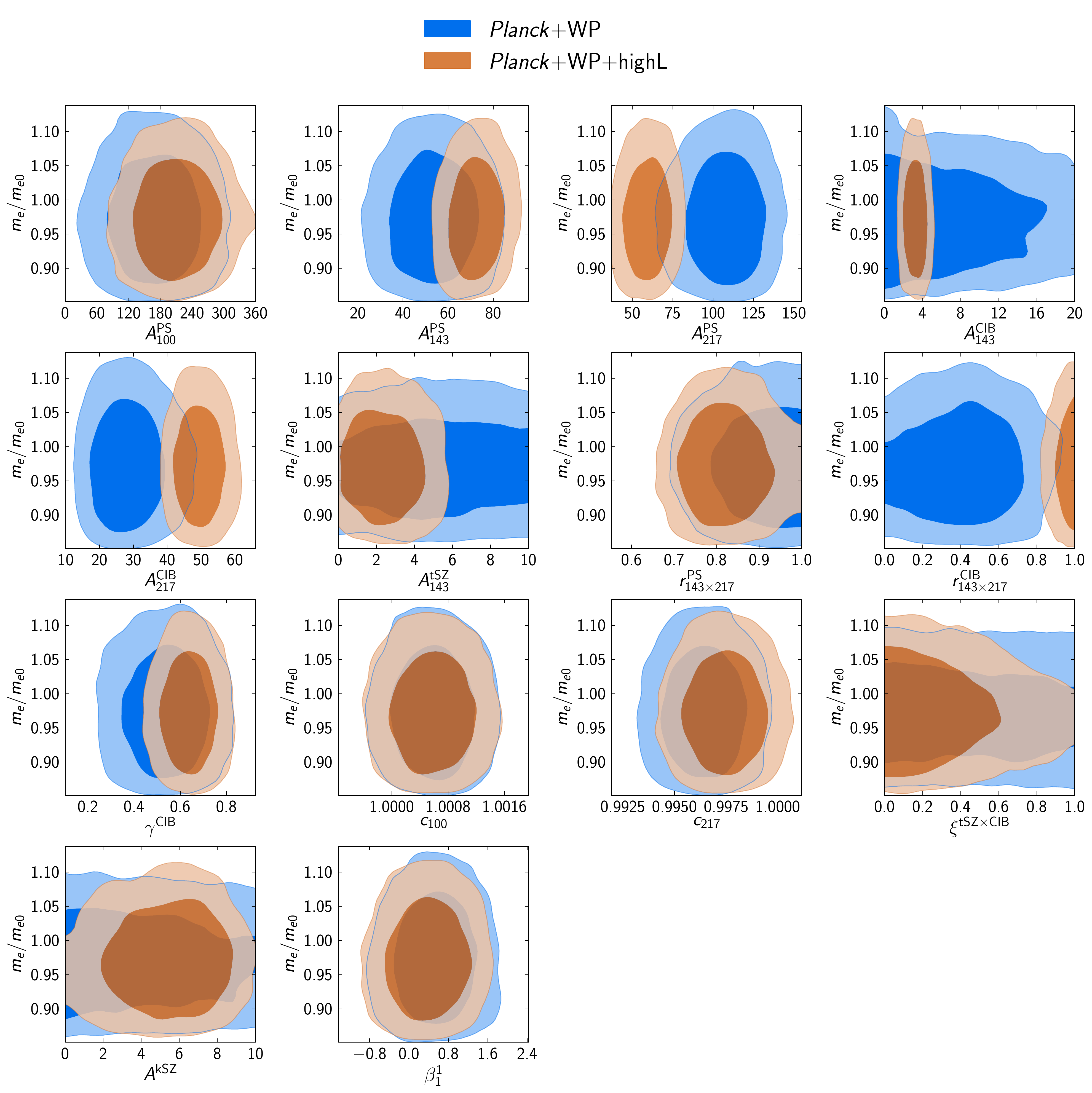}
\caption{Two-dimensional likelihood contours (68\,\% and 95\,\%) for $m_{\rm e}/m_{\rm e0}$ versus different foreground/beam/calibration parameters, as defined in table~4 in \cite{planck2013-p11}. We show results for \Planck+\WP\ (blue) and \Planck+\WP+\highL\ (orange) data combinations.}
\label{fig:me_fg}
\end{figure*}

In this Appendix, we collect some of the tables and figures already extensively described in the paper. We show the constraints on cosmological parameters for a $\Lambda$CDM$+\alpha+N_{\rm eff}$ model in Table~\ref{table:alpha-nnu}, a $\Lambda$CDM$+\alpha+\yhe$ model in Table~\ref{table:alpha-yhe}, a $\Lambda$CDM$+\alpha$ removing the low-$\ell$ data in Table~\ref{table:alpha-tau}, and for a $\Lambda$CDM+$m_{\rm e}$ model removing the low-$\ell$ data in Table~\ref{table:me-tau}. Furthermore, Fig.~\ref{fig:alpha_fg} shows the two-dimensional contour plots between $\alpha$ and foreground parameters, while Fig.~\ref{fig:me_fg} shows the same for $\me$.

\begin{table*}[htmb]                 
\caption{Constraints on the cosmological parameters for the base $\Lambda$CDM\ model with the addition
of a varying fine structure constant and number of relativistic species, $N_{\rm eff}$. We quote $68\,\%$ CL errors.}                        
\label{table:alpha-nnu}

\input table_alpha_nnu
\end{table*} 

\begin{table*}[htmb]                 
\caption{Constraints on the cosmological parameters for the base $\Lambda$CDM\ model with the addition
of a varying fine structure constant and helium abundance, $Y_{\rm p}$. We quote $68\,\%$ CL errors.}                        
\label{table:alpha-yhe}

\input table_alpha_yhe
\end{table*} 

\begin{table*}[htmb]                 
\caption{Constraints on the cosmological parameters for the base $\Lambda$CDM\ model with the addition
of a varying fine structure constant for \Planck\ data, removing the low-$\ell$ multipoles and placing a Gaussian prior on $\tau$. We quote $68\,\%$ CL errors.
}                        
\label{table:alpha-tau}

\input table_alpha_tauprior
\end{table*} 

\begin{table*}[htmb]                 
\caption{Constraints on the cosmological parameters for the base $\Lambda$CDM\ model with the addition
of the mass of the electron for \Planck\ data, removing the low-$\ell$ multipoles and placing a Gaussian prior on $\tau$. We quote $68\,\%$ CL errors. In this case we find that the Hubble parameter is unconstrained within the uniform prior we set, i.e., $40<(H_0/{\rm km}\,{\rm s}^{-1}\,{\rm Mpc}^{-1})<100$.
}                        
\label{table:me-tau}                            
\input table_me_tauprior
\end{table*}

\raggedright

\end{document}

%% file: Planck.tex
\def\setsymbol#1#2{\expandafter\def\csname #1\endcsname{#2}}
\def\getsymbol#1{\csname #1\endcsname}

\def\Planck{\textit{Planck}}

\def\all2013resultspapers{\nocite{planck2013-p01, planck2013-p02, planck2013-p02a, planck2013-p02d, planck2013-p02b, planck2013-p03, planck2013-p03c, planck2013-p03f, planck2013-p03d, planck2013-p03e, planck2013-p01a, planck2013-p06, planck2013-p03a, planck2013-pip88, planck2013-p08, planck2013-p11, planck2013-p12, planck2013-p13, planck2013-p14, planck2013-p15, planck2013-p05b, planck2013-p17, planck2013-p09, planck2013-p09a, planck2013-p20, planck2013-p19, planck2013-pipaberration, planck2013-p05, planck2013-p05a, planck2013-pip56, planck2013-p06b}}

\newbox\tablebox    \newdimen\tablewidth
\def\leaderfil{\leaders\hbox to 5pt{\hss.\hss}\hfil}
\def\endPlancktable{\tablewidth=\columnwidth 
    $$\hss\copy\tablebox\hss$$
    \vskip-\lastskip\vskip -2pt}
\def\endPlancktablewide{\tablewidth=\textwidth 
    $$\hss\copy\tablebox\hss$$
    \vskip-\lastskip\vskip -2pt}
\def\tablenote#1 #2\par{\begingroup \parindent=0.8em
    \abovedisplayshortskip=0pt\belowdisplayshortskip=0pt
    \noindent
    $$\hss\vbox{\hsize\tablewidth \hangindent=\parindent \hangafter=1 \noindent
    \hbox to \parindent{$^#1$\hss}\strut#2\strut\par}\hss$$
    \endgroup}
\def\doubleline{\vskip 3pt\hrule \vskip 1.5pt \hrule \vskip 5pt}

\def\L2{\ifmmode L_2\else $L_2$\fi}

\def\DeltaT{\ifmmode \Delta T\else $\Delta T$\fi}
\def\deltat{\ifmmode \Delta t\else $\Delta t$\fi}
\def\fknee{\ifmmode f_{\rm knee}\else $f_{\rm knee}$\fi}
\def\Fmax{\ifmmode F_{\rm max}\else $F_{\rm max}$\fi}
\def\solar{\ifmmode{\rm M}_{\mathord\odot}\else${\rm M}_{\mathord\odot}$\fi}
\def\Msolar{\ifmmode{\rm M}_{\mathord\odot}\else${\rm M}_{\mathord\odot}$\fi}
\def\Lsolar{\ifmmode{\rm L}_{\mathord\odot}\else${\rm L}_{\mathord\odot}$\fi}

\def\inv{\ifmmode^{-1}\else$^{-1}$\fi}
\def\mo{\ifmmode^{-1}\else$^{-1}$\fi}
\def\sup#1{\ifmmode ^{\rm #1}\else $^{\rm #1}$\fi}
\def\expo#1{\ifmmode \times 10^{#1}\else $\times 10^{#1}$\fi}
\def\,{\thinspace}
\def\lsim{\mathrel{\raise .4ex\hbox{\rlap{$<$}\lower 1.2ex\hbox{$\sim$}}}}
\def\gsim{\mathrel{\raise .4ex\hbox{\rlap{$>$}\lower 1.2ex\hbox{$\sim$}}}}

\def\simprop{\mathrel{\raise .4ex\hbox{\rlap{$\propto$}\lower 1.2ex\hbox{$\sim$}}}}
\def\deg{\ifmmode^\circ\else$^\circ$\fi}
\def\pdeg{\ifmmode $\setbox0=\hbox{$^{\circ}$}\rlap{\hskip.11\wd0 .}$^{\circ}
          \else \setbox0=\hbox{$^{\circ}$}\rlap{\hskip.11\wd0 .}$^{\circ}$\fi}
\def\arcs{\ifmmode {^{\scriptstyle\prime\prime}}
          \else $^{\scriptstyle\prime\prime}$\fi}
\def\arcm{\ifmmode {^{\scriptstyle\prime}}
          \else $^{\scriptstyle\prime}$\fi}
\newdimen\sa  \newdimen\sb
\def\parcs{\sa=.07em \sb=.03em
     \ifmmode \hbox{\rlap{.}}^{\scriptstyle\prime\kern -\sb\prime}\hbox{\kern -\sa}
     \else \rlap{.}$^{\scriptstyle\prime\kern -\sb\prime}$\kern -\sa\fi}
\def\parcm{\sa=.08em \sb=.03em
     \ifmmode \hbox{\rlap{.}\kern\sa}^{\scriptstyle\prime}\hbox{\kern-\sb}
     \else \rlap{.}\kern\sa$^{\scriptstyle\prime}$\kern-\sb\fi}
\def\ra[#1 #2 #3.#4]{#1\sup{h}#2\sup{m}#3\sup{s}\llap.#4}
\def\dec[#1 #2 #3.#4]{#1\deg#2\arcm#3\arcs\llap.#4}
\def\deco[#1 #2 #3]{#1\deg#2\arcm#3\arcs}
\def\rra[#1 #2]{#1\sup{h}#2\sup{m}}

\def\dots{\relax\ifmmode \ldots\else $\ldots$\fi}
\def\WHzsr{\ifmmode $W\,Hz\mo\,sr\mo$\else W\,Hz\mo\,sr\mo\fi}
\def\mHz{\ifmmode $\,mHz$\else \,mHz\fi}
\def\GHz{\ifmmode $\,GHz$\else \,GHz\fi}
\def\mKs{\ifmmode $\,mK\,s$^{1/2}\else \,mK\,s$^{1/2}$\fi}
\def\muKs{\ifmmode \,\mu$K\,s$^{1/2}\else \,$\mu$K\,s$^{1/2}$\fi}
\def\muKRJs{\ifmmode \,\mu$K$_{\rm RJ}$\,s$^{1/2}\else \,$\mu$K$_{\rm RJ}$\,s$^{1/2}$\fi}
\def\muKHz{\ifmmode \,\mu$K\,Hz$^{-1/2}\else \,$\mu$K\,Hz$^{-1/2}$\fi}
\def\MJysr{\ifmmode \,$MJy\,sr\mo$\else \,MJy\,sr\mo\fi}
\def\MJysrmK{\ifmmode \,$MJy\,sr\mo$\,mK$_{\rm CMB}\mo\else \,MJy\,sr\mo\,mK$_{\rm CMB}\mo$\fi}
\def\microns{\ifmmode \,\mu$m$\else \,$\mu$m\fi}

\def\muK{\ifmmode \,\mu$K$\else \,$\mu$\hbox{K}\fi}
\def\microK{\ifmmode \,\mu$K$\else \,$\mu$\hbox{K}\fi}
\def\muW{\ifmmode \,\mu$W$\else \,$\mu$\hbox{W}\fi}
\def\kms{\ifmmode $\,km\,s$^{-1}\else \,km\,s$^{-1}$\fi}
\def\kmsMpc{\ifmmode $\,\kms\,Mpc\mo$\else \,\kms\,Mpc\mo\fi}
\setsymbol{LFI:center:frequency:70GHz:units}{70.3\,GHz}
\setsymbol{LFI:center:frequency:44GHz:units}{44.1\,GHz}
\setsymbol{LFI:center:frequency:30GHz:units}{28.5\,GHz}

\setsymbol{LFI:center:frequency:70GHz}{70.3}
\setsymbol{LFI:center:frequency:44GHz}{44.1}
\setsymbol{LFI:center:frequency:30GHz}{28.5}

\setsymbol{LFI:center:frequency:LFI18:Rad:M:units}{71.7\GHz}
\setsymbol{LFI:center:frequency:LFI19:Rad:M:units}{67.5\GHz}
\setsymbol{LFI:center:frequency:LFI20:Rad:M:units}{69.2\GHz}
\setsymbol{LFI:center:frequency:LFI21:Rad:M:units}{70.4\GHz}
\setsymbol{LFI:center:frequency:LFI22:Rad:M:units}{71.5\GHz}
\setsymbol{LFI:center:frequency:LFI23:Rad:M:units}{70.8\GHz}
\setsymbol{LFI:center:frequency:LFI24:Rad:M:units}{44.4\GHz}
\setsymbol{LFI:center:frequency:LFI25:Rad:M:units}{44.0\GHz}
\setsymbol{LFI:center:frequency:LFI26:Rad:M:units}{43.9\GHz}
\setsymbol{LFI:center:frequency:LFI27:Rad:M:units}{28.3\GHz}
\setsymbol{LFI:center:frequency:LFI28:Rad:M:units}{28.8\GHz}
\setsymbol{LFI:center:frequency:LFI18:Rad:S:units}{70.1\GHz}
\setsymbol{LFI:center:frequency:LFI19:Rad:S:units}{69.6\GHz}
\setsymbol{LFI:center:frequency:LFI20:Rad:S:units}{69.5\GHz}
\setsymbol{LFI:center:frequency:LFI21:Rad:S:units}{69.5\GHz}
\setsymbol{LFI:center:frequency:LFI22:Rad:S:units}{72.8\GHz}
\setsymbol{LFI:center:frequency:LFI23:Rad:S:units}{71.3\GHz}
\setsymbol{LFI:center:frequency:LFI24:Rad:S:units}{44.1\GHz}
\setsymbol{LFI:center:frequency:LFI25:Rad:S:units}{44.1\GHz}
\setsymbol{LFI:center:frequency:LFI26:Rad:S:units}{44.1\GHz}
\setsymbol{LFI:center:frequency:LFI27:Rad:S:units}{28.5\GHz}
\setsymbol{LFI:center:frequency:LFI28:Rad:S:units}{28.2\GHz}

\setsymbol{LFI:center:frequency:LFI18:Rad:M}{71.7}
\setsymbol{LFI:center:frequency:LFI19:Rad:M}{67.5}
\setsymbol{LFI:center:frequency:LFI20:Rad:M}{69.2}
\setsymbol{LFI:center:frequency:LFI21:Rad:M}{70.4}
\setsymbol{LFI:center:frequency:LFI22:Rad:M}{71.5}
\setsymbol{LFI:center:frequency:LFI23:Rad:M}{70.8}
\setsymbol{LFI:center:frequency:LFI24:Rad:M}{44.4}
\setsymbol{LFI:center:frequency:LFI25:Rad:M}{44.0}
\setsymbol{LFI:center:frequency:LFI26:Rad:M}{43.9}
\setsymbol{LFI:center:frequency:LFI27:Rad:M}{28.3}
\setsymbol{LFI:center:frequency:LFI28:Rad:M}{28.8}
\setsymbol{LFI:center:frequency:LFI18:Rad:S}{70.1}
\setsymbol{LFI:center:frequency:LFI19:Rad:S}{69.6}
\setsymbol{LFI:center:frequency:LFI20:Rad:S}{69.5}
\setsymbol{LFI:center:frequency:LFI21:Rad:S}{69.5}
\setsymbol{LFI:center:frequency:LFI22:Rad:S}{72.8}
\setsymbol{LFI:center:frequency:LFI23:Rad:S}{71.3}
\setsymbol{LFI:center:frequency:LFI24:Rad:S}{44.1}
\setsymbol{LFI:center:frequency:LFI25:Rad:S}{44.1}
\setsymbol{LFI:center:frequency:LFI26:Rad:S}{44.1}
\setsymbol{LFI:center:frequency:LFI27:Rad:S}{28.5}
\setsymbol{LFI:center:frequency:LFI28:Rad:S}{28.2}

\setsymbol{LFI:white:noise:sensitivity:70GHz:units}{134.7\muKs}
\setsymbol{LFI:white:noise:sensitivity:44GHz:units}{164.7\muKs}
\setsymbol{LFI:white:noise:sensitivity:30GHz:units}{143.4\muKs}

\setsymbol{LFI:white:noise:sensitivity:70GHz}{134.7}
\setsymbol{LFI:white:noise:sensitivity:44GHz}{164.7}
\setsymbol{LFI:white:noise:sensitivity:30GHz}{143.4}

\setsymbol{LFI:white:noise:sensitivity:LFI18:Rad:M:units}{512.0\muKs}
\setsymbol{LFI:white:noise:sensitivity:LFI19:Rad:M:units}{581.4\muKs}
\setsymbol{LFI:white:noise:sensitivity:LFI20:Rad:M:units}{590.8\muKs}
\setsymbol{LFI:white:noise:sensitivity:LFI21:Rad:M:units}{455.2\muKs}
\setsymbol{LFI:white:noise:sensitivity:LFI22:Rad:M:units}{492.0\muKs}
\setsymbol{LFI:white:noise:sensitivity:LFI23:Rad:M:units}{507.7\muKs}
\setsymbol{LFI:white:noise:sensitivity:LFI24:Rad:M:units}{462.2\muKs}
\setsymbol{LFI:white:noise:sensitivity:LFI25:Rad:M:units}{413.6\muKs}
\setsymbol{LFI:white:noise:sensitivity:LFI26:Rad:M:units}{478.6\muKs}
\setsymbol{LFI:white:noise:sensitivity:LFI27:Rad:M:units}{277.7\muKs}
\setsymbol{LFI:white:noise:sensitivity:LFI28:Rad:M:units}{312.3\muKs}
\setsymbol{LFI:white:noise:sensitivity:LFI18:Rad:S:units}{465.7\muKs}
\setsymbol{LFI:white:noise:sensitivity:LFI19:Rad:S:units}{555.6\muKs}
\setsymbol{LFI:white:noise:sensitivity:LFI20:Rad:S:units}{623.2\muKs}
\setsymbol{LFI:white:noise:sensitivity:LFI21:Rad:S:units}{564.1\muKs}
\setsymbol{LFI:white:noise:sensitivity:LFI22:Rad:S:units}{534.4\muKs}
\setsymbol{LFI:white:noise:sensitivity:LFI23:Rad:S:units}{542.4\muKs}
\setsymbol{LFI:white:noise:sensitivity:LFI24:Rad:S:units}{399.2\muKs}
\setsymbol{LFI:white:noise:sensitivity:LFI25:Rad:S:units}{392.6\muKs}
\setsymbol{LFI:white:noise:sensitivity:LFI26:Rad:S:units}{418.6\muKs}
\setsymbol{LFI:white:noise:sensitivity:LFI27:Rad:S:units}{302.9\muKs}
\setsymbol{LFI:white:noise:sensitivity:LFI28:Rad:S:units}{285.3\muKs}

\setsymbol{LFI:white:noise:sensitivity:LFI18:Rad:M}{512.0}
\setsymbol{LFI:white:noise:sensitivity:LFI19:Rad:M}{581.4}
\setsymbol{LFI:white:noise:sensitivity:LFI20:Rad:M}{590.8}
\setsymbol{LFI:white:noise:sensitivity:LFI21:Rad:M}{455.2}
\setsymbol{LFI:white:noise:sensitivity:LFI22:Rad:M}{492.0}
\setsymbol{LFI:white:noise:sensitivity:LFI23:Rad:M}{507.7}
\setsymbol{LFI:white:noise:sensitivity:LFI24:Rad:M}{462.2}
\setsymbol{LFI:white:noise:sensitivity:LFI25:Rad:M}{413.6}
\setsymbol{LFI:white:noise:sensitivity:LFI26:Rad:M}{478.6}
\setsymbol{LFI:white:noise:sensitivity:LFI27:Rad:M}{277.7}
\setsymbol{LFI:white:noise:sensitivity:LFI28:Rad:M}{312.3}
\setsymbol{LFI:white:noise:sensitivity:LFI18:Rad:S}{465.7}
\setsymbol{LFI:white:noise:sensitivity:LFI19:Rad:S}{555.6}
\setsymbol{LFI:white:noise:sensitivity:LFI20:Rad:S}{623.2}
\setsymbol{LFI:white:noise:sensitivity:LFI21:Rad:S}{564.1}
\setsymbol{LFI:white:noise:sensitivity:LFI22:Rad:S}{534.4}
\setsymbol{LFI:white:noise:sensitivity:LFI23:Rad:S}{542.4}
\setsymbol{LFI:white:noise:sensitivity:LFI24:Rad:S}{399.2}
\setsymbol{LFI:white:noise:sensitivity:LFI25:Rad:S}{392.6}
\setsymbol{LFI:white:noise:sensitivity:LFI26:Rad:S}{418.6}
\setsymbol{LFI:white:noise:sensitivity:LFI27:Rad:S}{302.9}
\setsymbol{LFI:white:noise:sensitivity:LFI28:Rad:S}{285.3}

\setsymbol{LFI:knee:frequency:70GHz:units}{29.5\mHz}
\setsymbol{LFI:knee:frequency:44GHz:units}{56.2\mHz}
\setsymbol{LFI:knee:frequency:30GHz:units}{113.7\mHz}

\setsymbol{LFI:knee:frequency:70GHz}{29.5}
\setsymbol{LFI:knee:frequency:44GHz}{56.2}
\setsymbol{LFI:knee:frequency:30GHz}{113.7}

\setsymbol{LFI:knee:frequency:LFI18:Rad:M:units}{16.3\mHz}
\setsymbol{LFI:knee:frequency:LFI19:Rad:M:units}{15.1\mHz}
\setsymbol{LFI:knee:frequency:LFI20:Rad:M:units}{18.7\mHz}
\setsymbol{LFI:knee:frequency:LFI21:Rad:M:units}{37.2\mHz}
\setsymbol{LFI:knee:frequency:LFI22:Rad:M:units}{12.7\mHz}
\setsymbol{LFI:knee:frequency:LFI23:Rad:M:units}{34.6\mHz}
\setsymbol{LFI:knee:frequency:LFI24:Rad:M:units}{46.2\mHz}
\setsymbol{LFI:knee:frequency:LFI25:Rad:M:units}{24.9\mHz}
\setsymbol{LFI:knee:frequency:LFI26:Rad:M:units}{67.6\mHz}
\setsymbol{LFI:knee:frequency:LFI27:Rad:M:units}{187.4\mHz}
\setsymbol{LFI:knee:frequency:LFI28:Rad:M:units}{122.2\mHz}
\setsymbol{LFI:knee:frequency:LFI18:Rad:S:units}{17.7\mHz}
\setsymbol{LFI:knee:frequency:LFI19:Rad:S:units}{22.0\mHz}
\setsymbol{LFI:knee:frequency:LFI20:Rad:S:units}{8.7\mHz}
\setsymbol{LFI:knee:frequency:LFI21:Rad:S:units}{25.9\mHz}
\setsymbol{LFI:knee:frequency:LFI22:Rad:S:units}{15.8\mHz}
\setsymbol{LFI:knee:frequency:LFI23:Rad:S:units}{129.8\mHz}
\setsymbol{LFI:knee:frequency:LFI24:Rad:S:units}{100.9\mHz}
\setsymbol{LFI:knee:frequency:LFI25:Rad:S:units}{38.9\mHz}
\setsymbol{LFI:knee:frequency:LFI26:Rad:S:units}{58.9\mHz}
\setsymbol{LFI:knee:frequency:LFI27:Rad:S:units}{104.4\mHz}
\setsymbol{LFI:knee:frequency:LFI28:Rad:S:units}{40.7\mHz}

\setsymbol{LFI:knee:frequency:LFI18:Rad:M}{16.3}
\setsymbol{LFI:knee:frequency:LFI19:Rad:M}{15.1}
\setsymbol{LFI:knee:frequency:LFI20:Rad:M}{18.7}
\setsymbol{LFI:knee:frequency:LFI21:Rad:M}{37.2}
\setsymbol{LFI:knee:frequency:LFI22:Rad:M}{12.7}
\setsymbol{LFI:knee:frequency:LFI23:Rad:M}{34.6}
\setsymbol{LFI:knee:frequency:LFI24:Rad:M}{46.2}
\setsymbol{LFI:knee:frequency:LFI25:Rad:M}{24.9}
\setsymbol{LFI:knee:frequency:LFI26:Rad:M}{67.6}
\setsymbol{LFI:knee:frequency:LFI27:Rad:M}{187.4}
\setsymbol{LFI:knee:frequency:LFI28:Rad:M}{122.2}
\setsymbol{LFI:knee:frequency:LFI18:Rad:S}{17.7}
\setsymbol{LFI:knee:frequency:LFI19:Rad:S}{22.0}
\setsymbol{LFI:knee:frequency:LFI20:Rad:S}{8.7}
\setsymbol{LFI:knee:frequency:LFI21:Rad:S}{25.9}
\setsymbol{LFI:knee:frequency:LFI22:Rad:S}{15.8}
\setsymbol{LFI:knee:frequency:LFI23:Rad:S}{129.8}
\setsymbol{LFI:knee:frequency:LFI24:Rad:S}{100.9}
\setsymbol{LFI:knee:frequency:LFI25:Rad:S}{38.9}
\setsymbol{LFI:knee:frequency:LFI26:Rad:S}{58.9}
\setsymbol{LFI:knee:frequency:LFI27:Rad:S}{104.4}
\setsymbol{LFI:knee:frequency:LFI28:Rad:S}{40.7}

\setsymbol{LFI:slope:70GHz:units}{$-1.03$\mHz}
\setsymbol{LFI:slope:44GHz:units}{$-0.89$\mHz}
\setsymbol{LFI:slope:30GHz:units}{$-0.87$\mHz}

\setsymbol{LFI:slope:70GHz}{$-1.03$}
\setsymbol{LFI:slope:44GHz}{$-0.89$}
\setsymbol{LFI:slope:30GHz}{$-0.87$}

\setsymbol{LFI:slope:LFI18:Rad:M:units}{$-1.04$\mHz}
\setsymbol{LFI:slope:LFI19:Rad:M:units}{$-1.09$\mHz}
\setsymbol{LFI:slope:LFI20:Rad:M:units}{$-0.69$\mHz}
\setsymbol{LFI:slope:LFI21:Rad:M:units}{$-1.56$\mHz}
\setsymbol{LFI:slope:LFI22:Rad:M:units}{$-1.01$\mHz}
\setsymbol{LFI:slope:LFI23:Rad:M:units}{$-0.96$\mHz}
\setsymbol{LFI:slope:LFI24:Rad:M:units}{$-0.83$\mHz}
\setsymbol{LFI:slope:LFI25:Rad:M:units}{$-0.91$\mHz}
\setsymbol{LFI:slope:LFI26:Rad:M:units}{$-0.95$\mHz}
\setsymbol{LFI:slope:LFI27:Rad:M:units}{$-0.87$\mHz}
\setsymbol{LFI:slope:LFI28:Rad:M:units}{$-0.88$\mHz}
\setsymbol{LFI:slope:LFI18:Rad:S:units}{$-1.15$\mHz}
\setsymbol{LFI:slope:LFI19:Rad:S:units}{$-1.00$\mHz}
\setsymbol{LFI:slope:LFI20:Rad:S:units}{$-0.95$\mHz}
\setsymbol{LFI:slope:LFI21:Rad:S:units}{$-0.92$\mHz}
\setsymbol{LFI:slope:LFI22:Rad:S:units}{$-1.01$\mHz}
\setsymbol{LFI:slope:LFI23:Rad:S:units}{$-0.95$\mHz}
\setsymbol{LFI:slope:LFI24:Rad:S:units}{$-0.73$\mHz}
\setsymbol{LFI:slope:LFI25:Rad:S:units}{$-1.16$\mHz}
\setsymbol{LFI:slope:LFI26:Rad:S:units}{$-0.79$\mHz}
\setsymbol{LFI:slope:LFI27:Rad:S:units}{$-0.82$\mHz}
\setsymbol{LFI:slope:LFI28:Rad:S:units}{$-0.91$\mHz}

\setsymbol{LFI:slope:LFI18:Rad:M}{$-1.04$}
\setsymbol{LFI:slope:LFI19:Rad:M}{$-1.09$}
\setsymbol{LFI:slope:LFI20:Rad:M}{$-0.69$}
\setsymbol{LFI:slope:LFI21:Rad:M}{$-1.56$}
\setsymbol{LFI:slope:LFI22:Rad:M}{$-1.01$}
\setsymbol{LFI:slope:LFI23:Rad:M}{$-0.96$}
\setsymbol{LFI:slope:LFI24:Rad:M}{$-0.83$}
\setsymbol{LFI:slope:LFI25:Rad:M}{$-0.91$}
\setsymbol{LFI:slope:LFI26:Rad:M}{$-0.95$}
\setsymbol{LFI:slope:LFI27:Rad:M}{$-0.87$}
\setsymbol{LFI:slope:LFI28:Rad:M}{$-0.88$}
\setsymbol{LFI:slope:LFI18:Rad:S}{$-1.15$}
\setsymbol{LFI:slope:LFI19:Rad:S}{$-1.00$}
\setsymbol{LFI:slope:LFI20:Rad:S}{$-0.95$}
\setsymbol{LFI:slope:LFI21:Rad:S}{$-0.92$}
\setsymbol{LFI:slope:LFI22:Rad:S}{$-1.01$}
\setsymbol{LFI:slope:LFI23:Rad:S}{$-0.95$}
\setsymbol{LFI:slope:LFI24:Rad:S}{$-0.73$}
\setsymbol{LFI:slope:LFI25:Rad:S}{$-1.16$}
\setsymbol{LFI:slope:LFI26:Rad:S}{$-0.79$}
\setsymbol{LFI:slope:LFI27:Rad:S}{$-0.82$}
\setsymbol{LFI:slope:LFI28:Rad:S}{$-0.91$}

\setsymbol{LFI:FWHM:70GHz:units}{13\parcm01}
\setsymbol{LFI:FWHM:44GHz:units}{27\parcm92}
\setsymbol{LFI:FWHM:30GHz:units}{32\parcm65}

\setsymbol{LFI:FWHM:70GHz}{13.01}
\setsymbol{LFI:FWHM:44GHz}{27.92}
\setsymbol{LFI:FWHM:30GHz}{32.65}

\setsymbol{LFI:FWHM:LFI18:units}{13\parcm39}
\setsymbol{LFI:FWHM:LFI19:units}{13\parcm01}
\setsymbol{LFI:FWHM:LFI20:units}{12\parcm75}
\setsymbol{LFI:FWHM:LFI21:units}{12\parcm74}
\setsymbol{LFI:FWHM:LFI22:units}{12\parcm87}
\setsymbol{LFI:FWHM:LFI23:units}{13\parcm27}
\setsymbol{LFI:FWHM:LFI24:units}{22\parcm98}
\setsymbol{LFI:FWHM:LFI25:units}{30\parcm46}
\setsymbol{LFI:FWHM:LFI26:units}{30\parcm31}
\setsymbol{LFI:FWHM:LFI27:units}{32\parcm65}
\setsymbol{LFI:FWHM:LFI28:units}{32\parcm66}

\setsymbol{LFI:FWHM:LFI18}{13.39}
\setsymbol{LFI:FWHM:LFI19}{13.01}
\setsymbol{LFI:FWHM:LFI20}{12.75}
\setsymbol{LFI:FWHM:LFI21}{12.74}
\setsymbol{LFI:FWHM:LFI22}{12.87}
\setsymbol{LFI:FWHM:LFI23}{13.27}
\setsymbol{LFI:FWHM:LFI24}{22.98}
\setsymbol{LFI:FWHM:LFI25}{30.46}
\setsymbol{LFI:FWHM:LFI26}{30.31}
\setsymbol{LFI:FWHM:LFI27}{32.65}
\setsymbol{LFI:FWHM:LFI28}{32.66}

\setsymbol{LFI:FWHM:uncertainty:LFI18:units}{0.170\arcm}
\setsymbol{LFI:FWHM:uncertainty:LFI19:units}{0.174\arcm}
\setsymbol{LFI:FWHM:uncertainty:LFI20:units}{0.170\arcm}
\setsymbol{LFI:FWHM:uncertainty:LFI21:units}{0.156\arcm}
\setsymbol{LFI:FWHM:uncertainty:LFI22:units}{0.164\arcm}
\setsymbol{LFI:FWHM:uncertainty:LFI23:units}{0.171\arcm}
\setsymbol{LFI:FWHM:uncertainty:LFI24:units}{0.652\arcm}
\setsymbol{LFI:FWHM:uncertainty:LFI25:units}{1.075\arcm}
\setsymbol{LFI:FWHM:uncertainty:LFI26:units}{1.131\arcm}
\setsymbol{LFI:FWHM:uncertainty:LFI27:units}{1.266\arcm}
\setsymbol{LFI:FWHM:uncertainty:LFI28:units}{1.287\arcm}

\setsymbol{LFI:FWHM:uncertainty:LFI18}{0.170}
\setsymbol{LFI:FWHM:uncertainty:LFI19}{0.174}
\setsymbol{LFI:FWHM:uncertainty:LFI20}{0.170}
\setsymbol{LFI:FWHM:uncertainty:LFI21}{0.156}
\setsymbol{LFI:FWHM:uncertainty:LFI22}{0.164}
\setsymbol{LFI:FWHM:uncertainty:LFI23}{0.171}
\setsymbol{LFI:FWHM:uncertainty:LFI24}{0.652}
\setsymbol{LFI:FWHM:uncertainty:LFI25}{1.075}
\setsymbol{LFI:FWHM:uncertainty:LFI26}{1.131}
\setsymbol{LFI:FWHM:uncertainty:LFI27}{1.266}
\setsymbol{LFI:FWHM:uncertainty:LFI28}{1.287}

\setsymbol{HFI:center:frequency:100GHz:units}{100\,GHz}
\setsymbol{HFI:center:frequency:143GHz:units}{143\,GHz}
\setsymbol{HFI:center:frequency:217GHz:units}{217\,GHz}
\setsymbol{HFI:center:frequency:353GHz:units}{353\,GHz}
\setsymbol{HFI:center:frequency:545GHz:units}{545\,GHz}
\setsymbol{HFI:center:frequency:857GHz:units}{857\,GHz}

\setsymbol{HFI:center:frequency:100GHz}{100}
\setsymbol{HFI:center:frequency:143GHz}{143}
\setsymbol{HFI:center:frequency:217GHz}{217}
\setsymbol{HFI:center:frequency:353GHz}{353}
\setsymbol{HFI:center:frequency:545GHz}{545}
\setsymbol{HFI:center:frequency:857GHz}{857}

\setsymbol{HFI:Ndetectors:100GHz}{8}
\setsymbol{HFI:Ndetectors:143GHz}{11}
\setsymbol{HFI:Ndetectors:217GHz}{12}
\setsymbol{HFI:Ndetectors:353GHz}{12}
\setsymbol{HFI:Ndetectors:545GHz}{3}
\setsymbol{HFI:Ndetectors:857GHz}{4}

\setsymbol{HFI:FWHM:Maps:100GHz:units}{9\parcm88}
\setsymbol{HFI:FWHM:Maps:143GHz:units}{7\parcm18}
\setsymbol{HFI:FWHM:Maps:217GHz:units}{4\parcm87}
\setsymbol{HFI:FWHM:Maps:353GHz:units}{4\parcm65}
\setsymbol{HFI:FWHM:Maps:545GHz:units}{4\parcm72}
\setsymbol{HFI:FWHM:Maps:857GHz:units}{4\parcm39}
\setsymbol{HFI:FWHM:Maps:100GHz}{9.88}
\setsymbol{HFI:FWHM:Maps:143GHz}{7.18}
\setsymbol{HFI:FWHM:Maps:217GHz}{4.87}
\setsymbol{HFI:FWHM:Maps:353GHz}{4.65}
\setsymbol{HFI:FWHM:Maps:545GHz}{4.72}
\setsymbol{HFI:FWHM:Maps:857GHz}{4.39}

\setsymbol{HFI:beam:ellipticity:Maps:100GHz}{1.15}
\setsymbol{HFI:beam:ellipticity:Maps:143GHz}{1.01}
\setsymbol{HFI:beam:ellipticity:Maps:217GHz}{1.06}
\setsymbol{HFI:beam:ellipticity:Maps:353GHz}{1.05}
\setsymbol{HFI:beam:ellipticity:Maps:545GHz}{1.14}
\setsymbol{HFI:beam:ellipticity:Maps:857GHz}{1.19}

\setsymbol{HFI:FWHM:Mars:100GHz:units}{9\parcm37}
\setsymbol{HFI:FWHM:Mars:143GHz:units}{7\parcm04}
\setsymbol{HFI:FWHM:Mars:217GHz:units}{4\parcm68}
\setsymbol{HFI:FWHM:Mars:353GHz:units}{4\parcm43}
\setsymbol{HFI:FWHM:Mars:545GHz:units}{3\parcm80}
\setsymbol{HFI:FWHM:Mars:857GHz:units}{3\parcm67}

\setsymbol{HFI:FWHM:Mars:100GHz}{9.37}
\setsymbol{HFI:FWHM:Mars:143GHz}{7.04}
\setsymbol{HFI:FWHM:Mars:217GHz}{4.68}
\setsymbol{HFI:FWHM:Mars:353GHz}{4.43}
\setsymbol{HFI:FWHM:Mars:545GHz}{3.80}
\setsymbol{HFI:FWHM:Mars:857GHz}{3.67}

\setsymbol{HFI:beam:ellipticity:Mars:100GHz}{1.18}
\setsymbol{HFI:beam:ellipticity:Mars:143GHz}{1.03}
\setsymbol{HFI:beam:ellipticity:Mars:217GHz}{1.14}
\setsymbol{HFI:beam:ellipticity:Mars:353GHz}{1.09}
\setsymbol{HFI:beam:ellipticity:Mars:545GHz}{1.25}
\setsymbol{HFI:beam:ellipticity:Mars:857GHz}{1.03}

\setsymbol{HFI:CMB:relative:calibration:100GHz}{$\lsim 1\%$}
\setsymbol{HFI:CMB:relative:calibration:143GHz}{$\lsim 1\%$}
\setsymbol{HFI:CMB:relative:calibration:217GHz}{$\lsim 1\%$}
\setsymbol{HFI:CMB:relative:calibration:353GHz}{$\lsim 1\%$}
\setsymbol{HFI:CMB:relative:calibration:545GHz}{}
\setsymbol{HFI:CMB:relative:calibration:857GHz}{}

\setsymbol{HFI:CMB:absolute:calibration:100GHz}{$\lsim 2\%$}
\setsymbol{HFI:CMB:absolute:calibration:143GHz}{$\lsim 2\%$}
\setsymbol{HFI:CMB:absolute:calibration:217GHz}{$\lsim 2\%$}
\setsymbol{HFI:CMB:absolute:calibration:353GHz}{$\lsim 2\%$}
\setsymbol{HFI:CMB:absolute:calibration:545GHz}{}
\setsymbol{HFI:CMB:absolute:calibration:857GHz}{}

\setsymbol{HFI:FIRAS:gain:calibration:accuracy:statistical:100GHz}{}
\setsymbol{HFI:FIRAS:gain:calibration:accuracy:statistical:143GHz}{}
\setsymbol{HFI:FIRAS:gain:calibration:accuracy:statistical:217GHz}{}
\setsymbol{HFI:FIRAS:gain:calibration:accuracy:statistical:353GHz}{2.5\%}
\setsymbol{HFI:FIRAS:gain:calibration:accuracy:statistical:545GHz}{1\%}
\setsymbol{HFI:FIRAS:gain:calibration:accuracy:statistical:857GHz}{0.5\%}

\setsymbol{HFI:FIRAS:gain:calibration:accuracy:systematic:100GHz}{}
\setsymbol{HFI:FIRAS:gain:calibration:accuracy:systematic:143GHz}{}
\setsymbol{HFI:FIRAS:gain:calibration:accuracy:systematic:217GHz}{}
\setsymbol{HFI:FIRAS:gain:calibration:accuracy:systematic:353GHz}{}
\setsymbol{HFI:FIRAS:gain:calibration:accuracy:systematic:545GHz}{7\%}
\setsymbol{HFI:FIRAS:gain:calibration:accuracy:systematic:857GHz}{7\%}

\setsymbol{HFI:FIRAS:zero:point:accuracy:100GHz:units}{0.8\MJysr}
\setsymbol{HFI:FIRAS:zero:point:accuracy:143GHz:units}{}
\setsymbol{HFI:FIRAS:zero:point:accuracy:217GHz:units}{}
\setsymbol{HFI:FIRAS:zero:point:accuracy:353GHz:units}{1.4\MJysr}
\setsymbol{HFI:FIRAS:zero:point:accuracy:545GHz:units}{2.2\MJysr}
\setsymbol{HFI:FIRAS:zero:point:accuracy:857GHz:units}{1.7\MJysr}

\setsymbol{HFI:FIRAS:zero:point:accuracy:100GHz}{0.8}
\setsymbol{HFI:FIRAS:zero:point:accuracy:143GHz}{}
\setsymbol{HFI:FIRAS:zero:point:accuracy:217GHz}{}
\setsymbol{HFI:FIRAS:zero:point:accuracy:353GHz}{1.4}
\setsymbol{HFI:FIRAS:zero:point:accuracy:545GHz}{2.2}
\setsymbol{HFI:FIRAS:zero:point:accuracy:857GHz}{1.7}

\setsymbol{HFI:unit:conversion:100GHz:units}{0.2415\MJysrmK}
\setsymbol{HFI:unit:conversion:143GHz:units}{0.3694\MJysrmK}
\setsymbol{HFI:unit:conversion:217GHz:units}{0.4811\MJysrmK}
\setsymbol{HFI:unit:conversion:353GHz:units}{0.2883\MJysrmK}
\setsymbol{HFI:unit:conversion:545GHz:units}{0.05826\MJysrmK}
\setsymbol{HFI:unit:conversion:857GHz:units}{0.002238\MJysrmK}

\setsymbol{HFI:unit:conversion:100GHz}{0.2415}
\setsymbol{HFI:unit:conversion:143GHz}{0.3694}
\setsymbol{HFI:unit:conversion:217GHz}{0.4811}
\setsymbol{HFI:unit:conversion:353GHz}{0.2883}
\setsymbol{HFI:unit:conversion:545GHz}{0.05826}
\setsymbol{HFI:unit:conversion:857GHz}{0.002238}

\setsymbol{HFI:colour:correction:alpha=-2:V1.01:100GHz}{0.9893}
\setsymbol{HFI:colour:correction:alpha=-2:V1.01:143GHz}{0.9759}
\setsymbol{HFI:colour:correction:alpha=-2:V1.01:217GHz}{1.0007}
\setsymbol{HFI:colour:correction:alpha=-2:V1.01:353GHz}{1.0028}
\setsymbol{HFI:colour:correction:alpha=-2:V1.01:545GHz}{1.0019}
\setsymbol{HFI:colour:correction:alpha=-2:V1.01:857GHz}{0.9889}

\setsymbol{HFI:colour:correction:alpha=0:V1.01:100GHz}{1.0008}
\setsymbol{HFI:colour:correction:alpha=0:V1.01:143GHz}{1.0148}
\setsymbol{HFI:colour:correction:alpha=0:V1.01:217GHz}{0.9909}
\setsymbol{HFI:colour:correction:alpha=0:V1.01:353GHz}{0.9888}
\setsymbol{HFI:colour:correction:alpha=0:V1.01:545GHz}{0.9878}
\setsymbol{HFI:colour:correction:alpha=0:V1.01:857GHz}{1.0014}

\providecommand{\sorthelp}[1]{}

%% file: PIP_100_Rocha_authors_and_institutes.tex
\author{\small
Planck Collaboration:
P.~A.~R.~Ade\inst{73}
\and
N.~Aghanim\inst{50}
\and
M.~Arnaud\inst{62}
\and
M.~Ashdown\inst{59, 5}
\and
J.~Aumont\inst{50}
\and
C.~Baccigalupi\inst{72}
\and
A.~J.~Banday\inst{80, 9}
\and
R.~B.~Barreiro\inst{56}
\and
E.~Battaner\inst{81, 82}
\and
K.~Benabed\inst{51, 79}
% ...
}
\institute{\small
APC, AstroParticule et Cosmologie, Universit\'{e} Paris Diderot, CNRS/IN2P3, CEA/lrfu, Observatoire de Paris, Sorbonne Paris Cit\'{e}, 10, rue Alice Domon et L\'{e}onie Duquet, 75205 Paris Cedex 13, France\\
\and
African Institute for Mathematical Sciences, 6-8 Melrose Road, Muizenberg, Cape Town, South Africa\\
\and
Agenzia Spaziale Italiana Science Data Center, Via del Politecnico snc, 00133, Roma, Italy\\
\and
Agenzia Spaziale Italiana, Viale Liegi 26, Roma, Italy\\
\and
Astrophysics Group, Cavendish Laboratory, University of Cambridge, J J Thomson Avenue, Cambridge CB3 0HE, U.K.\\
\and
Astrophysics \& Cosmology Research Unit, School of Mathematics, Statistics \& Computer Science, University of KwaZulu-Natal, Westville Campus, Private Bag X54001, Durban 4000, South Africa\\
\and
Atacama Large Millimeter/submillimeter Array, ALMA Santiago Central Offices, Alonso de Cordova 3107, Vitacura, Casilla 763 0355, Santiago, Chile\\
\and
CITA, University of Toronto, 60 St. George St., Toronto, ON M5S 3H8, Canada\\
\and
CNRS, IRAP, 9 Av. colonel Roche, BP 44346, F-31028 Toulouse cedex 4, France\\
\and
California Institute of Technology, Pasadena, California, U.S.A.\\
\and
Centro de Estudios de F\'{i}sica del Cosmos de Arag\'{o}n (CEFCA), Plaza San Juan, 1, planta 2, E-44001, Teruel, Spain\\
\and
Computational Cosmology Center, Lawrence Berkeley National Laboratory, Berkeley, California, U.S.A.\\
\and
Consejo Superior de Investigaciones Cient\'{\i}ficas (CSIC), Madrid, Spain\\
\and
DSM/Irfu/SPP, CEA-Saclay, F-91191 Gif-sur-Yvette Cedex, France\\
\and
DTU Space, National Space Institute, Technical University of Denmark, Elektrovej 327, DK-2800 Kgs. Lyngby, Denmark\\
\and
D\'{e}partement de Physique Th\'{e}orique, Universit\'{e} de Gen\`{e}ve, 24, Quai E. Ansermet,1211 Gen\`{e}ve 4, Switzerland\\
\and
Departamento de F\'{\i}sica, Universidad de Oviedo, Avda. Calvo Sotelo s/n, Oviedo, Spain\\
\and
Department of Astrophysics/IMAPP, Radboud University Nijmegen, P.O. Box 9010, 6500 GL Nijmegen, The Netherlands\\
\and
Department of Physics \& Astronomy, University of British Columbia, 6224 Agricultural Road, Vancouver, British Columbia, Canada\\
\and
Department of Physics and Astronomy, Dana and David Dornsife College of Letter, Arts and Sciences, University of Southern California, Los Angeles, CA 90089, U.S.A.\\
\and
Department of Physics and Astronomy, University College London, London WC1E 6BT, U.K.\\
\and
Department of Physics, Florida State University, Keen Physics Building, 77 Chieftan Way, Tallahassee, Florida, U.S.A.\\
\and
Department of Physics, Gustaf H\"{a}llstr\"{o}min katu 2a, University of Helsinki, Helsinki, Finland\\
\and
Department of Physics, Princeton University, Princeton, New Jersey, U.S.A.\\
\and
Department of Physics, University of California, Santa Barbara, California, U.S.A.\\
\and
Dipartimento di Fisica e Astronomia G. Galilei, Universit\`{a} degli Studi di Padova, via Marzolo 8, 35131 Padova, Italy\\
\and
Dipartimento di Fisica e Scienze della Terra, Universit\`{a} di Ferrara, Via Saragat 1, 44122 Ferrara, Italy\\
\and
Dipartimento di Fisica, Universit\`{a} La Sapienza, P. le A. Moro 2, Roma, Italy\\
\and
Dipartimento di Fisica, Universit\`{a} degli Studi di Milano, Via Celoria, 16, Milano, Italy\\
\and
Dipartimento di Fisica, Universit\`{a} degli Studi di Trieste, via A. Valerio 2, Trieste, Italy\\
\and
Dipartimento di Fisica, Universit\`{a} di Roma Tor Vergata, Via della Ricerca Scientifica, 1, Roma, Italy\\
\and
Discovery Center, Niels Bohr Institute, Blegdamsvej 17, Copenhagen, Denmark\\
\and
Dpto. Astrof\'{i}sica, Universidad de La Laguna (ULL), E-38206 La Laguna, Tenerife, Spain\\
\and
European Southern Observatory, ESO Vitacura, Alonso de Cordova 3107, Vitacura, Casilla 19001, Santiago, Chile\\
\and
European Space Agency, ESAC, Planck Science Office, Camino bajo del Castillo, s/n, Urbanizaci\'{o}n Villafranca del Castillo, Villanueva de la Ca\~{n}ada, Madrid, Spain\\
\and
European Space Agency, ESTEC, Keplerlaan 1, 2201 AZ Noordwijk, The Netherlands\\
\and
Helsinki Institute of Physics, Gustaf H\"{a}llstr\"{o}min katu 2, University of Helsinki, Helsinki, Finland\\
\and
INAF - Osservatorio Astronomico di Padova, Vicolo dell'Osservatorio 5, Padova, Italy\\
\and
INAF - Osservatorio Astronomico di Roma, via di Frascati 33, Monte Porzio Catone, Italy\\
\and
INAF - Osservatorio Astronomico di Trieste, Via G.B. Tiepolo 11, Trieste, Italy\\
\and
INAF/IASF Bologna, Via Gobetti 101, Bologna, Italy\\
\and
INAF/IASF Milano, Via E. Bassini 15, Milano, Italy\\
\and
INFN, Sezione di Bologna, Via Irnerio 46, I-40126, Bologna, Italy\\
\and
INFN, Sezione di Roma 1, Universit\`{a} di Roma Sapienza, Piazzale Aldo Moro 2, 00185, Roma, Italy\\
\and
INFN/National Institute for Nuclear Physics, Via Valerio 2, I-34127 Trieste, Italy\\
\and
IPAG: Institut de Plan\'{e}tologie et d'Astrophysique de Grenoble, Universit\'{e} Joseph Fourier, Grenoble 1 / CNRS-INSU, UMR 5274, Grenoble, F-38041, France\\
\and
Imperial College London, Astrophysics group, Blackett Laboratory, Prince Consort Road, London, SW7 2AZ, U.K.\\
\and
Infrared Processing and Analysis Center, California Institute of Technology, Pasadena, CA 91125, U.S.A.\\
\and
Institut Universitaire de France, 103, bd Saint-Michel, 75005, Paris, France\\
\and
Institut d'Astrophysique Spatiale, CNRS (UMR8617) Universit\'{e} Paris-Sud 11, B\^{a}timent 121, Orsay, France\\
\and
Institut d'Astrophysique de Paris, CNRS (UMR7095), 98 bis Boulevard Arago, F-75014, Paris, France\\
\and
Institute for Space Sciences, Bucharest-Magurale, Romania\\
\and
Institute of Astronomy, University of Cambridge, Madingley Road, Cambridge CB3 0HA, U.K.\\
\and
Institute of Theoretical Astrophysics, University of Oslo, Blindern, Oslo, Norway\\
\and
Instituto de Astrof\'{\i}sica de Canarias, C/V\'{\i}a L\'{a}ctea s/n, La Laguna, Tenerife, Spain\\
\and
Instituto de F\'{\i}sica de Cantabria (CSIC-Universidad de Cantabria), Avda. de los Castros s/n, Santander, Spain\\
\and
Jet Propulsion Laboratory, California Institute of Technology, 4800 Oak Grove Drive, Pasadena, California, U.S.A.\\
\and
Jodrell Bank Centre for Astrophysics, Alan Turing Building, School of Physics and Astronomy, The University of Manchester, Oxford Road, Manchester, M13 9PL, U.K.\\
\and
Kavli Institute for Cosmology Cambridge, Madingley Road, Cambridge, CB3 0HA, U.K.\\
\and
LAL, Universit\'{e} Paris-Sud, CNRS/IN2P3, Orsay, France\\
\and
LERMA, CNRS, Observatoire de Paris, 61 Avenue de l'Observatoire, Paris, France\\
\and
Laboratoire AIM, IRFU/Service d'Astrophysique - CEA/DSM - CNRS - Universit\'{e} Paris Diderot, B\^{a}t. 709, CEA-Saclay, F-91191 Gif-sur-Yvette Cedex, France\\
\and
Laboratoire Univers et Th\'{e}ories (LUTh), UMR 8102 CNRS, Observatoire de Paris, Universit\'{e} Paris Diderot, 5 Place Jules Janssen, 92190 Meudon, France\\
\and
Laboratoire de Physique Subatomique et de Cosmologie, Universit\'{e} Joseph Fourier Grenoble I, CNRS/IN2P3, Institut National Polytechnique de Grenoble, 53 rue des Martyrs, 38026 Grenoble cedex, France\\
\and
Laboratoire de Physique Th\'{e}orique, Universit\'{e} Paris-Sud 11 \& CNRS, B\^{a}timent 210, 91405 Orsay, France\\
\and
Max-Planck-Institut f\"{u}r Astrophysik, Karl-Schwarzschild-Str. 1, 85741 Garching, Germany\\
\and
McGill Physics, Ernest Rutherford Physics Building, McGill University, 3600 rue University, Montr\'{e}al, QC, H3A 2T8, Canada\\
\and
National University of Ireland, Department of Experimental Physics, Maynooth, Co. Kildare, Ireland\\
\and
Niels Bohr Institute, Blegdamsvej 17, Copenhagen, Denmark\\
\and
Optical Science Laboratory, University College London, Gower Street, London, U.K.\\
\and
SB-ITP-LPPC, EPFL, CH-1015, Lausanne, Switzerland\\
\and
SISSA, Astrophysics Sector, via Bonomea 265, 34136, Trieste, Italy\\
\and
School of Physics and Astronomy, Cardiff University, Queens Buildings, The Parade, Cardiff, CF24 3AA, U.K.\\
\and
School of Physics and Astronomy, University of Nottingham, Nottingham NG7 2RD, U.K.\\
\and
Space Sciences Laboratory, University of California, Berkeley, California, U.S.A.\\
\and
Special Astrophysical Observatory, Russian Academy of Sciences, Nizhnij Arkhyz, Zelenchukskiy region, Karachai-Cherkessian Republic, 369167, Russia\\
\and
Sub-Department of Astrophysics, University of Oxford, Keble Road, Oxford OX1 3RH, U.K.\\
\and
Theory Division, PH-TH, CERN, CH-1211, Geneva 23, Switzerland\\
\and
UPMC Univ Paris 06, UMR7095, 98 bis Boulevard Arago, F-75014, Paris, France\\
\and
Universit\'{e} de Toulouse, UPS-OMP, IRAP, F-31028 Toulouse cedex 4, France\\
\and
University of Granada, Departamento de F\'{\i}sica Te\'{o}rica y del Cosmos, Facultad de Ciencias, Granada, Spain\\
\and
University of Granada, Instituto Carlos I de F\'{\i}sica Te\'{o}rica y Computacional, Granada, Spain\\
% ...
}

%% file: table_alpha
\begingroup
\openup 5pt
\newdimen\tblskip \tblskip=5pt
\nointerlineskip
\vskip -3mm
\setbox\tablebox=\vbox{
    \newdimen\digitwidth
    \setbox0=\hbox{\rm 0}
    \digitwidth=\wd0
    \catcode`"=\active
    \def"{\kern\digitwidth}
    \newdimen\signwidth
    \setbox0=\hbox{+}
    \signwidth=\wd0
    \catcode`!=\active
    \def!{\kern\signwidth}
\halign{
\hbox to 0.8in{$#$\leaderfil}\tabskip=0.75em&$#$\hfil&$#$\hfil&$#$\hfil&$#$\hfil&$#$\hfil&\hfil$#$\hfil\tabskip=0pt\cr
\noalign{\doubleline}
\multispan1\hfil \hfil&\multispan1\hfil \Planck+\WP\hfil&\multispan1\hfil \Planck+\WP+\HighL\hfil&\multispan1\hfil \Planck+\WP+BAO\hfil&\multispan1\hfil \Planck+\WP+\HST\hfil&\multispan1\hfil \Planck+\WP+lensing\hfil&\multispan1\hfil \WMAP-9\hfil\cr
\noalign{\vskip -3pt}
\omit\hfil Parameter\hfil&\omit\hfil 68\% limits\hfil&\omit\hfil 68\% limits\hfil&\omit\hfil 68\% limits\hfil&\omit\hfil 68\% limits\hfil&\omit\hfil 68\% limits\hfil&\omit\hfil 68\% limits\hfil\cr
\noalign{\vskip 3pt\hrule\vskip 5pt}
\Omega_{\mathrm{b}} h^2&0.02207\pm 0.00028&0.02212\pm 0.00028&0.02220\pm 0.00026&0.02226\pm 0.00028&0.02218\pm 0.00027&0.0231\pm 0.0013\cr
\Omega_{\mathrm{c}} h^2&0.1173\pm 0.0031&0.1183\pm 0.0030&0.1160\pm 0.0029&0.1167\pm 0.0031&0.1162\pm 0.0027&0.1145\pm 0.0048\cr
H_0&65.1^{+1.7}_{-1.9}&66.2\pm 1.7&66.8\pm 1.2&68.3\pm 1.5&65.9\pm 1.7&74^{+10}_{-10}\cr
\tau&0.095^{+0.013}_{-0.016}&0.095^{+0.013}_{-0.016}&0.097^{+0.014}_{-0.016}&0.095^{+0.013}_{-0.016}&0.095^{+0.013}_{-0.015}&0.089^{+0.013}_{-0.015}\cr
\alpha/\alpha_0&0.9934\pm 0.0042&0.9964\pm 0.0037&0.9955\pm 0.0039&0.9991\pm 0.0039&0.9938\pm 0.0043&1.007\pm 0.020\cr
n_\mathrm{s}&0.975\pm 0.012&0.967\pm 0.011&0.975\pm 0.012&0.969\pm 0.012&0.977\pm 0.011&0.974\pm 0.014\cr
\ln(10^{10} A_\mathrm{s})&3.106^{+0.027}_{-0.033}&3.101^{+0.026}_{-0.031}&3.104^{+0.028}_{-0.033}&3.095^{+0.027}_{-0.031}&3.102^{+0.026}_{-0.029}&3.090\pm 0.038\cr
\noalign{\vskip 5pt\hrule\vskip 3pt}
} 
} 
\endPlancktable
\endgroup

%% file: table_me
\begingroup
\openup 5pt
\newdimen\tblskip \tblskip=5pt
\nointerlineskip
\vskip -3mm
\setbox\tablebox=\vbox{
    \newdimen\digitwidth
    \setbox0=\hbox{\rm 0}
    \digitwidth=\wd0
    \catcode`"=\active
    \def"{\kern\digitwidth}
    \newdimen\signwidth
    \setbox0=\hbox{+}
    \signwidth=\wd0
    \catcode`!=\active
    \def!{\kern\signwidth}
\halign{
\hbox to 0.8in{$#$\leaderfil}\tabskip=1.0em&$#$\hfil&$#$\hfil&$#$\hfil&$#$\hfil&$#$\hfil&\hfil$#$\hfil\tabskip=0pt\cr
\noalign{\doubleline}
\multispan1\hfil \hfil&\multispan1\hfil \Planck+\WP\hfil&\multispan1\hfil \Planck+\WP+\HighL\hfil&\multispan1\hfil \Planck+\WP+BAO\hfil&\multispan1\hfil \Planck+\WP+\HST\hfil&\multispan1\hfil \Planck+\WP+lensing\hfil&\multispan1\hfil \WMAP-9\hfil\cr
\noalign{\vskip -3pt}
\omit\hfil Parameter\hfil&\omit\hfil 68\% limits\hfil&\omit\hfil 68\% limits\hfil&\omit\hfil 68\% limits\hfil&\omit\hfil 68\% limits\hfil&\omit\hfil 68\% limits\hfil&\omit\hfil 68\% limits\hfil\cr
\noalign{\vskip 3pt\hrule\vskip 5pt}
\Omega_{\mathrm{b}} h^2&0.0215^{+0.0013}_{-0.0016}&0.0215^{+0.0012}_{-0.0015}&0.02216\pm 0.00027&0.02270\pm 0.00033&0.0215^{+0.0013}_{-0.0014}&0.0229^{+0.0019}_{-0.0015}\cr
\Omega_{\mathrm{c}} h^2&0.1172^{+0.0070}_{-0.0087}&0.1171^{+0.0067}_{-0.0080}&0.1201\pm 0.0036&0.1229\pm 0.0035&0.1147^{+0.0069}_{-0.0078}&0.1154\pm 0.0083\cr
H_0&63^{+10}_{-20}&63^{+10}_{-10}&68.4\pm 1.7&73.5\pm 2.4&62^{+10}_{-10}&72\pm 10\cr
\tau&0.088^{+0.012}_{-0.014}&0.090^{+0.012}_{-0.014}&0.091^{+0.012}_{-0.014}&0.091^{+0.013}_{-0.014}&0.088^{+0.012}_{-0.014}&0.089^{+0.013}_{-0.015}\cr
m_e/m_{e0}&0.977^{+0.055}_{-0.070}&0.976^{+0.053}_{-0.064}&1.004\pm 0.011&1.027\pm 0.012&0.969\pm 0.055&1.011^{+0.077}_{-0.057}\cr
n_\mathrm{s}&0.9584\pm 0.0083&0.9565\pm 0.0077&0.9614\pm 0.0068&0.9628\pm 0.0072&0.9618\pm 0.0070&0.975\pm 0.014\cr
\ln(10^{10} A_\mathrm{s})&3.084\pm 0.027&3.085^{+0.024}_{-0.026}&3.091\pm 0.025&3.093^{+0.024}_{-0.027}&3.079\pm 0.025&3.097\pm 0.032\cr
\noalign{\vskip 5pt\hrule\vskip 3pt}
} 
} 
\endPlancktable
\endgroup

%% file: table_alpha_me
\begingroup
\openup 5pt
\newdimen\tblskip \tblskip=5pt
\nointerlineskip
\vskip -3mm
\setbox\tablebox=\vbox{
    \newdimen\digitwidth
    \setbox0=\hbox{\rm 0}
    \digitwidth=\wd0
    \catcode`"=\active
    \def"{\kern\digitwidth}
    \newdimen\signwidth
    \setbox0=\hbox{+}
    \signwidth=\wd0
    \catcode`!=\active
    \def!{\kern\signwidth}
\halign{
\hbox to 0.9in{$#$\leaderfil}\tabskip=1.5em&$#$\hfil&$#$\hfil&\hfil$#$\hfil\tabskip=0pt\cr
\noalign{\doubleline}
\multispan1\hfil \hfil&\multispan1\hfil \Planck+\WP\hfil&\multispan1\hfil \Planck+\WP+\HighL\hfil&\multispan1\hfil \WMAP-9\hfil\cr
\noalign{\vskip -3pt}
\omit\hfil Parameter\hfil&\omit\hfil 68\% limits\hfil&\omit\hfil 68\% limits\hfil&\omit\hfil 68\% limits\hfil\cr
\noalign{\vskip 3pt\hrule\vskip 5pt}
\Omega_{\mathrm{b}} h^2&0.0219\pm 0.0014&0.0216^{+0.0013}_{-0.0016}&0.0230^{+0.0018}_{-0.0015}\cr
\Omega_{\mathrm{c}} h^2&0.1166\pm 0.0069&0.1156^{+0.0069}_{-0.0081}&0.115\pm 0.010\cr
H_0&64^{+10}_{-10}&62^{+10}_{-20}&73\pm 10\cr
\tau&0.095^{+0.014}_{-0.016}&0.093^{+0.013}_{-0.016}&0.090^{+0.014}_{-0.015}\cr
\alpha/\alpha_0&0.9933\pm 0.0045&0.9963\pm 0.0037&1.006^{+0.025}_{-0.034}\cr
m_e/m_{e0}&0.994\pm 0.059&0.976^{+0.057}_{-0.066}&1.004\pm 0.091\cr
n_\mathrm{s}&0.974\pm 0.014&0.965\pm 0.012&0.975\pm 0.018\cr
\ln(10^{10} A_\mathrm{s})&3.105^{+0.030}_{-0.034}&3.096\pm 0.030&3.093\pm 0.051\cr
\noalign{\vskip 5pt\hrule\vskip 3pt}
} 
} 
\endPlancktable
\endgroup

%% file: table_alpha_nnu
\begingroup
\openup 5pt
\newdimen\tblskip \tblskip=5pt
\nointerlineskip
\vskip -3mm
\setbox\tablebox=\vbox{
    \newdimen\digitwidth
    \setbox0=\hbox{\rm 0}
    \digitwidth=\wd0
    \catcode`"=\active
    \def"{\kern\digitwidth}
    \newdimen\signwidth
    \setbox0=\hbox{+}
    \signwidth=\wd0
    \catcode`!=\active
    \def!{\kern\signwidth}
\halign{
\hbox to 0.9in{$#$\leaderfil}\tabskip=1.5em&$#$\hfil&$#$\hfil&\hfil$#$\hfil\tabskip=0pt\cr
\noalign{\doubleline}
\multispan1\hfil \hfil&\multispan1\hfil \Planck+\WP\hfil&\multispan1\hfil \Planck+\WP+\HighL\hfil&\multispan1\hfil \WMAP-9\hfil\cr
\noalign{\vskip -3pt}
\omit\hfil Parameter\hfil&\omit\hfil 68\% limits\hfil&\omit\hfil 68\% limits\hfil&\omit\hfil 68\% limits\hfil\cr
\noalign{\vskip 3pt\hrule\vskip 5pt}
\Omega_{\mathrm{b}} h^2&0.02204^{+0.00052}_{-0.00061}&0.02229\pm 0.00054&0.0230^{+0.0012}_{-0.0010}\cr
\Omega_{\mathrm{c}} h^2&0.1174^{+0.0088}_{-0.012}&0.1227^{+0.0088}_{-0.013}&0.140^{+0.024}_{-0.044}\cr
H_0&65.2^{+5.8}_{-8.1}&69.0^{+6.0}_{-8.0}&79^{+20}_{-9}\cr
\tau&0.095^{+0.014}_{-0.016}&0.095^{+0.014}_{-0.016}&0.089\pm 0.014\cr
N_{\mathrm{eff}}&3.04^{+0.54}_{-0.73}&3.30^{+0.53}_{-0.76}&4.46^{+1.5}_{-2.4}\cr
\alpha/\alpha_0&0.9933^{+0.0073}_{-0.0083}&0.9988^{+0.0067}_{-0.0075}&1.006^{+0.020}_{-0.016}\cr
n_\mathrm{s}&0.974\pm 0.017&0.971\pm 0.017&0.989\pm 0.026\cr
\ln(10^{10} A_\mathrm{s})&3.105\pm 0.035&3.107^{+0.034}_{-0.038}&3.130^{+0.075}_{-0.063}\cr
\noalign{\vskip 5pt\hrule\vskip 3pt}
} 
} 
\endPlancktable
\endgroup

%% file: table_alpha_yhe
\begingroup
\openup 5pt
\newdimen\tblskip \tblskip=5pt
\nointerlineskip
\vskip -3mm
\setbox\tablebox=\vbox{
    \newdimen\digitwidth
    \setbox0=\hbox{\rm 0}
    \digitwidth=\wd0
    \catcode`"=\active
    \def"{\kern\digitwidth}
    \newdimen\signwidth
    \setbox0=\hbox{+}
    \signwidth=\wd0
    \catcode`!=\active
    \def!{\kern\signwidth}
\halign{
\hbox to 0.9in{$#$\leaderfil}\tabskip=1.5em&$#$\hfil&$#$\hfil&\hfil$#$\hfil\tabskip=0pt\cr
\noalign{\doubleline}
\multispan1\hfil \hfil&\multispan1\hfil \Planck+\WP\hfil&\multispan1\hfil \Planck+\WP+\HighL\hfil&\multispan1\hfil \WMAP-9\hfil\cr
\noalign{\vskip -3pt}
\omit\hfil Parameter\hfil&\omit\hfil 68\% limits\hfil&\omit\hfil 68\% limits\hfil&\omit\hfil 68\% limits\hfil\cr
\noalign{\vskip 3pt\hrule\vskip 5pt}
\Omega_{\mathrm{b}} h^2&0.02220^{+0.00073}_{-0.0013}&0.02245^{+0.00075}_{-0.0014}&0.0232\pm 0.0014\cr
\Omega_{\mathrm{c}} h^2&0.1179^{+0.0051}_{-0.0063}&0.1202^{+0.0057}_{-0.0069}&0.1166^{+0.0061}_{-0.0079}\cr
H_0&66.6^{+6.2}_{-12}&69.8^{+6.1}_{-13}&75\pm 10\cr
\tau&0.095^{+0.014}_{-0.016}&0.094^{+0.013}_{-0.016}&0.088^{+0.013}_{-0.015}\cr
Y_{\mathrm{p}}&< 0.306&< 0.325&< 0.345\cr
\alpha/\alpha_0&0.996^{+0.015}_{-0.024}&1.003^{+0.014}_{-0.026}&1.010\pm 0.023\cr
n_\mathrm{s}&0.976\pm 0.013&0.968\pm 0.012&0.976\pm 0.015\cr
\ln(10^{10} A_\mathrm{s})&3.107\pm 0.030&3.102\pm 0.028&3.096\pm 0.039\cr
\noalign{\vskip 5pt\hrule\vskip 3pt}
} 
} 
\endPlancktable
\endgroup

%% file: table_alpha_tauprior
\begingroup
\openup 5pt
\newdimen\tblskip \tblskip=5pt
\nointerlineskip
\vskip -3mm
\setbox\tablebox=\vbox{
    \newdimen\digitwidth
    \setbox0=\hbox{\rm 0}
    \digitwidth=\wd0
    \catcode`"=\active
    \def"{\kern\digitwidth}
    \newdimen\signwidth
    \setbox0=\hbox{+}
    \signwidth=\wd0
    \catcode`!=\active
    \def!{\kern\signwidth}
\halign{
\hbox to 0.9in{$#$\leaderfil}\tabskip=1.5em&$#$\hfil&\hfil$#$\hfil\tabskip=0pt\cr
\noalign{\doubleline}
\multispan1\hfil \hfil&\multispan1\hfil \Planck $-$ low l+$\tau$ prior\hfil&\multispan1\hfil \Planck $-$ low l+$\tau$ prior+\HighL\hfil\cr
\noalign{\vskip -3pt}
\omit\hfil Parameter\hfil&\omit\hfil 68\% limits\hfil&\omit\hfil 68\% limits\hfil\cr
\noalign{\vskip 3pt\hrule\vskip 5pt}
\Omega_{\mathrm{b}} h^2&0.02194\pm 0.00028&0.02194\pm 0.00028\cr
\Omega_{\mathrm{c}} h^2&0.1205\pm 0.0040&0.1228\pm 0.0038\cr
H_0&65.6\pm 1.9&66.6\pm 1.7\cr
\tau&0.094\pm 0.013&0.093\pm 0.012\cr
\alpha/\alpha_0&0.9970\pm 0.0054&1.0011\pm 0.0045\cr
n_\mathrm{s}&0.961\pm 0.016&0.948\pm 0.015\cr
\ln(10^{10} A_\mathrm{s})&3.105\pm 0.026&3.101\pm 0.025\cr
\noalign{\vskip 5pt\hrule\vskip 3pt}
} 
} 
\endPlancktable
\endgroup

%% file: table_me_tauprior
\begingroup
\openup 5pt
\newdimen\tblskip \tblskip=5pt
\nointerlineskip
\vskip -3mm
\setbox\tablebox=\vbox{
    \newdimen\digitwidth
    \setbox0=\hbox{\rm 0}
    \digitwidth=\wd0
    \catcode`"=\active
    \def"{\kern\digitwidth}
    \newdimen\signwidth
    \setbox0=\hbox{+}
    \signwidth=\wd0
    \catcode`!=\active
    \def!{\kern\signwidth}
\halign{
\hbox to 0.9in{$#$\leaderfil}\tabskip=1.5em&$#$\hfil&\hfil$#$\hfil\tabskip=0pt\cr
\noalign{\doubleline}
\multispan1\hfil \hfil&\multispan1\hfil \Planck $-$ low l+$\tau$ prior\hfil&\multispan1\hfil \Planck $-$ low l+$\tau$ prior+\HighL\hfil\cr
\noalign{\vskip -3pt}
\omit\hfil Parameter\hfil&\omit\hfil 68\% limits\hfil&\omit\hfil 68\% limits\hfil\cr
\noalign{\vskip 3pt\hrule\vskip 5pt}
\Omega_{\mathrm{b}} h^2&0.0221^{+0.0023}_{-0.0019}&0.0220^{+0.0017}_{-0.0021}\cr
\Omega_{\mathrm{c}} h^2&0.123^{+0.011}_{-0.012}&0.1224\pm 0.0098\cr
H_0&\dots&< 76.1\cr
\tau&0.092\pm 0.012&0.094\pm 0.012\cr
m_e/m_{e0}&1.007^{+0.099}_{-0.081}&1.001\pm 0.075\cr
n_\mathrm{s}&0.9527\pm 0.0082&0.9503\pm 0.0081\cr
\ln(10^{10} A_\mathrm{s})&3.101\pm 0.026&3.102\pm 0.025\cr
\noalign{\vskip 5pt\hrule\vskip 3pt}
} 
} 
\endPlancktable
\endgroup

%% file: PIP_100_Rocha_v2.bbl
\begin{thebibliography}{120}
\expandafter\ifx\csname natexlab\endcsname\relax\def\natexlab#1{#1}\fi

\bibitem[{{Ali-Ha{\"i}moud} \& {Hirata}(2011)}]{hyrec}
{Ali-Ha{\"i}moud}, Y. \& {Hirata}, C.~M. 2011, \prd, 83, 043513

\bibitem[{Anderson {et~al.}(2013)Anderson, Aubourg, Bailey, Bizyaev, Blanton,
  {et~al.}}]{Anderson:2012sa}
Anderson, L., Aubourg, E., Bailey, S., {et~al.} 2013, Mon.Not.Roy.Astron.Soc.,
  428, 1036

\bibitem[{{Austermann} {et~al.}(2012){Austermann}, {Aird}, {Beall}, {Becker},
  {Bender}, {Benson}, {Bleem}, {Britton}, {Carlstrom}, {Chang}, {Chiang},
  {Cho}, {Crawford}, {Crites}, {Datesman}, {de Haan}, {Dobbs}, {George},
  {Halverson}, {Harrington}, {Henning}, {Hilton}, {Holder}, {Holzapfel},
  {Hoover}, {Huang}, {Hubmayr}, {Irwin}, {Keisler}, {Kennedy}, {Knox}, {Lee},
  {Leitch}, {Li}, {Lueker}, {Marrone}, {McMahon}, {Mehl}, {Meyer}, {Montroy},
  {Natoli}, {Nibarger}, {Niemack}, {Novosad}, {Padin}, {Pryke}, {Reichardt},
  {Ruhl}, {Saliwanchik}, {Sayre}, {Schaffer}, {Shirokoff}, {Stark}, {Story},
  {Vanderlinde}, {Vieira}, {Wang}, {Williamson}, {Yefremenko}, {Yoon}, \&
  {Zahn}}]{sptpol}
{Austermann}, J.~E., {Aird}, K.~A., {Beall}, J.~A., {et~al.} 2012, in Society
  of Photo-Optical Instrumentation Engineers (SPIE) Conference Series, Vol.
  8452, Society of Photo-Optical Instrumentation Engineers (SPIE) Conference
  Series

\bibitem[{Avelino {et~al.}(2001)Avelino, Esposito, Mangano, \&
  Martins}]{cmb-avelino01}
Avelino, P., Esposito, S., Mangano, G., \& Martins, C. e.~a. 2001, Phys. Rev.
  D, 64, 103505

\bibitem[{Avelino {et~al.}(2000)Avelino, Martins, \& Rocha}]{cmb-avelino00}
Avelino, P., Martins, C., \& Rocha, G. 2000, Phys. Rev. D, 62, 123508

\bibitem[{{Aver} {et~al.}(2013){Aver}, {Olive}, {Porter}, \&
  {Skillman}}]{aver2013}
{Aver}, E., {Olive}, K.~A., {Porter}, R.~L., \& {Skillman}, E.~D. 2013, \jcap,
  11, 17

\bibitem[{Battye {et~al.}(2001)Battye, Crittenden, \& Weller}]{cmb-bat}
Battye, R., Crittenden, R., \& Weller, J. 2001, Phys. Rev. D, 63, 043505

\bibitem[{{Bennett} {et~al.}(2013{\natexlab{a}}){Bennett}, {Larson}, {Weiland},
  {Jarosik}, {Hinshaw}, {Odegard}, {Smith}, {Hill}, {Gold}, {Halpern},
  {Komatsu}, {Nolta}, {Page}, {Spergel}, {Wollack}, {Dunkley}, {Kogut},
  {Limon}, {Meyer}, {Tucker}, \& {Wright}}]{Bennett:2012fp}
{Bennett}, C.~L., {Larson}, D., {Weiland}, J.~L., {et~al.} 2013{\natexlab{a}},
  \apjs, 208, 20

\bibitem[{{Bennett} {et~al.}(2013{\natexlab{b}}){Bennett}, {Larson}, {Weiland},
  {Jarosik}, {Hinshaw}, {Odegard}, {Smith}, {Hill}, {Gold}, {Halpern},
  {Komatsu}, {Nolta}, {Page}, {Spergel}, {Wollack}, {Dunkley}, {Kogut},
  {Limon}, {Meyer}, {Tucker}, \& {Wright}}]{bennett2012}
{Bennett}, C.~L., {Larson}, D., {Weiland}, J.~L., {et~al.} 2013{\natexlab{b}},
  \apjs, 208, 20

\bibitem[{Berengut {et~al.}(2011)Berengut, Flambaum, King, Curran, \&
  Webb}]{dipole2}
Berengut, J., Flambaum, V., King, J., Curran, S., \& Webb, J. 2011, Phys. Rev.
  D, { 83}, 123506

\bibitem[{Bergstr{\"{o}}m {et~al.}(1999)Bergstr{\"{o}}m, Iguri, \&
  Rubinstein}]{berg}
Bergstr{\"{o}}m, L., Iguri, S., \& Rubinstein, H. 1999, Phys. Rev. D, 60

\bibitem[{{Beutler} {et~al.}(2011){Beutler}, {Blake}, {Colless}, {Jones},
  {Staveley-Smith}, {Campbell}, {Parker}, {Saunders}, \& {Watson}}]{Beutler:11}
{Beutler}, F., {Blake}, C., {Colless}, M., {et~al.} 2011, \mnras, 416, 3017

\bibitem[{Bize {et~al.}(2003)Bize, Diddams, Tanaka, Tanner, Oskay, Drullinger,
  Parker, Heavner, Jefferts, Hollberg, Itano, \& Bergquist}]{bize}
Bize, S., Diddams, S., Tanaka, U., {et~al.} 2003, Phys. Rev. Lett., 90

\bibitem[{{Bonifacio} {et~al.}(2013){Bonifacio}, {Rahmani}, {Whitmore},
  {Wendt}, {Centurion}, {Molaro}, {Srianand}, {Murphy}, {Petitjean},
  {Agafonova}, {D'Odorico}, {Evans}, {Levshakov}, {Lopez}, {Martins},
  {Reimers}, \& {Vladilo}}]{bonifacio13}
{Bonifacio}, P., {Rahmani}, H., {Whitmore}, J.~B., {et~al.} 2013, ArXiv
  e-prints

\bibitem[{Cameron \& Pettitt(2012)}]{cameron2012evidence}
Cameron, E. \& Pettitt, T. 2012, arXiv preprint arXiv:1207.6223

\bibitem[{Cameron \& Pettitt(2013)}]{cameron2013evidence}
Cameron, E. \& Pettitt, T. 2013, arXiv preprint arXiv:1309.2737

\bibitem[{{Carilli} {et~al.}(2001){Carilli}, {Menten}, {Stocke}, {Perlman},
  {Vermeulen}, {Briggs}, {de Bruyn}, {Conway}, \& {Moore}}]{carilli00}
{Carilli}, C.~L., {Menten}, K.~M., {Stocke}, J.~T., {et~al.} 2001, Phys. Rev.
  Lett., { 85}

\bibitem[{{Chluba} \& {Thomas}(2011)}]{cosmorec}
{Chluba}, J. \& {Thomas}, R.~M. 2011, \mnras, 412, 748

\bibitem[{Cing{\"{o}}z {et~al.}(2008)Cing{\"{o}}z, Lapierre, Nguyen, Leefer,
  Budker, Lamoreaux, \& Torgerson}]{cingoz}
Cing{\"{o}}z, A., Lapierre, A., Nguyen, A.-T., {et~al.} 2008, Phys. Rev. Lett.,
  98

\bibitem[{Coc {et~al.}(2012)Coc, Descouvemont, Uzan, \& Vangioni}]{cduv}
Coc, A., Descouvemont, P., Uzan, J.-P., \& Vangioni, E. 2012, PoS, NICXII, 073

\bibitem[{Coc {et~al.}(2009)Coc, Ekstr{\"{o}}m, Descouvemont, Meynet, Olive,
  Uzan, \& Vangioni}]{sylvia2}
Coc, A., Ekstr{\"{o}}m, S., Descouvemont, P., {et~al.} 2009, Mem. Soc. Astron.
  Ital., 80, 809

\bibitem[{Coc {et~al.}(2007)Coc, Nunes, Olive, Uzan, \& Vangioni}]{cnouv}
Coc, A., Nunes, N.~J., Olive, K.~A., Uzan, J.-P., \& Vangioni, E. 2007, Phys.
  Rev. D, {76}, 023511

\bibitem[{Coc {et~al.}(2006)Coc, Olive, Uzan, \& Vangioni}]{bbn-ST2}
Coc, A., Olive, K.~A., Uzan, J.-P., \& Vangioni, E. 2006, Phys. Rev. D, {73},
  083525

\bibitem[{Damour \& Dyson(1996)}]{dady}
Damour, T. \& Dyson, F. 1996, Nucl. Phys. B, 480, 37

\bibitem[{Damour \& Pichon(1999)}]{bbn-ST1}
Damour, T. \& Pichon, B. 1999, Phys. Rev. D, {59}, 123502

\bibitem[{{Das} {et~al.}(2013){Das}, {Louis}, {Nolta}, {Addison},
  {Battistelli}, {Bond}, {Calabrese}, {Devlin}, {Dicker}, {Dunkley},
  {D{\"u}nner}, {Fowler}, {Gralla}, {Hajian}, {Halpern}, {Hasselfield},
  {Hilton}, {Hincks}, {Hlozek}, {Huffenberger}, {Hughes}, {Irwin}, {Kosowsky},
  {Lupton}, {Marriage}, {Marsden}, {Menanteau}, {Moodley}, {Niemack}, {Page},
  {Partridge}, {Reese}, {Schmitt}, {Sehgal}, {Sherwin}, {Sievers}, {Spergel},
  {Staggs}, {Swetz}, {Switzer}, {Thornton}, {Trac}, \& {Wollack}}]{das13}
{Das}, S., {Louis}, T., {Nolta}, M.~R., {et~al.} 2013, ArXiv e-prints

\bibitem[{Dicke(1964)}]{dicke64}
Dicke, R. 1964, in Relativity, Groups and Topology. Relativit\'e, Groupes et
  Topologie, ed. C.~DeWitt \& B.~DeWitt (New York; London: Gordon and Breach),
  165--313

\bibitem[{{Dicke}(1962)}]{dicke62}
{Dicke}, R.~H. 1962, Physical Review, 125, 2163

\bibitem[{{Duff}(2002)}]{duff02}
{Duff}, M.~J. 2002, ArXiv High Energy Physics - Theory e-prints

\bibitem[{Dyson(1972)}]{dyson}
Dyson, F. 1972, in Aspects of Quantum Theory, ed. A.~Salam \& E.~Wigner
  (Cambridge; New York: Cambridge University Press), 213--236

\bibitem[{Ekstr{\"{o}}m {et~al.}(2010)Ekstr{\"{o}}m, Coc, Descouvemont, Meynet,
  Olive, Uzan, \& Vangioni}]{sylvia}
Ekstr{\"{o}}m, S., Coc, A., Descouvemont, P., {et~al.} 2010, A\&A, 514, A62

\bibitem[{Ellis \& Uzan(2005)}]{uellis}
Ellis, G.~F. \& Uzan, J.-P. 2005, Am. J. Phys., { 73}, 240

\bibitem[{{Flambaum}(2007)}]{Flambaum2007}
{Flambaum}, V.~V. 2007, International Journal of Modern Physics A, 22, 4937

\bibitem[{Fujii {et~al.}(2000{\natexlab{a}})Fujii, Iwamoto, Fukahori, Ohnuki,
  Nakagawa, Hidaka, Oura, \& M{\"{o}}ller}]{fujii}
Fujii, Y., Iwamoto, A., Fukahori, T., {et~al.} 2000{\natexlab{a}}, Nucl. Phys.
  B, 573, 377

\bibitem[{Fujii {et~al.}(2000{\natexlab{b}})Fujii, Iwamoto, Fukahori, Ohnuki,
  Nakagawa, Hidaka, Oura, \& M{\"{o}}ller}]{fujii2}
Fujii, Y., Iwamoto, A., Fukahori, T., {et~al.} 2000{\natexlab{b}}, Nucl. Phys.
  B, 573, 377

\bibitem[{{Galli}(2013)}]{galli13}
{Galli}, S. 2013, \prd, 87, 123516

\bibitem[{{Galli} {et~al.}(2010){Galli}, {Martinelli}, {Melchiorri}, {Pagano},
  {Sherwin}, \& {Spergel}}]{galli}
{Galli}, S., {Martinelli}, M., {Melchiorri}, A., {et~al.} 2010, \prd, 82,
  123504

\bibitem[{{Galli} {et~al.}(2009){Galli}, {Melchiorri}, {Smoot}, \&
  {Zahn}}]{gallig}
{Galli}, S., {Melchiorri}, A., {Smoot}, G.~F., \& {Zahn}, O. 2009, \prd, 80,
  023508

\bibitem[{Gelman \& Rubin(1992)}]{gelman}
Gelman, A. \& Rubin, D. 1992, Statistical Science, 7, 457

\bibitem[{Gorski {et~al.}(2005)Gorski, Hivon, Banday, Wandelt, Hansen,
  Reinecke, \& Bartelmann}]{gorski2005healpix}
Gorski, K.~M., Hivon, E., Banday, A., {et~al.} 2005, The Astrophysical Journal,
  622, 759

\bibitem[{Gould {et~al.}(2006)Gould, Sharapov, \& Lamoreaux}]{gould}
Gould, C., Sharapov, E., \& Lamoreaux, S. 2006, Phys. Rev. C, 74

\bibitem[{Hamann {et~al.}(2011)Hamann, Hannestad, Raffelt, \&
  Wong}]{Hamann:2011ge}
Hamann, J., Hannestad, S., Raffelt, G.~G., \& Wong, Y.~Y. 2011, JCAP, 1109, 034

\bibitem[{Hannestad(1999)}]{cmb-han}
Hannestad, S. 1999, Phys. Rev. D, 60, 023515

\bibitem[{Hanson \& Lewis(2009)}]{HansonLewis}
Hanson, D. \& Lewis, A. 2009, Phys. Rev. D, {80}, 063004

\bibitem[{{Hinshaw} {et~al.}(2013){Hinshaw}, {Larson}, {Komatsu}, {Spergel},
  {Bennett}, {Dunkley}, {Nolta}, {Halpern}, {Hill}, {Odegard}, {Page}, {Smith},
  {Weiland}, {Gold}, {Jarosik}, {Kogut}, {Limon}, {Meyer}, {Tucker}, {Wollack},
  \& {Wright}}]{hinshaw2012}
{Hinshaw}, G., {Larson}, D., {Komatsu}, E., {et~al.} 2013, \apjs, 208, 19

\bibitem[{{Hou} {et~al.}(2013){Hou}, {Keisler}, {Knox}, {Millea}, \&
  {Reichardt}}]{Hou:2011ec}
{Hou}, Z., {Keisler}, R., {Knox}, L., {Millea}, M., \& {Reichardt}, C. 2013,
  \prd, 87, 083008

\bibitem[{{Hu} \& {Sugiyama}(1995)}]{hu95}
{Hu}, W. \& {Sugiyama}, N. 1995, \apj, 444, 489

\bibitem[{Ichikawa {et~al.}(2006)Ichikawa, Kanzaki, \& Kawasaki}]{cmb2-2006}
Ichikawa, K., Kanzaki, T., \& Kawasaki, M. 2006, Phys. Rev. D, 74, 023515

\bibitem[{{Jain} {et~al.}(2013){Jain}, {Joyce}, {Thompson}, {Upadhye},
  {Battat}, {Brax}, {Davis}, {de Rham}, {Dodelson}, {Erickcek}, {Gabadadze},
  {Hu}, {Hui}, {Huterer}, {Kamionkowski}, {Khoury}, {Koyama}, {Li}, {Linder},
  {Schmidt}, {Scoccimarro}, {Starkman}, {Stubbs}, {Takada}, {Tolley},
  {Trodden}, {Uzan}, {Vikram}, {Weltman}, {Wyman}, {Zaritsky}, \&
  {Zhao}}]{jain13}
{Jain}, B., {Joyce}, A., {Thompson}, R., {et~al.} 2013, ArXiv e-prints

\bibitem[{{Kaiser}(1983)}]{kaiser83}
{Kaiser}, N. 1983, \mnras, 202, 1169

\bibitem[{{Kanekar} {et~al.}(2005){Kanekar}, {Carilli}, {Langston}, {Rocha},
  {Combes}, {Subrahmanyan}, {Stocke}, {Menten}, {Briggs}, \&
  {Wiklind}}]{kanekar05}
{Kanekar}, N., {Carilli}, C.~L., {Langston}, C.~I., {et~al.} 2005, Phys. Rev.
  Lett., { 95}

\bibitem[{Kaplinghat {et~al.}(1999)Kaplinghat, Scherrer, \& Turner}]{cmb-kap}
Kaplinghat, M., Scherrer, R., \& Turner, M. 1999, Phys. Rev. D, 60, 023516

\bibitem[{{Kermish} {et~al.}(2012){Kermish}, {Ade}, {Anthony}, {Arnold},
  {Barron}, {Boettger}, {Borrill}, {Chapman}, {Chinone}, {Dobbs}, {Errard},
  {Fabbian}, {Flanigan}, {Fuller}, {Ghribi}, {Grainger}, {Halverson},
  {Hasegawa}, {Hattori}, {Hazumi}, {Holzapfel}, {Howard}, {Hyland}, {Jaffe},
  {Keating}, {Kisner}, {Lee}, {Le Jeune}, {Linder}, {Lungu}, {Matsuda},
  {Matsumura}, {Meng}, {Miller}, {Morii}, {Moyerman}, {Myers}, {Nishino},
  {Paar}, {Quealy}, {Reichardt}, {Richards}, {Ross}, {Shimizu}, {Shimon},
  {Shimmin}, {Sholl}, {Siritanasak}, {Spieler}, {Stebor}, {Steinbach},
  {Stompor}, {Suzuki}, {Tomaru}, {Tucker}, \& {Zahn}}]{polarbear}
{Kermish}, Z.~D., {Ade}, P., {Anthony}, A., {et~al.} 2012, in Society of
  Photo-Optical Instrumentation Engineers (SPIE) Conference Series, Vol. 8452,
  Society of Photo-Optical Instrumentation Engineers (SPIE) Conference Series

\bibitem[{King {et~al.}(2012)King, Webb, Murphy, Flambaum, Carswell,
  {et~al.}}]{dipole3}
King, J.~A., Webb, J.~K., Murphy, M.~T., {et~al.} 2012, {\tt arXiv:1202.4758}

\bibitem[{Kuroda(1956)}]{kuroda}
Kuroda, P. 1956, J. Chem. Phys., 25, 781

\bibitem[{Landau {et~al.}(2001)Landau, Harari, \& Zaldarriaga}]{cmb-landau01}
Landau, S., Harari, D., \& Zaldarriaga, M. 2001, Phys. Rev. D, 63, 083505

\bibitem[{Landau \& Sc\'occola(2010)}]{scoccola3}
Landau, S. \& Sc\'occola, C. 2010, A\&A, 517, A62

\bibitem[{Lewis \& Bridle(2002)}]{Lewis:2002ah}
Lewis, A. \& Bridle, S. 2002, Phys. Rev., D66, 103511

\bibitem[{Livio {et~al.}(1989)Livio, Hollowell, Weiss, \& Truran}]{livio}
Livio, M., Hollowell, D., Weiss, A., \& Truran, J. 1989, Nature, 340, 281

\bibitem[{Luo {et~al.}(2011)Luo, Olive, \& Uzan}]{luo}
Luo, F., Olive, K.~A., \& Uzan, J.-P. 2011, Phys. Rev. D, {84}, 096004

\bibitem[{Martins {et~al.}(2004{\natexlab{a}})Martins, Melchiorri, Rocha,
  Trotta, Avelino, \& Viana}]{cmb-martins}
Martins, C., Melchiorri, A., Rocha, G., {et~al.} 2004{\natexlab{a}}, Phys. Let.
  B, { 585}

\bibitem[{Martins {et~al.}(2004{\natexlab{b}})Martins, Melchiorri, Rocha, \&
  Trotta}]{cmb-rocha1}
Martins, C., Melchiorri, A., Rocha, G., \& Trotta, R. e.~a. 2004{\natexlab{b}},
  Phys. Lett. B, 585, 29

\bibitem[{{Martins}(2003)}]{Martins2003}
{Martins}, C.~J.~A.~P., ed. 2003, {The Cosmology of Extra Dimensions and
  Varying Fundamental Constants}

\bibitem[{Menegoni(2010)}]{menegoni2}
Menegoni, E. 2010, AIP Conf. Proc., {1256}, 288

\bibitem[{Menegoni {et~al.}(2012)Menegoni, Archidiacono, Calabrese, Galli,
  Martins, {et~al.}}]{Menegoni:2012tq}
Menegoni, E., Archidiacono, M., Calabrese, E., {et~al.} 2012, Phys.Rev., D85,
  107301

\bibitem[{Menegoni {et~al.}(2009)Menegoni, Galli, Bartlett, Martins, \&
  Melchiorri}]{menegoni1}
Menegoni, E., Galli, S., Bartlett, J., Martins, C., \& Melchiorri, A. 2009,
  Phys. Rev. D, 80, 087302

\bibitem[{Mohr {et~al.}(2008)Mohr, Taylor, \& Newell}]{codatalist}
Mohr, P., Taylor, B., \& Newell, D. 2008, Rev. Mod. Phys., 80, 633

\bibitem[{Moss {et~al.}(2011)Moss, Scott, Zibin, \& Battye}]{tilted}
Moss, A., Scott, D., Zibin, J.~P., \& Battye, R. 2011, Phys. Rev. D, { 84},
  023014

\bibitem[{Mukhanov {et~al.}(2012)Mukhanov, Kim, Naselsky, Trombetti, \&
  Burigana}]{mukhanov}
Mukhanov, V., Kim, J., Naselsky, P., Trombetti, T., \& Burigana, C. 2012, JCAP,
  {1206}, 040

\bibitem[{M{\"{u}}ller {et~al.}(2004)M{\"{u}}ller, Sch{\"{a}}fer, \&
  Wetterich}]{muller}
M{\"{u}}ller, C., Sch{\"{a}}fer, G., \& Wetterich, C. 2004, Phys. Rev. D, 70

\bibitem[{Nakashima {et~al.}(2010)Nakashima, Ichikawa, Nagata, \&
  Yokoyama}]{naka00}
Nakashima, M., Ichikawa, K., Nagata, R., \& Yokoyama, J. 2010, JCAP, 1001, 030

\bibitem[{Nakashima {et~al.}(2008)Nakashima, Nagata, \& Yokoyama}]{wmap-alpha}
Nakashima, M., Nagata, R., \& Yokoyama, J. 2008, Prog. Theor. Phys., 120, 1207

\bibitem[{{Narimani} {et~al.}(2012){Narimani}, {Moss}, \& {Scott}}]{ali}
{Narimani}, A., {Moss}, A., \& {Scott}, D. 2012, \apss, 341, 617

\bibitem[{{Niemack} {et~al.}(2010){Niemack}, {Ade}, {Aguirre}, {Barrientos},
  {Beall}, {Bond}, {Britton}, {Cho}, {Das}, {Devlin}, {Dicker}, {Dunkley},
  {D{\"u}nner}, {Fowler}, {Hajian}, {Halpern}, {Hasselfield}, {Hilton},
  {Hilton}, {Hubmayr}, {Hughes}, {Infante}, {Irwin}, {Jarosik}, {Klein},
  {Kosowsky}, {Marriage}, {McMahon}, {Menanteau}, {Moodley}, {Nibarger},
  {Nolta}, {Page}, {Partridge}, {Reese}, {Sievers}, {Spergel}, {Staggs},
  {Thornton}, {Tucker}, {Wollack}, \& {Yoon}}]{actpol}
{Niemack}, M.~D., {Ade}, P.~A.~R., {Aguirre}, J., {et~al.} 2010, in Society of
  Photo-Optical Instrumentation Engineers (SPIE) Conference Series, Vol. 7741,
  Society of Photo-Optical Instrumentation Engineers (SPIE) Conference Series

\bibitem[{{O'Bryan} {et~al.}(2013){O'Bryan}, {Smidt}, {De Bernardis}, \&
  {Cooray}}]{bryan}
{O'Bryan}, J., {Smidt}, J., {De Bernardis}, F., \& {Cooray}, A. 2013, ArXiv
  e-prints

\bibitem[{Olive {et~al.}(2002)Olive, Pospelov, Qian, Coc, Cass{\'{e}}, \&
  Vangioni-Flam}]{kids}
Olive, K., Pospelov, M., Qian, Y.-Z., {et~al.} 2002, Phys. Rev. D, 66

\bibitem[{Olive {et~al.}(2012)Olive, Peloso, \& Peterson}]{peloso}
Olive, K.~A., Peloso, M., \& Peterson, A.~J. 2012, Phys. Rev. D, { 86}, 043501

\bibitem[{Olive {et~al.}(2011)Olive, Peloso, \& Uzan}]{wall}
Olive, K.~A., Peloso, M., \& Uzan, J.-P. 2011, Phys. Rev. D, { 83}, 043509

\bibitem[{Olive \& Pospelov(2002)}]{xiN}
Olive, K.~A. \& Pospelov, M. 2002, Phys. Rev. D, { 65}, 085044

\bibitem[{Padmanabhan {et~al.}(2012)Padmanabhan, Xu, Eisenstein, Scalzo,
  Cuesta, {et~al.}}]{Padmanabhan:2012hf}
Padmanabhan, N., Xu, X., Eisenstein, D.~J., {et~al.} 2012, \mnras, 427, 2132

\bibitem[{Peik {et~al.}(2008)Peik, Lipphardt, Schnatz, Tamm, Weyers, \&
  Wynands}]{peik}
Peik, E., Lipphardt, B., Schnatz, H., {et~al.} 2008, in The Eleventh Marcel
  Grossmann Meeting on General Relativity, ed. H.~Kleinert, R.~Jantzen, \&
  R.~Ruffini (Singapore; Hackensack, NJ: World Scientific), 941--951

\bibitem[{P\'equignot {et~al.}(1991)P\'equignot, Petitjean, \&
  Boisson}]{pequignot}
P\'equignot, D., Petitjean, P., \& Boisson, C. 1991, A\&A, 251, 680

\bibitem[{Pisanti {et~al.}(2008)Pisanti, Cirillo, Esposito, Iocco, Mangano,
  {et~al.}}]{Pisanti:2007hk}
Pisanti, O., Cirillo, A., Esposito, S., {et~al.} 2008, Comput.Phys.Commun.,
  178, 956

\bibitem[{Pitrou {et~al.}(2008)Pitrou, Uzan, \& Bernardeau}]{Pitrou08}
Pitrou, C., Uzan, J.-P., \& Bernardeau, F. 2008, Phys.Rev., D78, 063526

\bibitem[{{\sorthelp{Planck Collaboration 2013A}}{Planck Collaboration
  I}(2014)}]{planck2013-p01}
{\sorthelp{Planck Collaboration 2013A}}{Planck Collaboration I}. 2014, \aap, in
  press

\bibitem[{{\sorthelp{Planck Collaboration 2013L}}{Planck Collaboration
  XII}(2014)}]{planck2013-p06}
{\sorthelp{Planck Collaboration 2013L}}{Planck Collaboration XII}. 2014, \aap,
  in press

\bibitem[{{\sorthelp{Planck Collaboration 2013O}}{Planck Collaboration
  XV}(2014)}]{planck2013-p08}
{\sorthelp{Planck Collaboration 2013O}}{Planck Collaboration XV}. 2014, \aap,
  in press

\bibitem[{{\sorthelp{Planck Collaboration 2013P}}{Planck Collaboration
  XVI}(2014)}]{planck2013-p11}
{\sorthelp{Planck Collaboration 2013P}}{Planck Collaboration XVI}. 2014, \aap,
  in press

\bibitem[{{\sorthelp{Planck Collaboration 2013Q}}{Planck Collaboration
  XVII}(2014)}]{planck2013-p12}
{\sorthelp{Planck Collaboration 2013Q}}{Planck Collaboration XVII}. 2014, \aap,
  in press

\bibitem[{{\sorthelp{Planck Collaboration 2013ZB}}{Planck Collaboration
  XXVII}(2014)}]{planck2013-pipaberration}
{\sorthelp{Planck Collaboration 2013ZB}}{Planck Collaboration XXVII}. 2014,
  \aap, in press

\bibitem[{Prunet {et~al.}(2005)Prunet, Uzan, Bernardeau, \&
  Brunier}]{prunet2005}
Prunet, S., Uzan, J.-P., Bernardeau, F., \& Brunier, T. 2005, Phys. Rev. D,
  {71}, 083508

\bibitem[{{Reichardt} {et~al.}(2012){Reichardt}, {Shaw}, {Zahn}, {Aird},
  {Benson}, {Bleem}, {Carlstrom}, {Chang}, {Cho}, {Crawford}, {Crites}, {de
  Haan}, {Dobbs}, {Dudley}, {George}, {Halverson}, {Holder}, {Holzapfel},
  {Hoover}, {Hou}, {Hrubes}, {Joy}, {Keisler}, {Knox}, {Lee}, {Leitch},
  {Lueker}, {Luong-Van}, {McMahon}, {Mehl}, {Meyer}, {Millea}, {Mohr},
  {Montroy}, {Natoli}, {Padin}, {Plagge}, {Pryke}, {Ruhl}, {Schaffer},
  {Shirokoff}, {Spieler}, {Staniszewski}, {Stark}, {Story}, {van Engelen},
  {Vanderlinde}, {Vieira}, \& {Williamson}}]{Reichardt:12}
{Reichardt}, C.~L., {Shaw}, L., {Zahn}, O., {et~al.} 2012, \apj, 755, 70

\bibitem[{Riazuelo \& Uzan(2002)}]{cmb-G1}
Riazuelo, A. \& Uzan, J.-P. 2002, Phys. Rev. D, {66}, 023525

\bibitem[{Riess {et~al.}(2011)Riess, Macri, Casertano, Lampeitl, Ferguson,
  {et~al.}}]{Riess:2011yx}
Riess, A.~G., Macri, L., Casertano, S., {et~al.} 2011, Astrophys.J., 730, 119

\bibitem[{Rocha {et~al.}(2004)Rocha, Trotta, Martins, \&
  Melchiorri}]{cmb-rocha}
Rocha, G., Trotta, R., Martins, C., \& Melchiorri, A. e.~a. 2004, MNRAS, 32, 20

\bibitem[{Rosenband {et~al.}(2008)Rosenband, Hume, Schmidt, Chou, Brusch,
  Lorini, Oskay, Drullinger, Fortier, Stalnaker, Diddams, Swann, Newbury,
  Itano, Wineland, \& Bergquist}]{rosenband}
Rosenband, T., Hume, D., Schmidt, P., {et~al.} 2008, Science, 319, 1808

\bibitem[{Savedoff(1956)}]{savedoff}
Savedoff, M. 1956, Nature, 178, 688

\bibitem[{Scoccola {et~al.}(2008)Scoccola, Landau, \& Vucetich}]{scoccola}
Scoccola, C., Landau, S., \& Vucetich, H. 2008, Phys. Lett. B, 669, 212

\bibitem[{Scoccola {et~al.}(2009)Scoccola, Landau, \& Vucetich}]{scoccola2}
Scoccola, C., Landau, S., \& Vucetich, H. 2009, Mem. Soc. Astron. Ital., 80,
  814

\bibitem[{Scoccola(2009)}]{scoccolaPHD}
Scoccola, C.~G. 2009, PhD. thesis, {\tt arXiv:0906.0329}

\bibitem[{{Sc{\'o}ccola} {et~al.}(2013){Sc{\'o}ccola}, {S{\'a}nchez},
  {Rubi{\~n}o-Mart{\'{\i}}n}, {G{\'e}nova-Santos}, {Rebolo}, {Ross},
  {Percival}, {Manera}, {Bizyaev}, {Brownstein}, {Ebelke}, {Malanushenko},
  {Malanushenko}, {Oravetz}, {Pan}, {Schneider}, \& {Simmons}}]{scoccola2013}
{Sc{\'o}ccola}, C.~G., {S{\'a}nchez}, A.~G., {Rubi{\~n}o-Mart{\'{\i}}n}, J.~A.,
  {et~al.} 2013, \mnras, 434, 1792

\bibitem[{Seager {et~al.}(1999)Seager, Davelov, \& Scott}]{recfast}
Seager, S., Davelov, D.~D., \& Scott, D. 1999, ApJ, {523}, L1

\bibitem[{{Seager} {et~al.}(2000){Seager}, {Sasselov}, \&
  {Scott}}]{recfast_long}
{Seager}, S., {Sasselov}, D.~D., \& {Scott}, D. 2000, \apjs, 128, 407

\bibitem[{Shlyakhter(1976)}]{shly}
Shlyakhter, A. 1976, Nature, 264, 340

\bibitem[{{Sievers} {et~al.}(2013){Sievers}, {Hlozek}, {Nolta}, {Acquaviva},
  {Addison}, {Ade}, {Aguirre}, {Amiri}, {Appel}, {Barrientos}, {Battistelli},
  {Battaglia}, {Bond}, {Brown}, {Burger}, {Calabrese}, {Chervenak}, {Crichton},
  {Das}, {Devlin}, {Dicker}, {Bertrand Doriese}, {Dunkley}, {D{\"u}nner},
  {Essinger-Hileman}, {Faber}, {Fisher}, {Fowler}, {Gallardo}, {Gordon},
  {Gralla}, {Hajian}, {Halpern}, {Hasselfield}, {Hern{\'a}ndez-Monteagudo},
  {Hill}, {Hilton}, {Hilton}, {Hincks}, {Holtz}, {Huffenberger}, {Hughes},
  {Hughes}, {Infante}, {Irwin}, {Jacobson}, {Johnstone}, {Baptiste Juin},
  {Kaul}, {Klein}, {Kosowsky}, {Lau}, {Limon}, {Lin}, {Louis}, {Lupton},
  {Marriage}, {Marsden}, {Martocci}, {Mauskopf}, {McLaren}, {Menanteau},
  {Moodley}, {Moseley}, {Netterfield}, {Niemack}, {Page}, {Page}, {Parker},
  {Partridge}, {Plimpton}, {Quintana}, {Reese}, {Reid}, {Rojas}, {Sehgal},
  {Sherwin}, {Schmitt}, {Spergel}, {Staggs}, {Stryzak}, {Swetz}, {Switzer},
  {Thornton}, {Trac}, {Tucker}, {Uehara}, {Visnjic}, {Warne}, {Wilson},
  {Wollack}, {Zhao}, \& {Zunckel}}]{sievers}
{Sievers}, J.~L., {Hlozek}, R.~A., {Nolta}, M.~R., {et~al.} 2013, \jcap, 10, 60

\bibitem[{Sigurdson {et~al.}(2003)Sigurdson, Kurylov, \&
  Kamionkowski}]{Sigurdson}
Sigurdson, K., Kurylov, A., \& Kamionkowski, M. 2003, Phys.Rev., D68, 103509

\bibitem[{{Silk}(1968)}]{silk68}
{Silk}, J. 1968, \apj, 151, 459

\bibitem[{Srianand {et~al.}(2004)Srianand, Chand, Petitjean, \& Aracil}]{vlt1}
Srianand, R., Chand, H., Petitjean, P., \& Aracil, B. 2004, Phys. Rev. Lett.,
  92, 121302

\bibitem[{Srianand {et~al.}(2007)Srianand, Chand, Petitjean, \& Aracil}]{vlt2}
Srianand, R., Chand, H., Petitjean, P., \& Aracil, B. 2007, Phys. Rev. Lett.,
  99, 239002

\bibitem[{Stefanecsu(2007)}]{cmb3-2007}
Stefanecsu, P. 2007, NewA, 12, 635

\bibitem[{Uzan(2003)}]{jpu-revue}
Uzan, J.-P. 2003, Rev. Mod. Phys., { 75}, 403

\bibitem[{Uzan(2007)}]{ugrg}
Uzan, J.-P. 2007, Gen. Rel. Grav., { 39}, 307

\bibitem[{Uzan(2011)}]{jpu-llr}
Uzan, J.-P. 2011, Living Rev. Rel., { 14}, 2

\bibitem[{Webb {et~al.}(2011)Webb, King, Murphy, Flambaum, Carswell,
  {et~al.}}]{dipole1}
Webb, J., King, J., Murphy, M., {et~al.} 2011, Phys. Rev. Lett., { 107}, 191101

\bibitem[{Webb {et~al.}(2001)Webb, Murphy, Flambaum, \& Dzuba}]{webb01}
Webb, J., Murphy, M., Flambaum, V., \& Dzuba, V. e.~a. 2001, Phys. Rev. Lett.,
  87, 091301

\bibitem[{Wilkinson(1958)}]{wilk}
Wilkinson, D. 1958, Philos. Mag., 3, 582

\bibitem[{{Will}(1981)}]{will81}
{Will}, C.~M. 1981, {Theory and experiment in gravitational physics} (Cambridge
  University Press)

\bibitem[{{Wong} {et~al.}(2008){Wong}, {Moss}, \& {Scott}}]{wong2008}
{Wong}, W.~Y., {Moss}, A., \& {Scott}, D. 2008, \mnras, 386, 1023

\bibitem[{{Zahn} \& {Zaldarriaga}(2003)}]{zahn}
{Zahn}, O. \& {Zaldarriaga}, M. 2003, \prd, 67, 063002

\bibitem[{{Zaldarriaga} \& {Harari}(1995)}]{zaldarriaga95}
{Zaldarriaga}, M. \& {Harari}, D.~D. 1995, \prd, 52, 3276

\end{thebibliography}
